\documentclass[final,12p]{article}
\usepackage[utf8]{inputenc}
\usepackage{amsmath, amssymb, amsthm}
\usepackage{graphicx}
\usepackage{authblk}
\usepackage{tabularx}
\usepackage[hidelinks]{hyperref}
\usepackage{subcaption}
\usepackage{cite}
\usepackage{newtxtext}
\graphicspath{{./Images/}}
\usepackage{multirow}
\usepackage{verbatim}
\usepackage{wrapfig}
\usepackage{parskip}
\usepackage{txfonts}
\usepackage{csquotes}
\usepackage[numbers, square]{natbib}
\usepackage[top=2cm, bottom=2cm, left=2cm, right=2cm]{geometry}

\newcommand{\blfootnote}[1]{%
  \begingroup
  \renewcommand\thefootnote{}\footnote{#1}%
  \addtocounter{footnote}{-1}%
  \endgroup
}

\title{Flow-induced vibration of an array of circular cylinders in a square enclosure due to thermal buoyancy}
\author[1,2]{Amit Pramod Joshi}
\author[1]{Nishit Gupta}
\author[1]{Unnat Agarwal}
\author[1,*]{Vaibhav Joshi}
\affil[1]{Department of Mechanical Engineering, BITS Pilani, K.~K.~Birla Goa Campus, Zuarinagar, Goa 403726, India}
\affil[2]{Thermax Ltd., Pune, India}
\date{}

\begin{document}

\maketitle
\blfootnote{*Corresponding author: vaibhavj@goa.bits-pilani.ac.in}

\vspace{-1.5cm}
\begin{abstract}
Flow-induced oscillations of heated cylinder arrays in a square enclosure driven solely by natural convection are investigated in the current study at Prandtl number of 7 and Raleigh number of $10^7$ (based on side of square enclosure). This is accomplished by a coupled thermal-fluid-structure-interaction formulation based on body-fitted arbitrary Lagrangian-Eulerian framework, implemented using variational finite element methodology. Study on stationary array shows asymmetry in the $3\times 3$ configuration, along with a non-dimensional wall-gap plume oscillation frequency of 0.183 on the left column cylinders. When elastically mounted, these cylinders synchronize and lock-in with the frequency at reduced velocity of 15.7. The remaining cylinders undergo broadband buffeting that grows monotonically with reduced velocity and exceeds the synchronized response. The $4\times 4$ array shows symmetric behavior with no synchronization. The response is observed to be forced buffeting for all the cylinders, and dominated by the slow plume forcing for the inner column cylinders. The motion of the cylinders also influence the heat transfer characteristics. The $3\times 3$ array shows negligible change in heat transfer for the whole array, while measurable enhancement is observed for $4\times 4$ array.
\end{abstract}


\section{Introduction}
Fluid motion driven purely by density differences due to temperature variation as a consequence of buoyancy results in natural convection. It is broadly of two types, viz., external and internal. External natural convection occurs between a solid surface and the surrounding fluid where the fluid is unbounded and unconfined, for example, oceanic air circulations, atmospheric air interacting with large structures, to name a few. On the other hand, internal natural convection is observed within enclosed or semi-enclosed spaces where the boundaries significantly influence the fluid dynamics and heat transfer. Examples include steam boilers, heat exchangers for air conditioning and refrigeration, pin–fin type heat sinks for electronics cooling and solar heating systems. Buoyancy-driven natural convection can also be  categorized into two types: flow over vertical surfaces, where the temperature gradient is typically orthogonal to the direction of gravity \cite{Fan2021, Guha2017}, as observed in various heated vertical enclosures, and flow over horizontal surfaces, where the temperature gradient aligns with the direction of gravity, i.e. the Rayleigh–Bénard convection problem \cite{Bodenschatz2000}. These studies are imperative for the design and optimization of various thermal devices, which has motivated many researchers to explore them from both numerical and experimental perspectives. Heat transfer and fluid dynamics under natural convection are influenced when a solid body comes in the path of the fluid flow, which is often one of the methods of enhancing heat transfer. A comprehensive review of buoyancy-driven convection with and without internal bodies has been carried out by Pandey et al. \cite{PANDEY2019762}. The geometry, number and arrangement of these bodies significantly influence the convective heat transfer. Cylindrical bodies in thermal systems, owing to their balance of simplicity and efficiency, have drawn considerable attention from researchers. Their geometrical properties allow easier analysis and optimization of thermal systems. 

One of the critical applications of such systems is shell-and-tube heat exchangers and boilers, which consist of an array of tubes immersed in a fluid such as water or air. The thermal and fluid flow behavior around cylinder array arrangements immersed in a fluid exhibits plume-type phenomena under natural convection, which are called buoyant plumes. These buoyancy-driven thermal plumes originate at the surface at higher temperature (typically hot cylinder arrays in heat exchangers or boilers) and traverse opposite to the direction of gravity. In case of confined domain, the plumes are redirected when they reach the top and result in continuous circulation inside the enclosure. The interaction between these plumes with multiple bodies in the cylindrical array can be quite complex, and may result in flow-induced vibrations (FIV) of the array system. It is therefore essential to understand the plume interactions to assess the forces exerted on the bodies and their influence on the response of the cylindrical arrays, purely from natural convection. Furthermore, the variation in the overall heat transfer in the vibrating cylindrical arrays is also important. 

The parameters that principally affect convective heat transfer and the nature of the buoyant plumes are the Rayleigh number and the geometry, number and arrangement of the cylinders. The Rayleigh number is a non-dimensional number that measures the strength of buoyancy-driven natural convection relative to the viscous and thermal conduction effects. It is one of the main parameters used to predict the dominance of convection in the heat transfer as well as classify the mode of natural convection as laminar or turbulent. Fiscaletti et al. \cite{FISCALETTI2013619} demonstrated a critical Rayleigh number of $5.88\times 10^4$ above which the plume generated from a single cylinder begins to oscillate. Eckert and Soehngen \cite{Eckert1948StudiesOH} investigated the interaction of the developing thermal plumes and thermal boundary layers of two vertically aligned cylinders in an unconfined domain. They showed that the plume generated from the lower cylinder could either enhance or diminish the heat transfer performance of the upper cylinder depending on the spacing between the cylinders. Sparrow and Niethammer \cite{Sparrow1981} studied natural convection of a vertical cylinder pair in air. They found that the lower cylinder was unaffected by the upper cylinder in all cases. This result was corroborated by Reymond et al. \cite{REYMOND20081702}. Cianfrini et al. \cite{CIANFRINI2006923} found a minimum critical spacing below which the upper cylinder decreases the heat transfer performance of the lower cylinder. Persoons et al. \cite{PERSOONS20115163} investigated natural convection heat transfer and fluid dynamics for a pair of horizontal cylinders arranged inline to the direction of gravity. Generally, a heat transfer enhancement of the upper cylinder was observed in comparison to a single cylinder. Their simultaneous particle image velocimetry and heat transfer data demonstrated two separate alternating flow regimes around the upper cylinder. Pelletier et al. \cite{PELLETIER2016693} numerically studied the interaction between the two buoyant plumes and the relationship between unsteady fluid flow and heat transfer enhancement. They analyzed the variation of the Nusselt number on the upper cylinder at different Rayleigh numbers and quantified the non-dimensional frequency of the plume shedding. 
Corcione et al. \cite{Corcione2008} also investigated the thermal characteristics of cylinders by changing the alignment angle from vertically aligned to horizontally aligned. Their observations reveal that the contribution of the chimney effect increases and that of the plume effect decreases. Various studies \cite{Yuncu1994, CHOUIKH1999329, Park01071988, SADEGHIPOUR1994103, PAYKOC1991362, CHOUIKH200065, GRAFSRONNINGEN20125552, CIANFRINI20192607} conducted on columns of cylinders reveal that, above a minimum spacing, the bottom-most cylinder behaves close to a single cylinder, the heat transfer performance increases up to an optimum spacing, and closer spacing reduces the performance relative to a single cylinder.

Beyond cylinders subjected to unconfined natural convection, there have been several studies focusing on single as well as multiple cylinder arrangements inside a square cavity \cite{KIM20081888, PARK2013295, Mukherjee18102025, Yoon01112012, CHO2017307, PARK2015490, MUN2017755, MUN20181365, Xu2025}, thus modeling convection in a confined enclosure. Here, the Rayleigh number is defined based on the enclosure size instead of the cylinder diameter. Almost all the numerical studies employed an immersed boundary approach for their fluid modeling. For a single cylinder, Kim et al. \cite{KIM20081888} moved a heated cylinder of radius $0.2L$ along the vertical centreline of a cooled square enclosure of side $L$, varying Rayleigh number in the range $10^3 \leq Ra \leq 10^6$. The flow settled to a steady state, symmetric about the vertical centreline for all the Rayleigh numbers. At low Rayleigh number, where conduction dominates, the position of the cylinder alone decided the flow pattern. 
At high Rayleigh number, where convection dominates, each side held a single circulation region regardless of position, but the plume rising from the cylinder created secondary vortices above it, joined by the tertiary vortices at $Ra=10^6$. 
Studies of circular cylinder with tilted square enclosure were conducted in \cite{PARK2013295} and those for a semi-circular cylinder inside a circular cavity was studied in \cite{Mukherjee18102025}.
Yoon et al. \cite{Yoon01112012} placed two equal cylinders one above the other, centred in the lower and upper halves of the enclosure, and varied their radii for $10^3 \leq Ra \leq 10^5$. 
Cho et al. \cite{CHO2017307} kept two cylinders of radius $0.1L$ and aligned them vertically, horizontally and diagonally, with spacings of $0.1L$ to $0.5L$ and $10^3 \leq Ra \leq 10^6$. At the closest spacing, the interaction between the cylinders raised the heat transfer of the horizontal and diagonal pairs but lowered it for the vertical pair, and at $Ra=10^6$ the closely spaced horizontal and diagonal pairs resulted in unsteady flow. For four cylinders of radius $0.1L$ on a square grid, Park et al. \cite{PARK2015490} found the flow to be unsteady at $Ra=10^6$ for the considered spacings. They showed that the plumes rising from the lower cylinders reduce the heat transfer from the upper cylinders, while the upper cylinders barely affect the lower. When the same four cylinders were arranged as a diamond, Mun et al. \cite{MUN2017755, MUN20181365} found the flow at $Ra=10^6$ to be unsteady only at intermediate spacings, returning to a steady state when the cylinders were either closer together or farther apart. Xu and Choi \cite{Xu2025} extended these square arrays to as many as 36 cylinders for $10^5 \leq Ra \leq 10^8$ and identified steady, periodic and chaotic regimes. Arrays with an odd number of cylinders per side resulted in inherently asymmetric flow patterns, and the flow turned periodic at $Ra=10^6$ while the flow for even-count arrays stayed steady. The heat transfer also increased with the Rayleigh number. Therefore, in the enclosed configuration, arrangement, symmetry and orientation were found to control the flow behavior.
It is evident that the flow and heat transfer characteristics of a cylinder array cannot be predicted by simple superposition of single-cylinder behavior. Strong interactions occur between the developing thermal plumes and thermal boundary layers of adjacent cylinders in the confined enclosure, directly impacting the flow and heat transfer. All of the above studies, however, treat the cylinders as stationary, whereas the tubes of boilers and heat exchangers are elastic structures that respond to the fluctuating forces exerted on them by the surrounding natural convection.

Uniform cross-flow interacting with stationary cylinder array has been extensively investigated in the literature \cite{HOSSEINI2020103103, KRISHNACHAITANYA2024112969, Fornarelli2017}. An array therefore has dynamics of its own that cannot be recovered by extending the reasoning of single isolated cylinders. Fornarelli et al \cite{Fornarelli2017} scaled the problem to a genuine array, taking a single in-line row of six isothermal cylinders aligned with gravity at Reynolds number $Re=100$, Prandtl number $Pr=0.7$ and Richardson number $-1 \leq Ri \leq 1$, with non-dimensional centre-to-centre spacings of $3.6$ and $4$, and found several transitions in the flow pattern as the buoyancy changed from aiding to opposing. 
Flow-induced oscillations of tube and cylinder arrays have been thoroughly investigated when the array is subjected to a cross-flow (no thermal coupling \cite{Zhao2012, Han2023, HOSSEINI2022108922, Tang2019, Wang2019, ZHAO2023112223, TANG2020107156} and with mixed convection \cite{Islam2023, Garg2020, Zou2022, Song2022, Ding2023}). For cross-flow, Weaver and Fitzpatrick \cite{Weaver1988} established the engineering framework for this problem, classifying the vibrations into four mechanisms: turbulent buffeting, vortex shedding, acoustic resonance and fluid-elastic instability. The first three generate bounded vibration amplitudes, so that wear and tear accumulates slowly and can result in the fatigue failure of the tubes. Fluid-elastic instability is qualitatively different. Beyond a critical flow velocity a tube extracts energy from the flow faster than damping can dissipate it, the amplitude grows rapidly and failure can follow within a short operating period. The mechanisms of fluid-elastic instabilities in cylinder arrays have been discussed in \cite{Paidoussis_book}. Hosseini et al. \cite{HOSSEINI2022108922} extended their tandem row to six equi-spaced but independent elastically mounted cylinders free to move transversely at $Re=200$. The convective instabilities identified in rigid rows persist and explain the response at moderate to high stiffness, while at low stiffness the row behaves more like a single flexible structure that exhibits waves. Beyond these studies, array-scale FIV of tube bundles has been pursued almost entirely in the turbulent regime. Tang et al. \cite{Tang2019} coupled a high-order fluid-structure scheme to a normal triangular tube array with one tube free to oscillate, and found that the displacement and velocity of that tube rise sharply once its natural frequency coincides with the dominant frequency of the flow.

Flow-induced vibrations of heated cylinder arrays subjected to uniform cross-flow have only recently begun to receive attention. Islam et al. \cite{Islam2023} simulated $3\times 3$ array of nine heated circular cylinders at $Re=100$ with six held stationary and three elastically supported in the transverse direction, considering forced convection alone with no buoyancy. They varied the transverse spacing, the cylinder diameters and the reduced velocity over $2 \leq U_r \leq 20$. The oscillation amplitudes of the elastically supported cylinders were lowest at the closest spacing and grew as the gap increased and vortex shedding developed, and unequal diameters helped heat transfer only at the closest spacing. Most significantly, the vibration altered the Nusselt number of the stationary cylinders as well as the moving ones. Thermal performance in an array therefore cannot be attributed to any single body, since vibration and heat transfer are entangled across the whole bundle. 
Three cylinders in an equilateral triangle arrangement was considered by \cite{Zou2022, Song2022, Ding2023} investigating effects of buoyancy in the direction of flow (aiding), buoyancy perpendicular to flow (cross), and buoyancy opposite of flow (opposing), respectively on the flow-induced vibrations.
Most industrial heat exchanger applications use an elastically mounted multiple-cylinder array geometry, and hence the study of their heat transfer characteristics is of utmost practical importance. Research has mainly focused on investigating flow-induced vibrations due to mixed or forced convection of a single cylinder or an array of cylinders. As per the knowledge of the authors, there exists no study which has investigated flow-induced vibrations of an array of elastically mounted cylinders in an enclosure.

The present study focuses on investigating the flow-induced vibrations and heat transfer characteristics of a circular cylinder array inside a square enclosure at Rayleigh number of $10^7$ (based on the side length of the square enclosure) and Prandtl number of $7$. The flow is purely buoyancy-driven with no externally imposed velocity. The cylinder array arrangement considered in the present study are $2\times 2$, $3 \times 3$ and $4\times 4$. Each of the cylinders is elastically mounted and is heated to a constant temperature above the surrounding fluid temperature. The numerical formulation employs a body-fitted arbitrary Lagrangian-Eulerian (ALE) framework which follows the cylinders as they displace under the action of the buoyancy-driven flow. First, the thermo-fluid characteristics of the stationary array are studied to identify the fluid forcing at high Prandtl number and isolate the fluid forcing frequencies. Thereafter, the added mass coefficients are computed for the cylinders which are submerged in a fluid such that their mass ratio is low. Finally, the elastically mounted cylinders in the $3\times 3$ and $4\times 4$ arrays are subjected to the buoyancy-induced fluid forcing and their response is investigated and analyzed.  

The rest of the article is organized as follows. Section \ref{num_formulation} describes the coupled thermal-fluid-structure interaction formulation consisting of governing equations and variational formulations as well as implementation details. The formulation has been benchmarked with the tests from the literature in Appendix \ref{benchmarking}. The description of the problem setup is given in Section \ref{problemDescp} where all the physical and geometrical parameters are defined. The stationary array of cylinders is studied first in Section \ref{sec:stat_array} to understand their thermo-fluid characteristics. The added mass coefficient is measured by carrying out the free decay test in Section \ref{sec:added_mass_damping}. The response of elastically mounted cylinders under the action of the buoyancy-driven flow are presented and analyzed in Sections \ref{sec:FSI-3by3} and \ref{sec:FSI-4by4} for $3\times 3$ and $4\times 4$ arrays, respectively. The effect of cylinder motion on the heat transfer is discussed in Section \ref{sec:effectNu}. The key findings of the work are then summarized in Section \ref{conclusion}.

\section{Numerical Framework}
\label{num_formulation}
In this section, the computational methodology utilized in the present work has been described. The coupled thermal-fluid-structure interaction framework solves the conservation of mass, momentum and energy equations in the moving-mesh reference frame. The structure is modelled using rigid body equation and its motion is tracked exactly in Lagrangian description. Nonlinear iterative relaxation of fluid forces is carried out for numerical stability for low mass ratio problems. The governing equations and their discretization using variational finite element method are presented along with the coupled implementation under the partitioned iterative approach.

\subsection{Flow modelling}
The incompressible Navier-Stokes equations under the Boussinesq approximation are solved to obtain a two-dimensional flow field around the moving rigid body. The governing equations in the arbitrary Lagrangian-Eulerian (ALE) framework for moving mesh in the fluid domain $\Omega^\mathrm{f}(t)$ can be written as follows:
\begin{align}
 \nabla \cdot \boldsymbol{v}^\mathrm{f} &= 0, \label{cont}\\
 \rho^\mathrm{f}\left(\frac{\partial \boldsymbol{v}^\mathrm{f}}{\partial t}\bigg|_{\hat{\boldsymbol{\chi}}}+\left(\boldsymbol{v}^\mathrm{f}-\boldsymbol{w}\right)\cdot\nabla \boldsymbol{v}^\mathrm{f}\right) &= \nabla \cdot \boldsymbol{\sigma}^\mathrm{f} - \rho^\mathrm{f}\boldsymbol{g}\beta(T - T_L), \label{mom}
\end{align}
where the fluid velocity and the mesh velocity are denoted by $\boldsymbol{v}^\mathrm{f}$ and $\boldsymbol{w}$, respectively. The time derivative is evaluated with the ALE referential coordinates $\hat{\boldsymbol{\chi}}$ held fixed in Eq. (\ref{mom}). The density of the fluid is $\rho^\mathrm{f}$ and $\boldsymbol{\sigma}^\mathrm{f}=-p\boldsymbol{I}+\mu^\mathrm{f}(\nabla \boldsymbol{v}^\mathrm{f}+(\nabla \boldsymbol{v}^\mathrm{f})^T)$ denotes the Cauchy stress tensor for the Newtonian fluid. The fluid pressure is given by $p$ and $\mu^\mathrm{f}$ is the dynamic viscosity of the fluid. The body force applied on the fluid consists of the contribution of buoyancy term where $\beta$ is the thermal expansion coefficient, $\boldsymbol{g}$ is the acceleration due to gravity and $T_L$ is the cold fluid (ambient) temperature. The governing equation is discretized with the help of variational multiscale based orthogonal subgrid stabilization technique \cite{CODINA20001579, CODINA20024295, CODINA20072413, Escobar2026}. Let $\mathcal{S}^{\boldsymbol{v}^\mathrm{f}}_h$ and $\mathcal{S}^p_h$ represent the function space of trial solution for velocity and pressure, respectively, and $\mathcal{V}^{\boldsymbol{v}^\mathrm{f}}_h$ and $\mathcal{V}^p_h$ denote the respective test function spaces. With the help of second-order backward difference approximation in time and $\Delta t$ denoting the time step, the variational statement for the flow equation can be written as: find $[\boldsymbol{v}_h^\mathrm{f,n+1}, p^{\mathrm{n+1}}_h]\in \mathcal{S}_h^{\boldsymbol{v}^\mathrm{f}}\times \mathcal{S}_h^{p}$ such that $\forall [\boldsymbol{\psi}_h^\mathrm{f}, q_h] \in \mathcal{V}_h^{\boldsymbol{v}^\mathrm{f}}\times \mathcal{V}_h^{p}$:
\begin{align}
    &\int_{\Omega^\mathrm{f}} \rho^\mathrm{f} \bigg( \frac{3\boldsymbol{v}^\mathrm{f,n+1}_h - 4\boldsymbol{v}^\mathrm{f,n}_h + \boldsymbol{v}^\mathrm{f,n-1}_h}{2\Delta t} + (\boldsymbol{v}^\mathrm{f,n+1}_h - \boldsymbol{w}_h)\cdot \nabla\boldsymbol{v}^\mathrm{f,n+1}_h \bigg) \cdot \boldsymbol{\psi}^\mathrm{f}_h ~d\Omega + \int_{\Omega^\mathrm{f}} \boldsymbol{\sigma}^\mathrm{f,n+1}_h : \nabla\boldsymbol{\psi}^\mathrm{f}_h ~d\Omega \notag \\
    & + \int_{\Omega^\mathrm{f}} (\nabla \cdot \boldsymbol{v}^\mathrm{f,n+1}_h) q_h ~d\Omega - \displaystyle\sum_{e=1}^{n_{el}} \int_{\Omega^e} \widetilde{\boldsymbol{v}}^{\mathrm{n+1}}_1\cdot (\rho^\mathrm{f} (\boldsymbol{v}^\mathrm{f,n+1}_h-\boldsymbol{w}_h)\cdot \nabla \boldsymbol{\psi}^\mathrm{f}_h)~ d\Omega - \displaystyle\sum_{e=1}^{n_{el}} \int_{\Omega^e} \widetilde{p}^{\mathrm{n+1}} (\nabla\cdot \boldsymbol{\psi}^\mathrm{f}_h) ~d\Omega  \notag \\
    & - \displaystyle\sum_{e=1}^{n_{el}} \int_{\Omega^e} \widetilde{\boldsymbol{v}}^{\mathrm{n+1}}_2\cdot (\nabla q_h) ~d\Omega = -\int_{\Omega^\mathrm{f}} (\rho^\mathrm{f} \boldsymbol{g}\beta(T^\mathrm{n+1}_h - T_L)) \cdot \boldsymbol{\psi}^\mathrm{f}_h ~d\Omega + \int_{\Gamma} (\boldsymbol{\sigma}^\mathrm{f,n+1}_h \boldsymbol{n})\cdot \boldsymbol{\psi}^\mathrm{f}_h ~d\Gamma, \label{NS-variational}
\end{align}
where $\boldsymbol{\psi}^\mathrm{f}_h$ and $q_h$ are the test functions for the momentum and continuity equations, respectively, $\widetilde{\boldsymbol{v}}_1^\mathrm{n+1}$, $\widetilde{\boldsymbol{v}}_2^\mathrm{n+1}$ and $\widetilde{p}^\mathrm{n+1}$ denote the subgrid scales in the variational multiscale framework. The velocity subscales are assumed to be dynamic and split into convective and pressure-gradient contributions as follows
\begin{align}
    \rho^\mathrm{f}\frac{\partial \widetilde{\boldsymbol{v}}_1}{\partial t} + \tau_1^{-1} \widetilde{\boldsymbol{v}}_1 &= - \rho^\mathrm{f}(\boldsymbol{v}^\mathrm{f,n+1}_h-\boldsymbol{w}_h)\cdot \nabla \boldsymbol{v}^\mathrm{f,n+1}_h + \mathcal{P}_h(\rho^\mathrm{f}(\boldsymbol{v}^\mathrm{f,n+1}_h-\boldsymbol{w}_h)\cdot \nabla \boldsymbol{v}^\mathrm{f,n+1}_h), \label{SGS1}\\
    \rho^\mathrm{f}\frac{\partial \widetilde{\boldsymbol{v}}_2}{\partial t} + \tau_1^{-1} \widetilde{\boldsymbol{v}}_2 &= - \nabla p^\mathrm{n+1}_h + \mathcal{P}_h(\nabla p^\mathrm{n+1}_h). \label{SGS2}
\end{align}
Here, the right-hand-side terms denote the orthogonal projections of the convective and pressure-gradient terms, and $\mathcal{P}_h(\cdot)$ is the $L^2$ projection in the discrete finite element space. Using the second-order backward difference time integration, Eqs. (\ref{SGS1}-\ref{SGS2}) can be written as
\begin{align}
    \bigg( \tau_1^{-1} + \frac{3\rho^\mathrm{f}}{2\Delta t} \bigg) \widetilde{\boldsymbol{v}}^{\mathrm{n+1}}_1 &= \rho^\mathrm{f} \bigg(\frac{4\widetilde{\boldsymbol{v}}^\mathrm{n}_1 - \widetilde{\boldsymbol{v}}^\mathrm{n-1}_1}{2\Delta t}\bigg) - \rho^\mathrm{f}(\boldsymbol{v}^\mathrm{f,n+1}_h-\boldsymbol{w}_h)\cdot \nabla \boldsymbol{v}^\mathrm{f,n+1}_h + \mathcal{P}_h(\rho^\mathrm{f}(\boldsymbol{v}^\mathrm{f,n+1}_h-\boldsymbol{w}_h)\cdot \nabla \boldsymbol{v}^\mathrm{f,n+1}_h),\label{SGS_Eqn1}\\
    \bigg( \tau_1^{-1} + \frac{3\rho^\mathrm{f}}{2\Delta t} \bigg) \widetilde{\boldsymbol{v}}^{\mathrm{n+1}}_2 &= \rho^\mathrm{f} \bigg(\frac{4\widetilde{\boldsymbol{v}}^\mathrm{n}_2 - \widetilde{\boldsymbol{v}}^\mathrm{n-1}_2}{2\Delta t}\bigg) - \nabla p^\mathrm{n+1}_h + \mathcal{P}_h(\nabla p^\mathrm{n+1}_h). \label{SGS_Eqn2}
\end{align}
The pressure subscale is assumed to be quasi-static and is defined as $\tau_2^{-1} \widetilde{p}^\mathrm{n+1} = -\nabla\cdot \boldsymbol{v}^\mathrm{f,n+1}_h + \mathcal{P}_h(\nabla\cdot\boldsymbol{v}^\mathrm{f,n+1}_h)$. The stabilization parameters $\tau_1$ and $\tau_2$ are computed as \cite{SHAKIB1991141, BROOKS1982199, TEZDUYAR1992221, FRANCA1992209}
\begin{align}
    \tau_1 &= \bigg[ \bigg(\frac{2\rho^\mathrm{f}}{\Delta t}\bigg)^2 + (\rho^\mathrm{f})^2(\boldsymbol{v}^\mathrm{f,n+1}_h-\boldsymbol{w}_h)\cdot \boldsymbol{G}(\boldsymbol{v}^\mathrm{f,n+1}_h-\boldsymbol{w}_h) + 36(\mu^\mathrm{f})^2 \boldsymbol{G}:\boldsymbol{G} \bigg]^{-1/2},\\
    \tau_2 &= \frac{1}{\mathrm{tr}({\boldsymbol{G}}) \tau_1},
\end{align}
where $\boldsymbol{G}$ is the element contravariant metric tensor.

The above subscale quantities $\widetilde{\boldsymbol{v}}_1^{\mathrm{n+1}}$, $\widetilde{\boldsymbol{v}}_2^{\mathrm{n+1}}$ and $\widetilde{p}^\mathrm{n+1}$ are substituted in Eq. (\ref{NS-variational}) and the equation is solved to get the discrete velocity $(\boldsymbol{v}^{\mathrm{f,n+1}}_h)$ and pressure $(p^\mathrm{n+1}_h)$ at a particular time step.

\subsection{Thermal modelling}
The fluid temperature $T$ is solved by the energy conservation equation in the ALE framework as
\begin{align}
    \rho^\mathrm{f}C_p\frac{\partial T}{\partial t}\bigg|_{\hat{\boldsymbol{\chi}}} + \rho^\mathrm{f}C_p\big( \boldsymbol{v}^\mathrm{f} - \boldsymbol{w} \big)\cdot\nabla T &= \nabla \cdot(k^\mathrm{f} \nabla T),
\end{align}
where $k^\mathrm{f}$ and $C_p$ are the thermal conductivity and specific heat capacity of the fluid, respectively. Consider the trial solution and test function spaces as $\mathcal{S}^T_h$ and $\mathcal{V}^T_h$, respectively, with the help of second-order accurate backward-difference approximation in time, the variational statement for the thermal equation can be written as: find $T^{\mathrm{n+1}}_h \in \mathcal{S}^T_h$ such that $\forall \phi_h\in \mathcal{V}^T_h$:
\begin{align}
    &\int_{\Omega^\mathrm{f}} \rho^\mathrm{f}C_p\bigg( \frac{3T^\mathrm{n+1}_h - 4T^\mathrm{n}_h+T^\mathrm{n-1}_h}{2\Delta t} + (\boldsymbol{v}^\mathrm{f,n+1}_h-\boldsymbol{w}_h)\cdot \nabla T^\mathrm{n+1}_h \bigg) \phi_h ~d\Omega + \int_{\Omega^\mathrm{f}} (k^\mathrm{f}\nabla T^\mathrm{n+1}_h)\cdot\nabla\phi_h ~d\Omega \notag \\
    -&\displaystyle\sum_{e=1}^{n_{el}}\int_{\Omega^e} \widetilde{T}^\mathrm{n+1}\bigg( \rho^\mathrm{f}C_p (\boldsymbol{v}^\mathrm{f,n+1}_h-\boldsymbol{w}_h)\cdot\nabla\phi_h + k^\mathrm{f}\nabla^2\phi_h \bigg) ~d\Omega = \int_\Gamma (k^\mathrm{f}\nabla T^\mathrm{n+1}_h\cdot \boldsymbol{n})\phi_h ~d\Gamma, \label{thermal_variational}
\end{align}
where $\phi_h$ is the test function for the equation and $\widetilde{T}^\mathrm{n+1}$ denotes the temperature subgrid scale which is assumed to be quasi-static and is defined as \cite{Codina_ThermalOSS, OSSES2021114099}
\begin{align}
    \tau_3^{-1} \widetilde{T}^\mathrm{n+1} = -\mathcal{R}_{T^\mathrm{n+1}} + \mathcal{P}_h(\mathcal{R}_{T^\mathrm{n+1}}),
\end{align}
where $\mathcal{R}_{T^\mathrm{n+1}}$ is the residual of the thermal equation and the stabilization parameter is computed as
\begin{align}
    \tau_3 &= \bigg[ (\rho^\mathrm{f}C_p)^2(\boldsymbol{v}^\mathrm{f,n+1}_h-\boldsymbol{w}_h)\cdot \boldsymbol{G}(\boldsymbol{v}^\mathrm{f,n+1}_h-\boldsymbol{w}_h) + 36(k^\mathrm{f})^2 \boldsymbol{G}:\boldsymbol{G} \bigg]^{-1/2}.
\end{align}

\subsection{Rigid body structural modelling}

The motion of the structure is governed by the two-degree of freedom rigid body translational equation:
\begin{align}
\boldsymbol{M}^\mathrm{s} \frac{\partial^2 \boldsymbol{\varphi}^\mathrm{s}}{\partial t^2}+ \boldsymbol{C}^\mathrm{s} \frac{\partial \boldsymbol{\varphi}^\mathrm{s}}{\partial t} + \boldsymbol{K}^\mathrm{s} (\boldsymbol{\varphi}^\mathrm{s} - \boldsymbol{X}_0) &= \boldsymbol{F}^\mathrm{s} + \boldsymbol{P}^\mathrm{s}, \label{MBD}
\end{align}
where $\boldsymbol{M}^\mathrm{s}$, $\boldsymbol{C}^\mathrm{s}$ and $\boldsymbol{K}^\mathrm{s}$ represent the mass, damping, and stiffness matrices, respectively. $\boldsymbol{\varphi}^\mathrm{s}(\boldsymbol{X}^\mathrm{s},t)$ is a one-to-one mapping function which maps the undeformed reference configuration $\Omega^\mathrm{s}$ with material coordinates $\boldsymbol{X}^\mathrm{s}$ of the structure at $t=0$ to the deformed configuration $\Omega^\mathrm{s}(t)$ at time $t$. The centre of mass of the rigid body has the location $\boldsymbol{X}_0$ in the reference configuration. The fluid force integrated over the surface of the rigid body is represented by $\boldsymbol{F}^\mathrm{s}$. Any external preloads on the rigid body are denoted by $\boldsymbol{P}^\mathrm{s}$.


The rigid body equation solves for the structural velocity $\boldsymbol{v}^\mathrm{s} = \partial\boldsymbol{\varphi}^\mathrm{s}/\partial t$. The displacement of the structure is given by $\boldsymbol{u}^\mathrm{s} = \boldsymbol{\varphi}^\mathrm{s} - \boldsymbol{X}_0$.
The above equation is discretized in time using second-order accurate Crank-Nicolson scheme, thus giving
\begin{align}
    \bigg( \frac{1}{\Delta t} \boldsymbol{M}^\mathrm{s} + \frac{1}{2} \boldsymbol{C}^\mathrm{s} + \frac{\Delta t}{4}\boldsymbol{K}^\mathrm{s}\bigg) \boldsymbol{v}^\mathrm{s,n+1} &= \boldsymbol{F}^\mathrm{s,n+1/2} + \bigg( \frac{1}{\Delta t} \boldsymbol{M}^\mathrm{s} - \frac{1}{2} \boldsymbol{C}^\mathrm{s} - \frac{\Delta t}{4}\boldsymbol{K}^\mathrm{s} \bigg) \boldsymbol{v}^\mathrm{s,n} - \boldsymbol{K}^\mathrm{s} \boldsymbol{u}^\mathrm{s,n}, \label{discrete-structure-eqn-1}\\
    \boldsymbol{u}^\mathrm{s,n+1} &= \frac{\Delta t}{2}(\boldsymbol{v}^\mathrm{s,n+1} + \boldsymbol{v}^\mathrm{s,n}) + \boldsymbol{u}^\mathrm{s,n}, \label{discrete-structure-eqn-2}
\end{align}
where the fluid force $\boldsymbol{F}^\mathrm{s,n+1/2} = (1/2)(\boldsymbol{F}^\mathrm{s,n+1} + \boldsymbol{F}^\mathrm{s,n})$. The rigid body in the computational domain is displaced based on the calculated displacement. 

\subsection{Conditions at the fluid-structure interface}
For the coupled fluid–structure system, both the no-slip condition and the continuity of traction forces must be satisfied at the interface $\Gamma^\mathrm{fs}$ as follows
\begin{align}
   \boldsymbol{v}^\mathrm{f,n+1} (\boldsymbol{\varphi}^\mathrm{s,n+1}(\boldsymbol{X}^\mathrm{s},t),t) &= \boldsymbol{v}^\mathrm{s,n+1}(\boldsymbol{X}^\mathrm{s},t), \forall \boldsymbol{X}^\mathrm{s} \in \Gamma^\mathrm{fs}\\
 \int_{\boldsymbol{\varphi}^\mathrm{s,n+1}(\gamma ,t)} \boldsymbol{\sigma}^\mathrm{f,n+1}(\boldsymbol{x},t)\cdot\boldsymbol{n}~ d\Gamma  + \boldsymbol{F}^\mathrm{s,n+1} &= 0 ~~~ \forall~ \gamma~ \subset ~\Gamma^\mathrm{fs}.
\end{align}
Here, $\boldsymbol{n}$ is the outer unit normal at the fluid–structure boundary, and $\gamma$ is an arbitrary segment of the overall interface $\Gamma^\mathrm{fs}$. $\boldsymbol{F}^\mathrm{s,n+1}$ is the contribution to the traction on the structure from $\gamma$ and $\boldsymbol{\varphi}^\mathrm{s,n+1}(\gamma,t)$ is the corresponding fluid part at time $t$.

\subsection{Moving mesh modelling}
In the moving mesh ALE framework, the fluid nodes are displaced by assuming the fluid mesh as a pseudo-elastic material. The fluid displacement denoted by $\boldsymbol{u}^\mathrm{f}$ is found by solving the following equation:
\begin{align}
    \nabla\cdot \boldsymbol{\sigma}^\mathrm{m} &= 0,
\end{align}
where $\boldsymbol{\sigma}^\mathrm{m}$ is the stress experienced by the fluid mesh due to the strain induced by the interface deformation of the structure, and is defined as $\boldsymbol{\sigma}^\mathrm{m} = (1+k_m)\bigg[ \nabla\boldsymbol{u}^\mathrm{f} + (\nabla\boldsymbol{u}^\mathrm{f})^T + (\nabla\cdot\boldsymbol{u}^\mathrm{f})\boldsymbol{I} \bigg]$. Here, $k_m$ represents a mesh stiffness variable chosen as a function of the element size to control the mesh distortion and is defined as $k_m = (\max |T_j|-\min |T_j|)/|T_j|$, where $T_j$ denotes the area of the $j$th element on the mesh in two-dimensions. 

Given the trial solution and test function space for the mesh equation as $\mathcal{S}^{\boldsymbol{u}^\mathrm{f}}_h$ and $\mathcal{V}^{\boldsymbol{u}^\mathrm{f}}_h$, respectively, the variational statement for the moving mesh equation can be written as: find $\boldsymbol{u}_h^\mathrm{f,n+1}\in \mathcal{S}_h^{\boldsymbol{u}^\mathrm{f}}$ such that $\forall \boldsymbol{\xi}_h^\mathrm{f} \in \mathcal{V}_h^{\boldsymbol{u}^\mathrm{f}}$:
\begin{align}
    \int_{\Omega^\mathrm{f}} \nabla\boldsymbol{\xi}^\mathrm{f}_h : \boldsymbol{\sigma}^\mathrm{m}_h ~d\Omega &= \int_{\Gamma} (\boldsymbol{\sigma}^\mathrm{m}_h \boldsymbol{n})\cdot\boldsymbol{\xi}^\mathrm{f}_h ~d\Gamma, \label{meshEqnVariational}
\end{align}
where $\boldsymbol{\xi}^\mathrm{f}_h$ is the test function. The boundary integral term on the right-hand-side of the above equation is zero as the fluid displacements at those boundaries satisfy the Dirichlet condition. Once the fluid nodal displacements are obtained, the mesh velocity of each fluid node is given by $\boldsymbol{w}_h = \partial \boldsymbol{u}^\mathrm{f}_h/\partial t = (3\boldsymbol{u}^\mathrm{f,n+1}_h-4\boldsymbol{u}^\mathrm{f,n}_h+\boldsymbol{u}^\mathrm{f,n-1}_h)/(2\Delta t)$. 

\subsection{Implementation details}
\begin{figure}
    \centering
    \includegraphics[width=\textwidth]{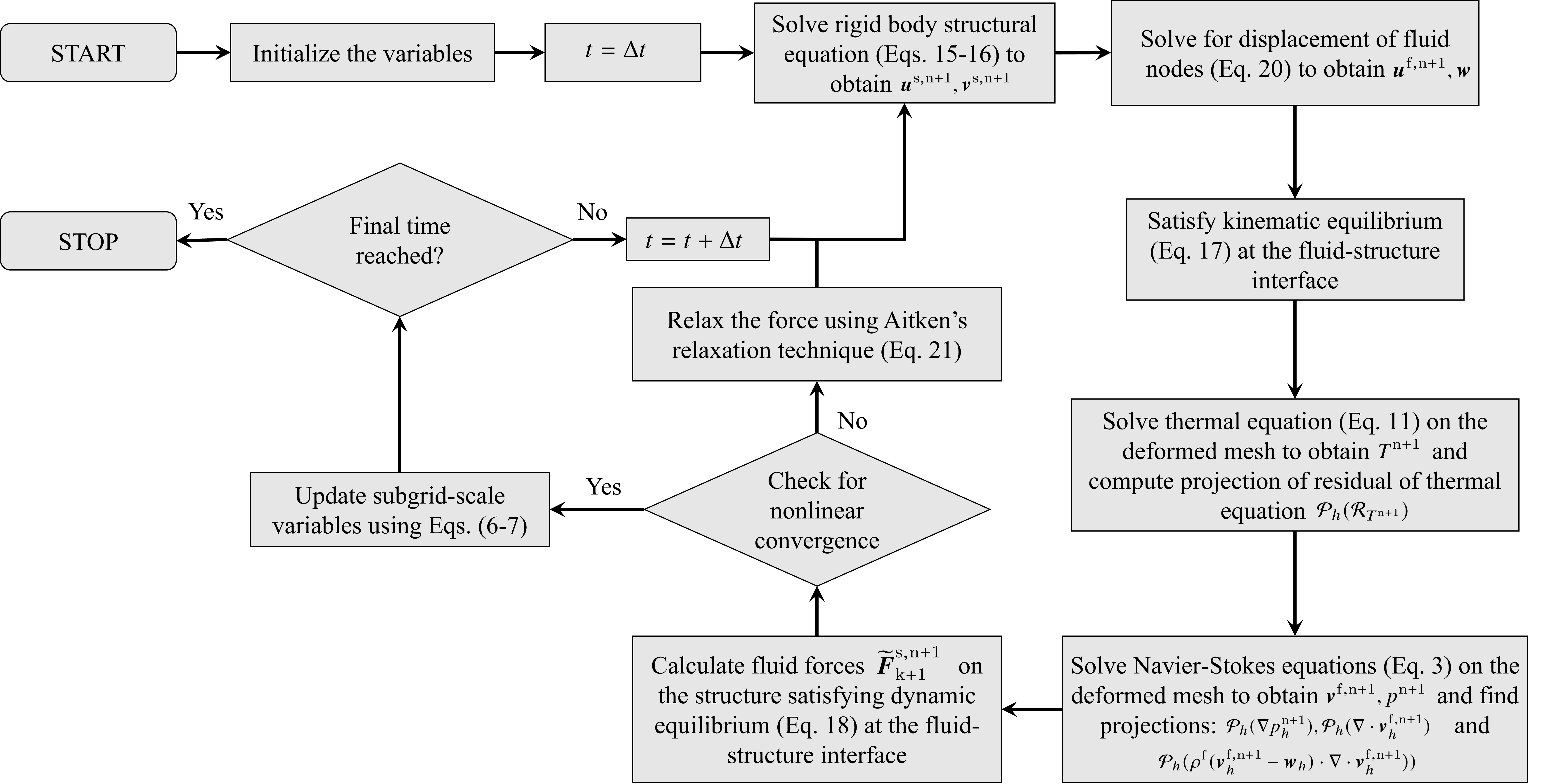}
    \caption{Schematic of the coupled thermal-fluid-structure interaction framework.}
    \label{fig:solver-schematic}
\end{figure}

The coupled thermal-fluid-structure interaction formulation is schematically depicted in Fig. \ref{fig:solver-schematic}. The coupled equations are solved iteratively through a partitioned approach, satisfying the boundary conditions at the fluid-structure interface. With the computed fluid force from the previous time step acting on the structure, the discretized structural equation (Eqs. \ref{discrete-structure-eqn-1}-\ref{discrete-structure-eqn-2}) is solved to give the structural velocity and displacements. The fluid nodes of the domain are then displaced by solving the mesh equation (Eq. (\ref{meshEqnVariational})) which evaluates the fluid node displacements and the mesh velocities. The kinematic condition at the fluid-structure interface is then satisfied by equating the fluid velocity at the interface with that of the computed structural velocity. Thereafter, thermal equation (Eq. (\ref{thermal_variational})) is solved to get the temperature at the displaced nodes. The flow equations (Eq. (\ref{NS-variational})) is solved on the deformed mesh with the updated temperature in the buoyancy source term, to get the fluid velocity and pressure. The fluid forces on the structural boundaries are subsequently computed using the updated fluid variables. Convergence is then checked to ensure that the coupled variables have achieved nonlinear convergence. If not, the fluid forces on the structure are relaxed using Aitken's technique before the structural equation is again solved and the process is repeated until convergence. The fluid equations are linearized using fixed point iterations where the convective velocity is evaluated at the previous iteration step.

For low mass ratio problems which may encounter convergence issues, the fluid force is relaxed with the help of Aitken's relaxation technique within the nonlinear iterations. In a nonlinear iteration $\mathrm{k}$, the fluid force is given as \cite{Kuttler_FSI, Irons1969}
\begin{align}
    \boldsymbol{F}^\mathrm{s,n+1}_\mathrm{k+1} &= \omega^\mathrm{n+1}_\mathrm{k} \widetilde{\boldsymbol{F}}^\mathrm{s,n+1}_\mathrm{k+1} + (1-\omega^\mathrm{n+1}_\mathrm{k})\boldsymbol{F}^\mathrm{s,n+1}_\mathrm{k},
\end{align}
where $\widetilde{\boldsymbol{F}}^\mathrm{s,n+1}_\mathrm{k+1}$ is the fluid force obtained after solving the Navier-Stokes equations (Eq. \ref{NS-variational}), and $\omega_\mathrm{k+1}$ is the dynamic relaxation factor given by
\begin{align}
    \omega^\mathrm{n+1}_\mathrm{k+1} &= -\omega^\mathrm{n+1}_\mathrm{k} \frac{\boldsymbol{r}_\mathrm{k+1}\cdot(\boldsymbol{r}_\mathrm{k+2}-\boldsymbol{r}_\mathrm{k+1})}{|\boldsymbol{r}_\mathrm{k+2}-\boldsymbol{r}_\mathrm{k+1}|^2}. \label{relaxationFactor}
\end{align}
Here, $\boldsymbol{r}_\mathrm{k+1} = \widetilde{\boldsymbol{F}}^\mathrm{s,n+1}_\mathrm{k+1} - \boldsymbol{F}^\mathrm{s,n+1}_\mathrm{k}$ is the residual of the forces computed after solving the flow equations. Note that two nonlinear iterations are required to compute the relaxation factor. For the first iteration, the relaxation factor is defined as $\omega_1 = \max(\omega^\mathrm{n+1}, 0.01)$ and is computed using Eq. (\ref{relaxationFactor}) for successive iterations.

Once the nonlinear convergence is achieved, the subgrid scale velocities are updated using Eqs. (\ref{SGS_Eqn1}-\ref{SGS_Eqn2}) and the solver proceeds to the next time step. All the required $L^2$ projections ($\mathcal{P}_h(\cdot)$) while solving the Navier-Stokes and thermal equations are evaluated using lumped mass matrix. The presented formulation has been extensively validated for a variety of problems by the works carried out by the authors \cite{Jaiman_2022a, Tandon2024}. It has also been benchmarked for relevant numerical problems to establish successful thermal-fluid-structure coupling, which has been presented rigorously in Appendix \ref{benchmarking}. 


\section{\textbf{Problem description}}
\label{problemDescp}
As the current study aims to investigate the vibrations in a circular cylinder array subjected to pure natural convection in an enclosure, a square cavity of side $L$ is considered consisting of $n_{cyl}$ circular cylinders arranged in an in-line array. To account for both the cavity confinement and array solidity, the governing fluid equations are non-dimensionalized using an aggregate length scale defined as $W = L\sqrt{V_f}$, where $V_f = V_\text{cyl}/V_\text{cavity}$ is the volume fraction giving the ratio of the volume occupied by the cylinders to that of the cavity. Here, $V_\text{cyl} = (\pi/4)D^2 n_\text{cyl}$ and $V_\text{cavity}=L^2$, thus giving $W = D\sqrt{\pi n_\text{cyl}}/2$. This scale inherently couples the tube count $n_\text{cyl}$ and cylinder diameter $D$, ensuring consistent scaling across different array configurations. The characteristic velocity for the natural convection is defined as $U_b = \sqrt{|\boldsymbol{g}|\beta\Delta T W}$, where $\Delta T = T_H - T_L$, where $T_H$ and $T_L$ are the hot and cold temperatures of the domain. This results in a time scale $W/U_b$. The dimensionless parameters governing the flow and thermal behavior are Prandtl number and Rayleigh number, respectively defined as
\begin{align}
    &Pr = \frac{\mu^\mathrm{f}}{\rho^\mathrm{f}\alpha},\quad  Ra = \frac{|\boldsymbol{g}|\beta\Delta T W^3 \rho^\text{f}}{\mu^\mathrm{f} \alpha},
\end{align}
where $\alpha = k^\mathrm{f}/(\rho^\mathrm{f}C_p)$ is the thermal diffusivity.

On the other hand, the structural equation is non-dimensionalized by the length scale local to the cylinders, $D$ with time scale same as $W/U_b$. Therefore, the resulting dimensionless numbers are mass ratio and reduced velocity, respectively written as
\begin{align}
    m^* = \frac{m}{\rho^\text{f}(\pi/4)D^2},\quad U_r = \frac{U_b}{f_{n,wet}D}.
\end{align}
where $f_{n,wet} = \frac{1}{2\pi}\sqrt{k/(m+m_a)}$ is the natural frequency of the elastically mounted structure in a fluid, $m_a$ is the added mass (measured by the decay test). No structural damping is considered in this study. For the current study, the volume fraction is chosen as $V_f = 0.2827$, $Pr = 7.0$, $Ra = 1.5\times 10^6$ ($Ra_L = 10^7$ based on the cavity length $L$), and $m^* = 2.0$. The reduced velocity is varied to investigate the lock-in/synchronization of the cylinder response with the plume frequency. Note that since the added mass of each cylinder may be different, each cylinder will have its own range of reduced velocities.

\begin{figure}
    \centering
    \begin{subfigure}{0.5\textwidth}
        \includegraphics[trim={1cm 0.1cm 2cm 1cm}, clip,width=\textwidth]{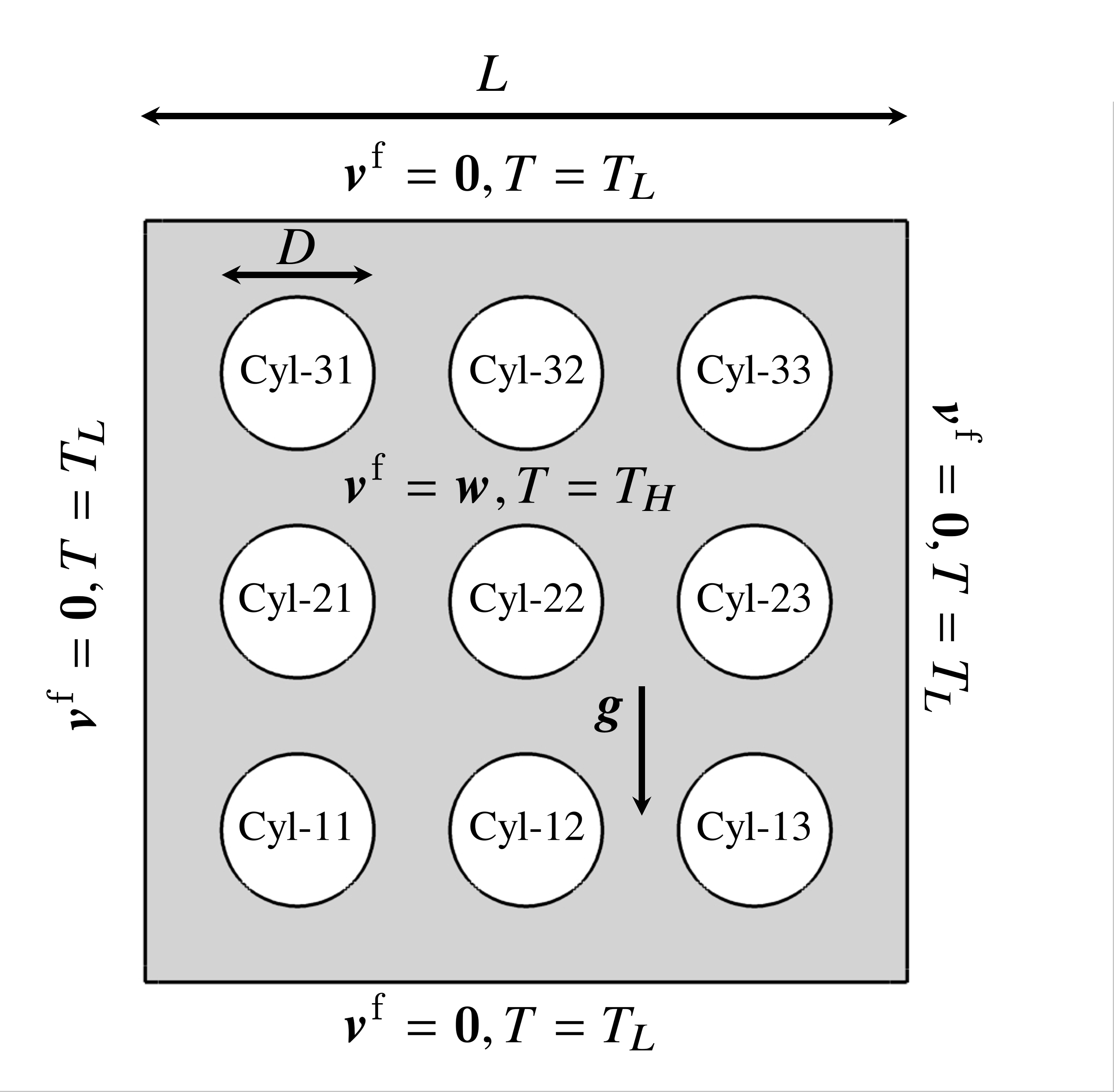}
        \caption{}
    \end{subfigure}%
    \begin{subfigure}{0.5\textwidth}
        \includegraphics[trim={1cm 0.1cm 2cm 1cm}, clip,width=\textwidth]{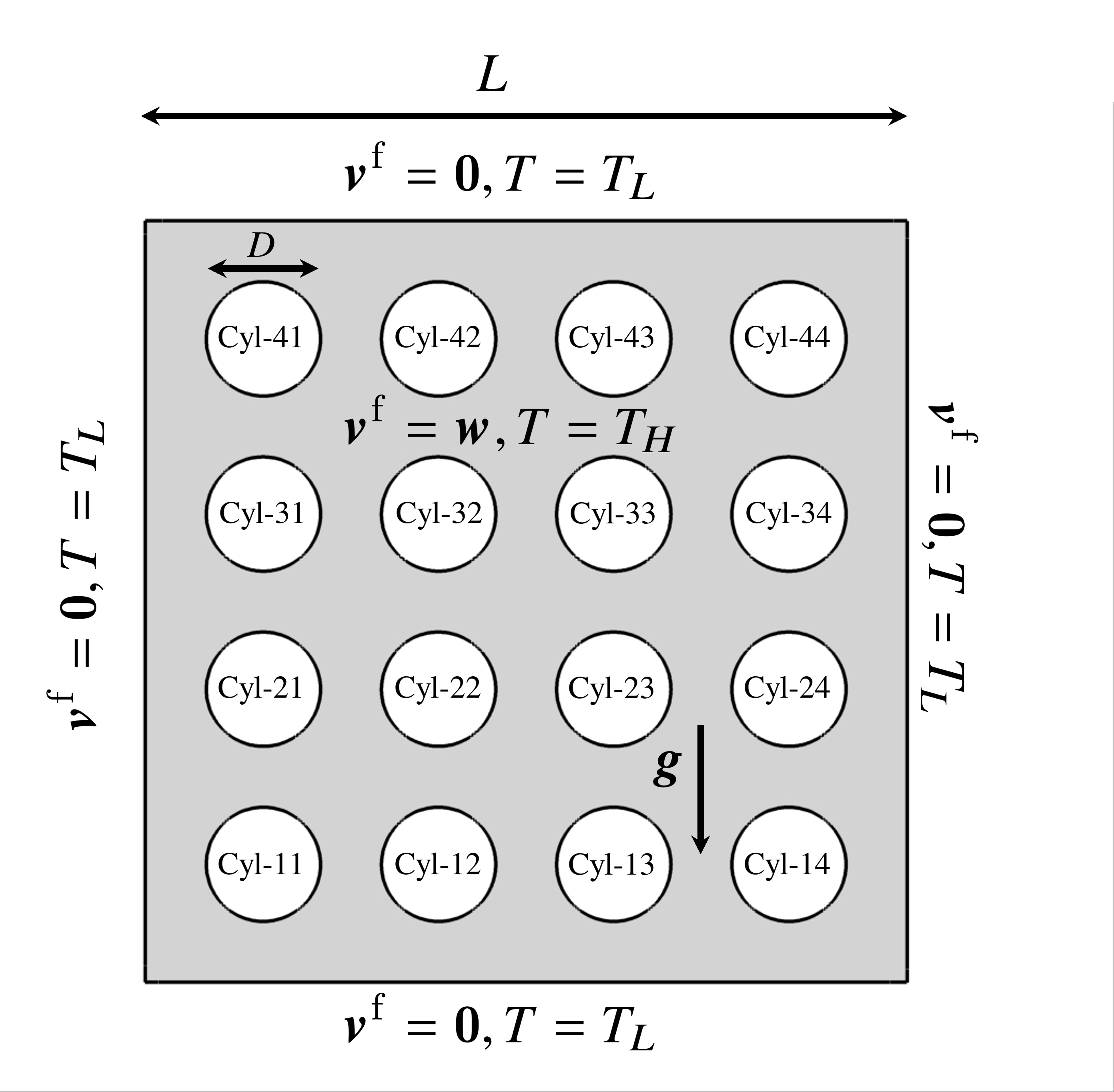}
        \caption{}
    \end{subfigure}
    \caption{Physical domain of the square cavity with cylinder array consisting of (a) $3\times 3$, and (b) $4\times 4$ cylinders. All the cylinders are elastically mounted individually and are free to oscillate in two degrees-of-freedom.}
    \label{fig:problem_description}
\end{figure}
The typical geometry of the array configuration is shown in Fig. \ref{fig:problem_description}. The centres of the cylinders are chosen such that the gap between the cylinders and the walls is equidistant, similar to the work by \cite{Xu2025}. Three array configurations are chosen $2\times 2$, $3\times 3$ and $4 \times 4$ which correspond to $D/L$ of $0.3$, $0.2$, and $0.15$, respectively. The centre-to-centre distance between inline cylinders is $0.434L$, $0.3L$, and $0.23L$, respectively. The enclosure walls are stationary with no-slip boundary condition and are at cold temperature $T_L$. The cylinders are heated to a constant temperature of $T_H$ and satisfy no-slip condition. The flow inside the cavity is purely buoyancy driven due to the temperature difference. The cylinders are identified as $\text{Cyl}$-$ij$ where $i$ and $j$ denote the row and column of the array starting from the bottom-left corner, respectively, as shown in Fig. \ref{fig:problem_description}.

\begin{figure}
    \centering
    \begin{subfigure}{0.5\textwidth}
        \includegraphics[trim={0.1cm 0.1cm 0.1cm 0.1cm}, clip,width=\textwidth]{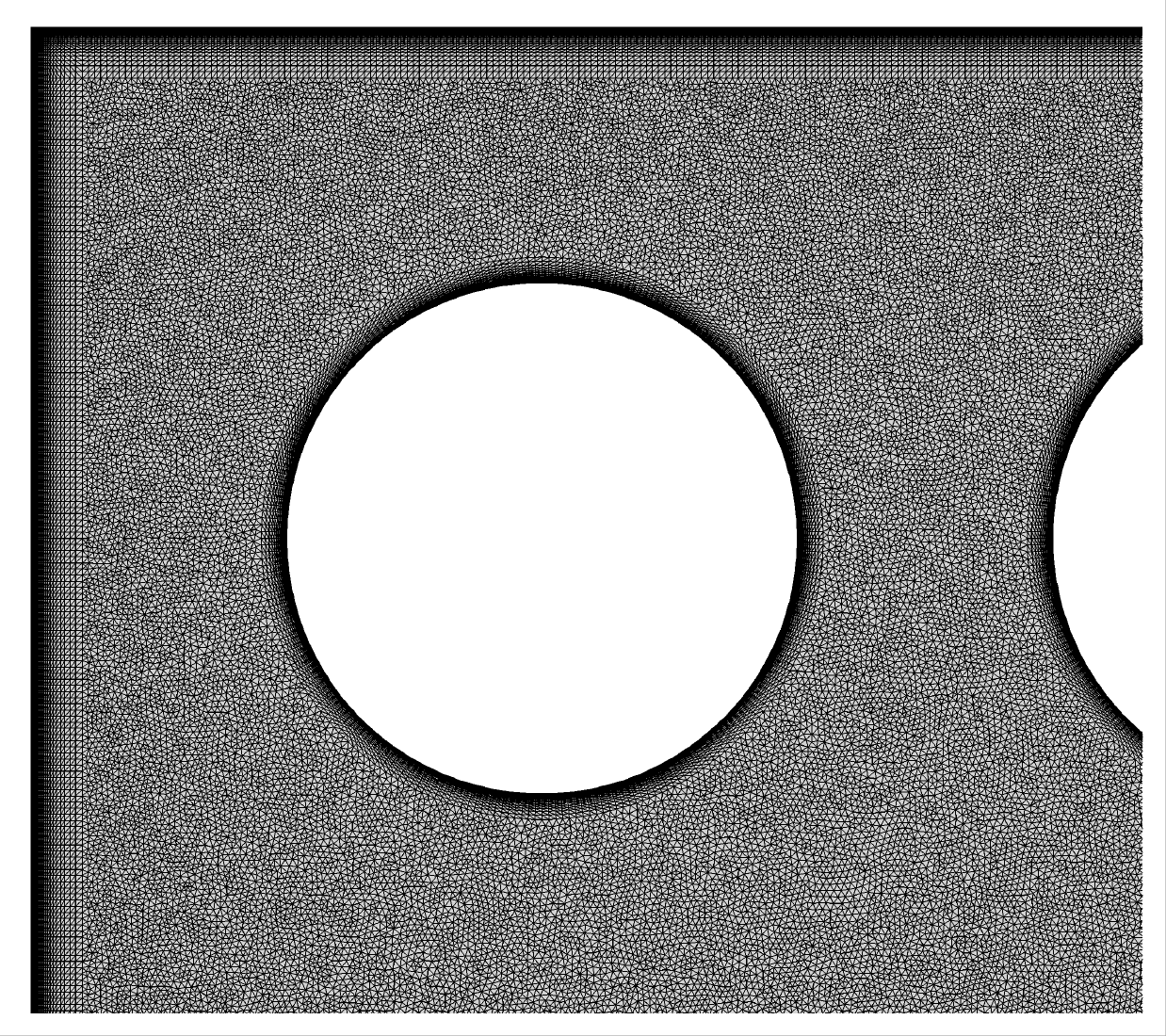}
        \caption{}
    \end{subfigure}%
    \begin{subfigure}{0.5\textwidth}
        \includegraphics[trim={0.1cm 0.1cm 0.1cm 0.1cm}, clip,width=\textwidth]{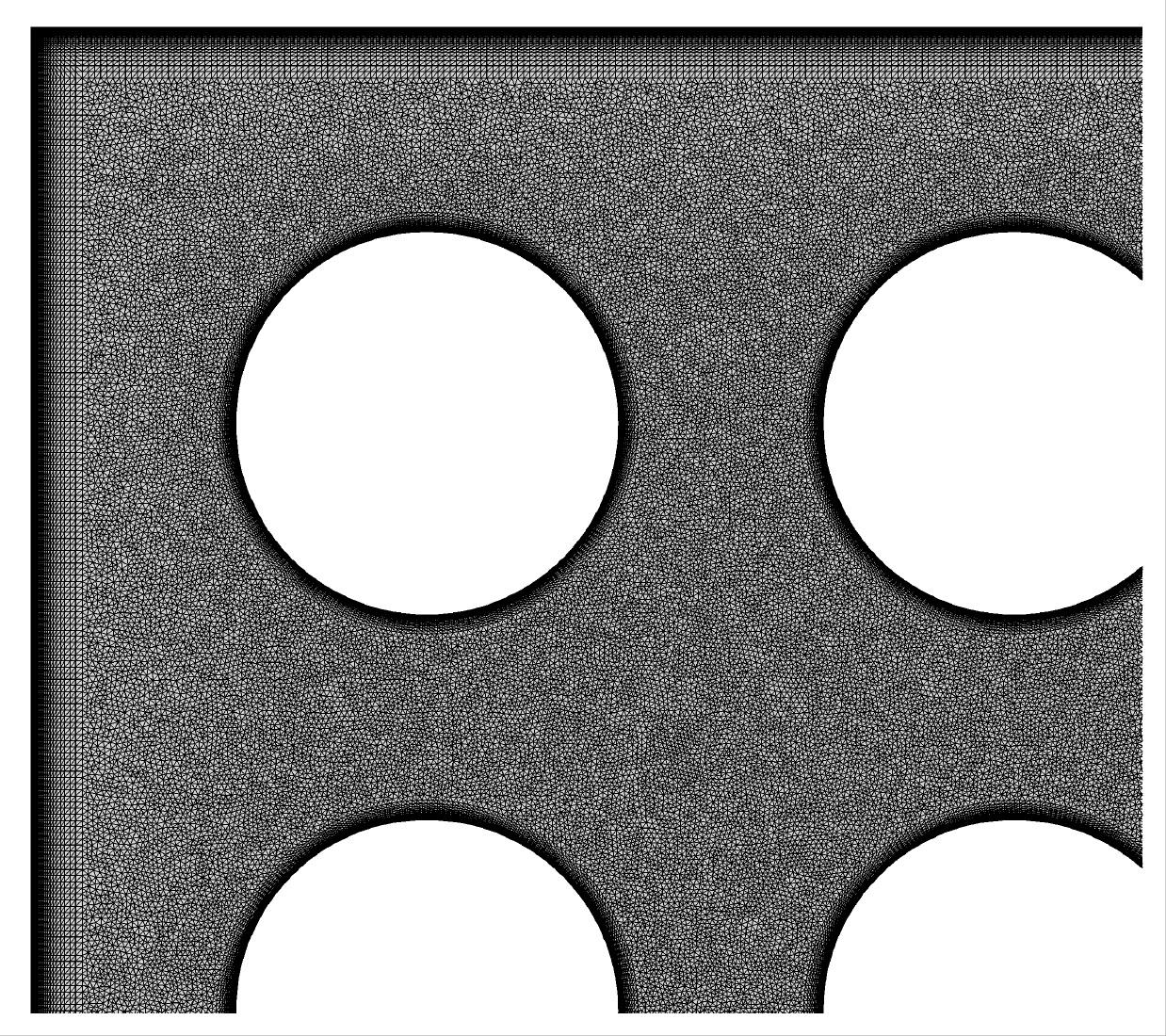}
        \caption{}
    \end{subfigure}
    \caption{Computational mesh (zoomed-in view) near the top-left corner of the domain for (a) $3\times 3$, and (b) $4\times 4$ cylinder arrays.}
    \label{fig:problem_mesh}
\end{figure}

The computational domain is discretized using three-node triangular elements. The constructed mesh is body-fitted which aligns with the cylinders to accurately capture the boundary layers and the elastic movement of the cylinders. A zoomed view of a typical mesh at the top-corner of the $3\times 3$ and $4\times 4$ array is shown in Fig. \ref{fig:problem_mesh}. Each cylinder is divided into 260 segments along the circumference. A boundary layer is constructed surrounding the cylinder with 60 layers perpendicular to the surface of the cylinder and progression ratio of 1.2 giving $y^+ << 1$ for capturing the thermal boundary layer at $Pr=7.0$. Furthermore, the cavity walls also have a boundary layer mesh of 40 layers with progression ratio of 1.15 giving $y^+ << 1$ for capturing of thermal boundary layer. The mesh enclosure is refined appropriately such that the mesh size transitions between the cylinder and wall boundary layers. The $2\times 2$, $3\times 3$, and $4\times 4$ arrays consist of 206203 nodes (409468 elements), 323481 nodes (642754 elements), and 492353 nodes (978720 elements), respectively. The time step size for the computations is selected as $\Delta t U_b/W = 0.041$. The mesh and temporal resolution is chosen based on the numerical tests carried out for benchmarks tests in Appendix \ref{benchmarking}. 


To analyze the results, the instantaneous force coefficients acting on a cylinder surface are defined as
\begin{align}
    C_{X} = \frac{1}{\frac{1}{2}\rho^\text{f}U_b^2 D} \int_{\Gamma^\text{fs}} (\boldsymbol{\sigma}^\text{f}\cdot \boldsymbol{n})\cdot \boldsymbol{n}_x ~d\Gamma, \quad C_Y = \frac{1}{\frac{1}{2}\rho^\text{f}U_b^2 D} \int_{\Gamma^\text{fs}} (\boldsymbol{\sigma}^\text{f}\cdot \boldsymbol{n})\cdot \boldsymbol{n}_y ~d\Gamma,
\end{align}
where $\boldsymbol{n}$ is the outward normal to the cylinder boundary $\Gamma^\mathrm{fs}$. The elastic displacements of cylinders are non-dimensionalized by the diameter of the cylinder in the array. The amplitude of oscillations of the cylinders are quantified as maximum and root-mean-square (RMS) amplitudes (vectors with components in X and Y directions) which are defined as 
\begin{align}
    \boldsymbol{A}_{\mathrm{max}} &= \frac{\max (\boldsymbol{u}^\mathrm{s}) - \min(\boldsymbol{u}^\mathrm{s})}{2},\\
    \boldsymbol{A}_{\mathrm{rms}} &= \sqrt{\frac{1}{n}\displaystyle\sum_{i=1}^n (\boldsymbol{u}^\mathrm{s}-\overline{\boldsymbol{u}}^\mathrm{s})^2},
\end{align}
where $n$ is the number of samples considered after the response has been stabilized.

Furthermore, the heat transfer inside the domain is characterized by evaluating local as well as surface averaged Nusselt numbers, respectively given below:
\begin{align}
    Nu_\mathrm{local} &= \frac{\partial \theta}{\partial \boldsymbol{n}}\bigg|_{\Gamma},\quad Nu = \frac{1}{A}\int_{\Gamma} Nu_\mathrm{local}~ d\Gamma,
\end{align}
where $\theta = (T-T_L)/(T_H-T_L)$ is the non-dimensional temperature, and $A$ is the area of the surface $\Gamma$ over which the integration of the local Nusselt number $Nu_\mathrm{local}$ is carried out. The time-averaging of a quantity $X$ is denoted by $X_\mathrm{mean}$.

In the upcoming sections, we begin with first understanding the dynamics of stationary array of cylinders at $Ra = 1.5 \times 10^6$. Next, we quantify the added mass coefficient for each cylinder in the array which will be crucial in defining the wet natural frequency in the later investigations. Then, we discuss the findings of the flow-induced oscillations experienced by the elastically mounted cylinders in the array subjected to pure natural convection. 

\section{Quantification of fluid characteristics for stationary array of cylinders}
\label{sec:stat_array}

At this juncture, it is imperative to understand the flow dynamics of the system when the cylinders are kept stationary. Such flow pattern behavior highlights possible flow instabilities which may trigger oscillations when the cylinders are elastically mounted. Several studies in the literature have categorized the flow dynamics into steady, periodic-unsteady and chaotic-unsteady states as well as identified asymmetries in the flow \cite{PARK2013295, PARK2015490, MUN2017755, CHO2017307, Xu2025}. However, all these studies were carried out at low Prandtl number of $0.7$. The present subsection discusses the flow and thermal behavior of stationary cylinders in various array configurations at a higher Prandtl number of 7.0. 

\subsection{Force coefficients and Nusselt number}

\begin{table}
\centering 
\caption{Time-averaged force coefficients and Nusselt number on the cylinders in $2\times 2$ array.} 
\label{tab:CXCYNu_Mean_2by2} 
\begin{tabular}{l|cc|cc|cc}
\hline
& \multicolumn{2}{c|}{$C_{X,\mathrm{mean}}$} & \multicolumn{2}{c|}{$C_{Y,\mathrm{mean}}$} & \multicolumn{2}{c}{$Nu_\mathrm{mean}$} \\ \hline
Row 2 (Cyl-21 to 22) & -0.0624 & 0.0624 & -0.5096 & -0.5096 & 2.8207 & 2.8208  \\ 
Row 1 (Cyl-11 to 12) & 0.0265 & -0.0265 & -0.2962 & -0.2962 & 6.5648 & 6.5649 \\ \hline
\end{tabular}
\end{table}

\begin{table}
\centering 
\caption{Time-averaged force coefficients and Nusselt number on the cylinders in $3\times 3$ array.} 
\label{tab:CXCYNu_Mean_3by3} 
\begin{tabular}{l|ccc|ccc|ccc}
\hline
& \multicolumn{3}{c|}{$C_{X,\mathrm{mean}}$} & \multicolumn{3}{c|}{$C_{Y,\mathrm{mean}}$} & \multicolumn{3}{c}{$Nu_{\mathrm{mean}}$} \\ \hline
Row 3 (Cyl-31 to 33) & -0.0003 & -0.0168 & 0.0206 & -0.4365 & -0.4359 & -0.4412 & 0.983 & 1.124 & 1.763  \\
Row 2 (Cyl-21 to 23) & 0.0029 & -0.0197 & 0.0039 & -0.3935 & -0.4003 & -0.4051 & 1.649 & 1.362 & 1.622 \\ 
Row 1 (Cyl-11 to 13) & 0.0603 & -0.0052 & -0.0605 & -0.2014 & -0.2127 & -0.2047 & 4.399 & 4.689 & 4.347 \\ \hline
\end{tabular}
\end{table}

\begin{table}
\centering 
\caption{Time-averaged force coefficients on the cylinders in $4\times 4$ array.} 
\label{tab:CXCY_Mean_4by4} 
\begin{tabular}{l|cccc|cccc}
\hline
& \multicolumn{4}{c|}{$C_{X,\mathrm{mean}}$} & \multicolumn{4}{c}{$C_{Y,\mathrm{mean}}$} \\ \hline
Row 4 (Cyl-41 to 44) & -0.0423 & 0.0008  & -0.0014 & 0.0438 & -0.3482 & -0.3511 & -0.3523 & -0.3483    \\
Row 3 (Cyl-31 to 34) & -0.0047 & -0.0004 & 0.0013 & 0.0039 & -0.3369 & -0.3284 & -0.3280 & -0.3363  \\
Row 2 (Cyl-21 to 24) & 0.0220 & 0.0091 & -0.0092 & -0.0228 & -0.2953 & -0.2959 & -0.2958 & -0.2952   \\ 
Row 1 (Cyl-11 to 14) & 0.0799 & 0.0267 & -0.0270 & -0.0805 & -0.1617 & -0.1674 & -0.1674  & -0.1618 \\ \hline
\end{tabular}
\end{table}

\begin{table}
\centering 
\caption{Time-averaged Nusselt number on the cylinders in $4\times 4$ array.} 
\label{tab:Nu_Mean_4by4} 
\begin{tabular}{l|cccc}
\hline
& \multicolumn{4}{c}{$Nu_{\mathrm{mean}}$} \\ \hline
Row 4 (Cyl-41 to 44) & 1.011 & 0.469 & 0.473 & 1.009\\
Row 3 (Cyl-31 to 34) & 0.572 & 0.477 & 0.479 & 0.577   \\
Row 2 (Cyl-21 to 24) & 1.243 & 1.097 & 1.101 & 1.247 \\ 
Row 1 (Cyl-11 to 14) & 3.265 & 3.608 & 3.607 & 3.264 \\ \hline
\end{tabular}
\end{table}

The time-averaged force coefficients and Nusselt number on each cylinder of the three array configurations ($2\times 2$, $3\times 3$ and $4\times 4$) are quantified in Tables \ref{tab:CXCYNu_Mean_2by2}, \ref{tab:CXCYNu_Mean_3by3}, \ref{tab:CXCY_Mean_4by4}, and \ref{tab:Nu_Mean_4by4}. It is found that the arrays $2\times 2$ and $4\times 4$ have mirror-symmetry about Y-axis as the cylinders on either side have opposite $C_{X,\mathrm{mean}}$. There is an asymmetric behavior for $3\times 3$ array especially near the top row with $C_{X,\mathrm{mean}}$ drastically different for Cyl-31 and Cyl-33. This asymmetry arises as a result of strong interaction between the large convection and the small gap between the cylinders and the walls of the enclosure, which was also pointed out in \cite{PARK2015490}. The symmetry in the force coefficient $C_{Y,\mathrm{mean}}$ exist in all the configurations with a net negative force trying to push the cylinders along the direction of gravity. Note that the mean force coefficient in Y direction is much higher than that in the X direction which highlights the effect of static force due to buoyancy. Generally, in all the three configurations, the time-averaged Nusselt number decreases from the bottom row to the top, similar to the observations in the literature \cite{PARK2013295, PARK2015490, MUN2017755, CHO2017307}. The bottom row is exposed to the colder fluid resulting in enhanced heat transfer due to convection. Transitioning to the upper rows, the cylinders are subjected to the hot plume generated from the bottom cylinders which reduces the heat transfer resulting in lower Nusselt numbers. The symmetry breaking in $3\times 3$ array also affects the heat transfer as reflected by the lower value of $Nu_{\mathrm{mean}}$ for Cyl-31 compared to that for Cyl-33. The Nusselt number reduces by 44.24\% for Cyl-31 in comparison with Cyl-33.

\begin{figure}
    \centering
    \begin{subfigure}{0.32\textwidth}
        \includegraphics[trim={12cm 0.5cm 12cm 1cm}, clip,width=\textwidth]{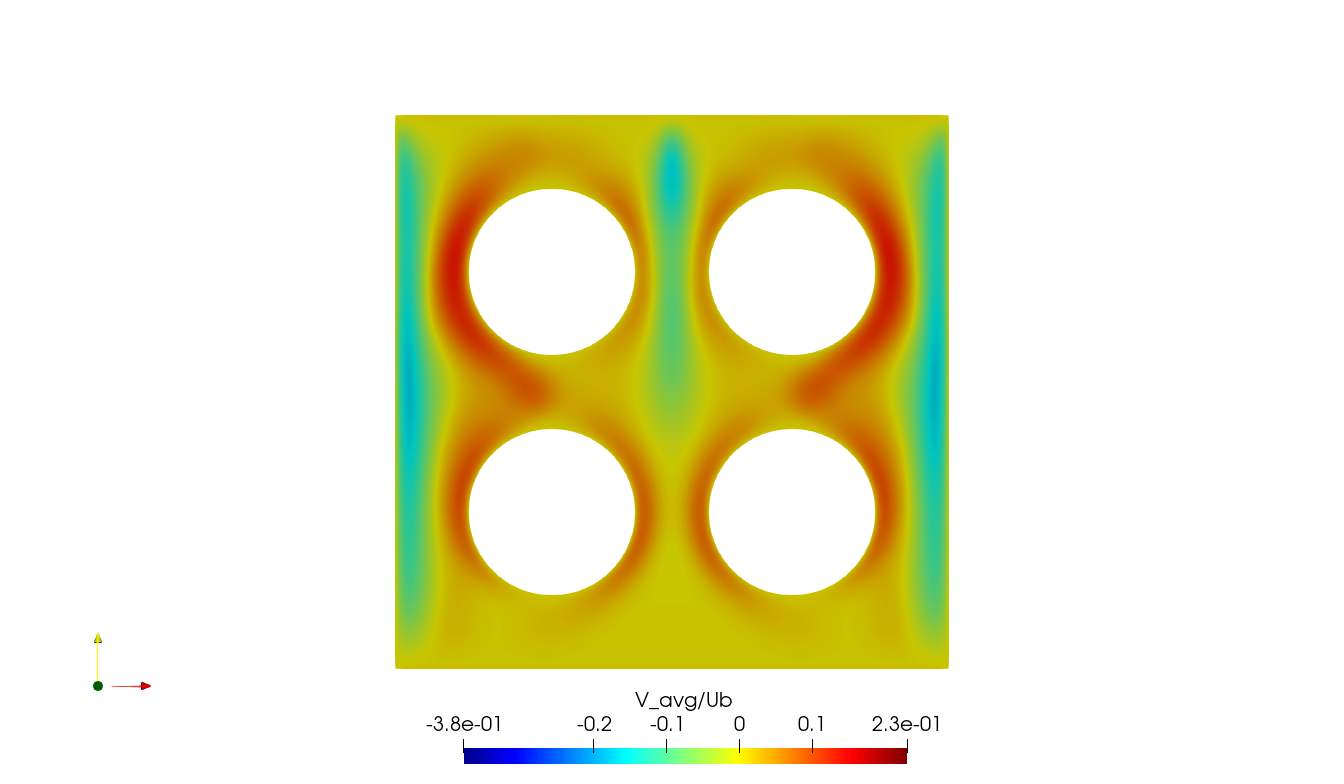}
        \caption{}
    \end{subfigure}%
    \begin{subfigure}{0.32\textwidth}
        \includegraphics[trim={12cm 0.5cm 12cm 1cm}, clip,width=\textwidth]{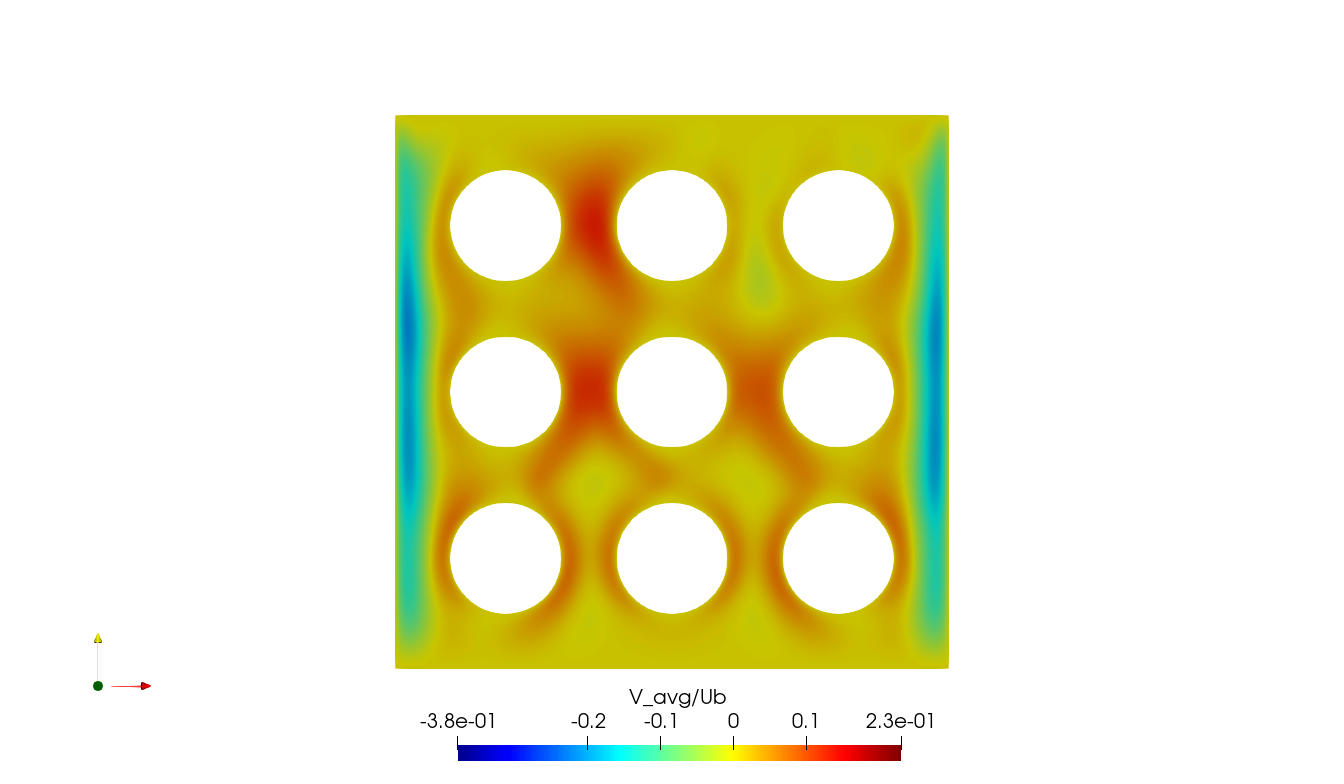}
        \caption{}
    \end{subfigure}%
    \begin{subfigure}{0.32\textwidth}
        \includegraphics[trim={12cm 0.5cm 12cm 1cm}, clip,width=\textwidth]{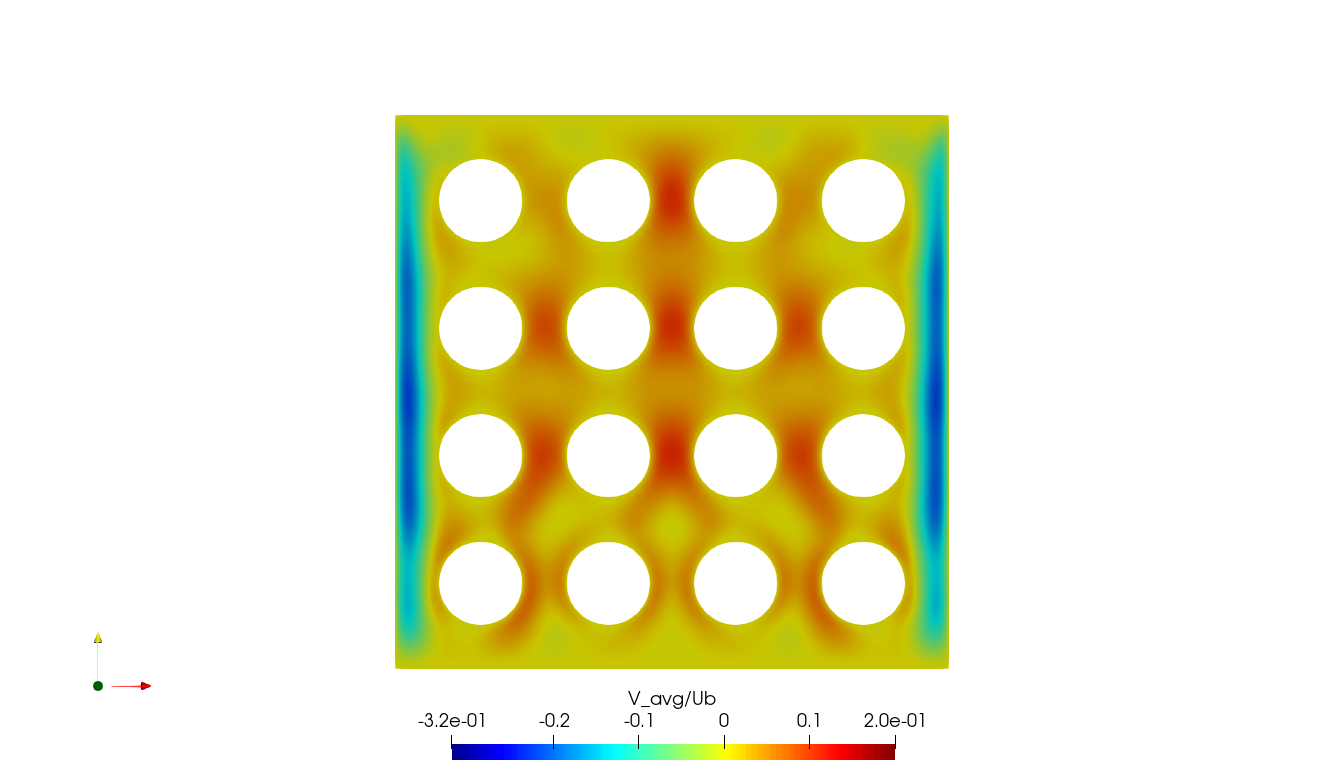}
        \caption{}
    \end{subfigure}%
    \caption{Contours of time-averaged non-dimensional vertical velocity $v^\mathrm{f}/U_b$ for (a) $2\times 2$, (b) $3\times 3$, and (c) $4\times 4$ arrays.}
    \label{fig:stationary_cyl_Vavg}
\end{figure}
\begin{figure}
    \centering
    \begin{subfigure}{0.32\textwidth}
        \includegraphics[trim={12cm 0.5cm 12cm 1cm}, clip,width=\textwidth]{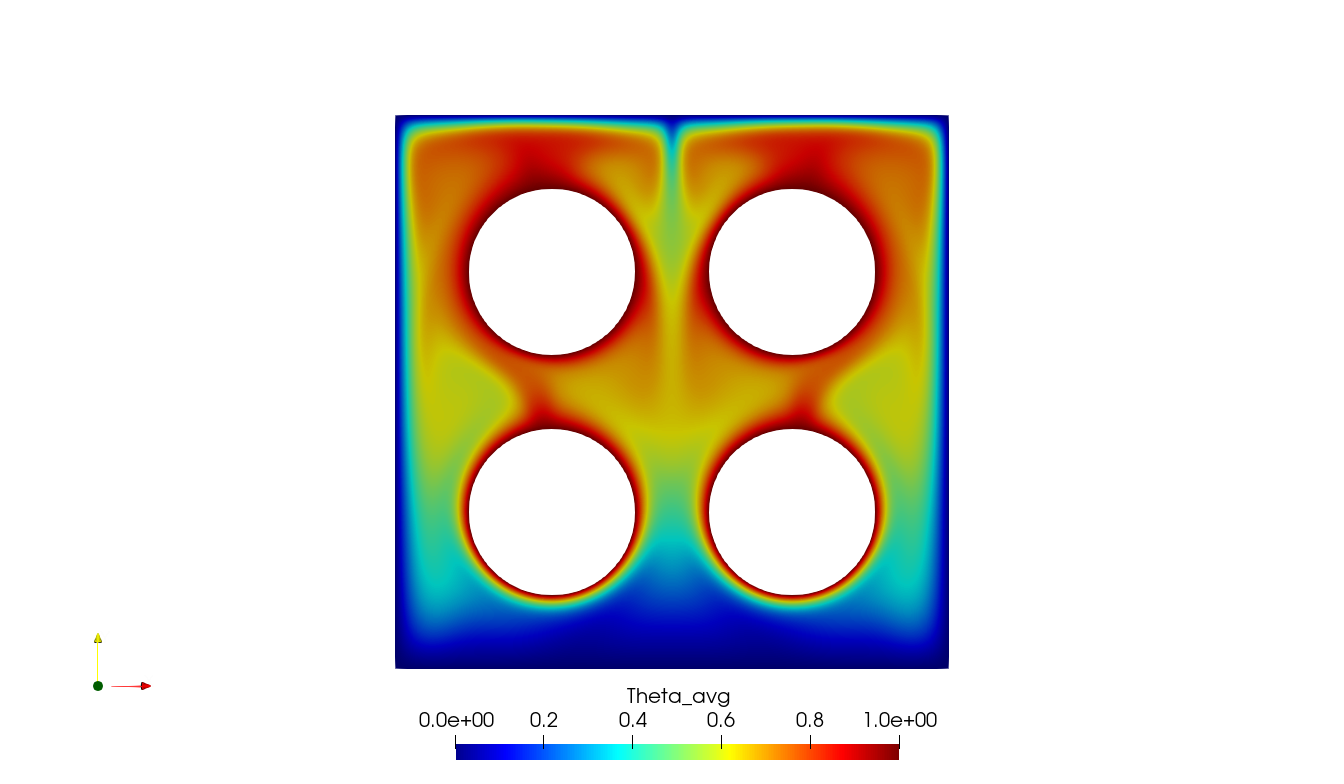}
        \caption{}
    \end{subfigure}%
    \begin{subfigure}{0.32\textwidth}
        \includegraphics[trim={12cm 0.5cm 12cm 1cm}, clip,width=\textwidth]{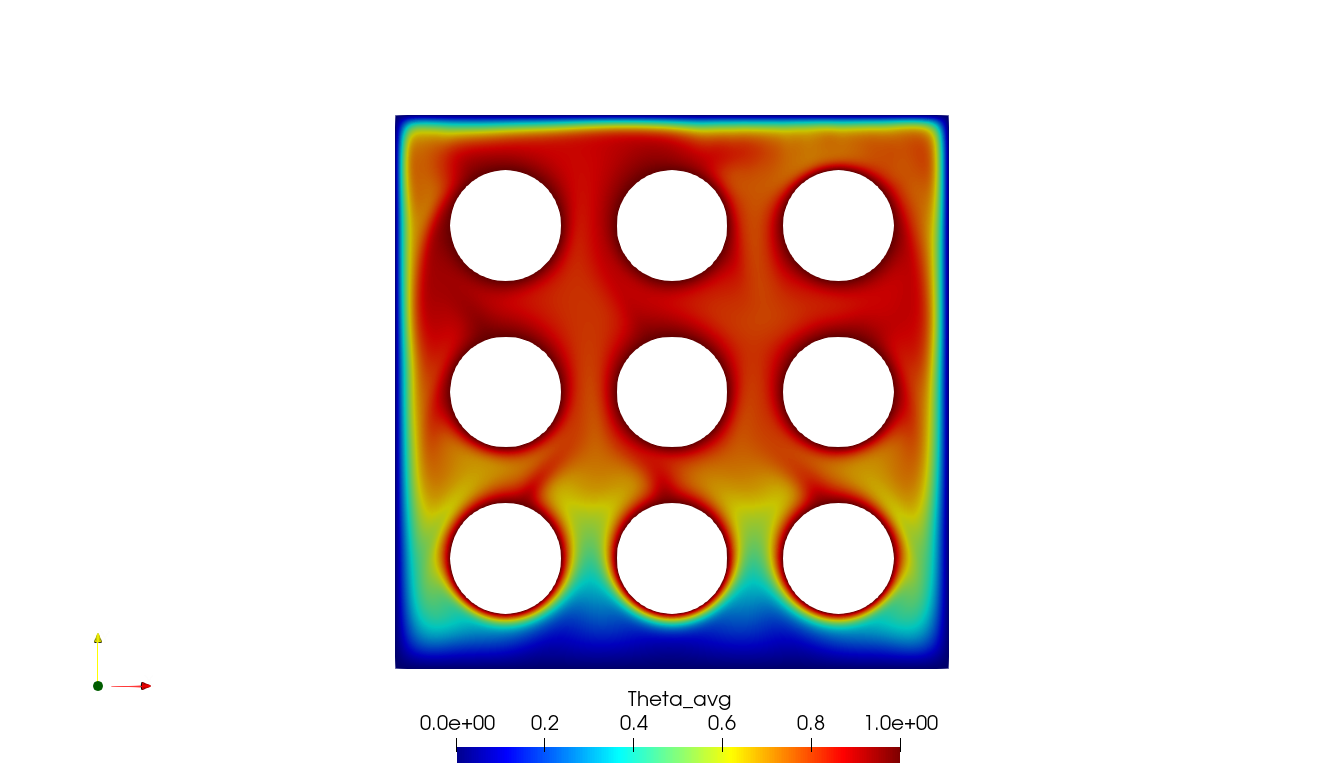}
        \caption{}
    \end{subfigure}%
    \begin{subfigure}{0.32\textwidth}
        \includegraphics[trim={12cm 0.5cm 12cm 1cm}, clip,width=\textwidth]{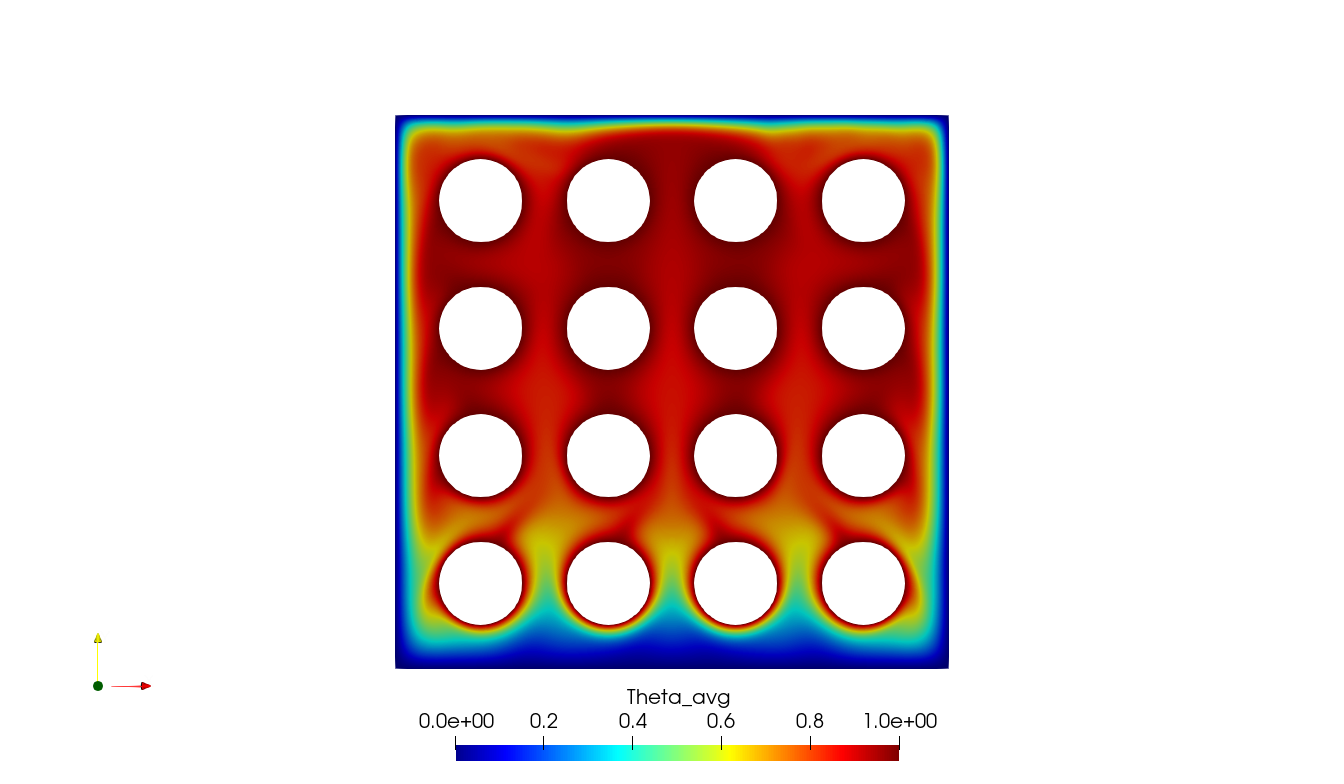}
        \caption{}
    \end{subfigure}%
    \caption{Contours of time-averaged non-dimensional temperature $\theta$ for (a) $2\times 2$, (b) $3\times 3$, and (c) $4\times 4$ arrays.}
    \label{fig:stationary_cyl_Theta}
\end{figure}

Geometrically, in the $3\times 3$ array, the centreline runs through the middle column of the cylinders, so the two interior gaps (between columns 1 and 2, and columns 2 and 3) are mirror images of each other. Two equal rising jets side by side tend to attract and merge, with one direction preferred by them, resulting in breaking of symmetry. In other configurations, the central gap lies on the centreline, so the dominant jet can exist there without breaking symmetry. This is confirmed by the time-averaged vertical velocity $(v^\mathrm{f}/U_b)$ and temperature contours shown in Figs. \ref{fig:stationary_cyl_Vavg} and \ref{fig:stationary_cyl_Theta}, respectively. The high-velocity jet is mirror-symmetric about centreline for $2\times 2$ and $4\times 4$ arrays, while it breaks symmetry in the $3\times 3$ configuration. The movement of the jet plume inside the cavity can also be visualized in the temperature distribution contour in Fig. \ref{fig:stationary_cyl_Theta}. For the $2\times 2$ array, the plume from the bottom row diverges towards the edge of the cavity resulting in the higher jet velocity in the gap between the cylinder and the wall. After interacting with the top row, the fluid circulates downward through the centre gap. In the $3\times 3$ array, the plumes from the bottom row converge towards the centreline. After interacting with the middle row, the plume breaks symmetry and is observed to be dominant on the left side of the cavity. On the other hand, the plume interaction is noticed to be symmetric for $4\times 4$ array with highest jet velocity at the centreline and circulation to the bottom of the cavity occurring along the side walls of the cavity.

\begin{figure}
    \centering
    \begin{subfigure}{0.4\textwidth}
        \includegraphics[trim={9cm 0.1cm 8cm 1cm}, clip,width=\textwidth]{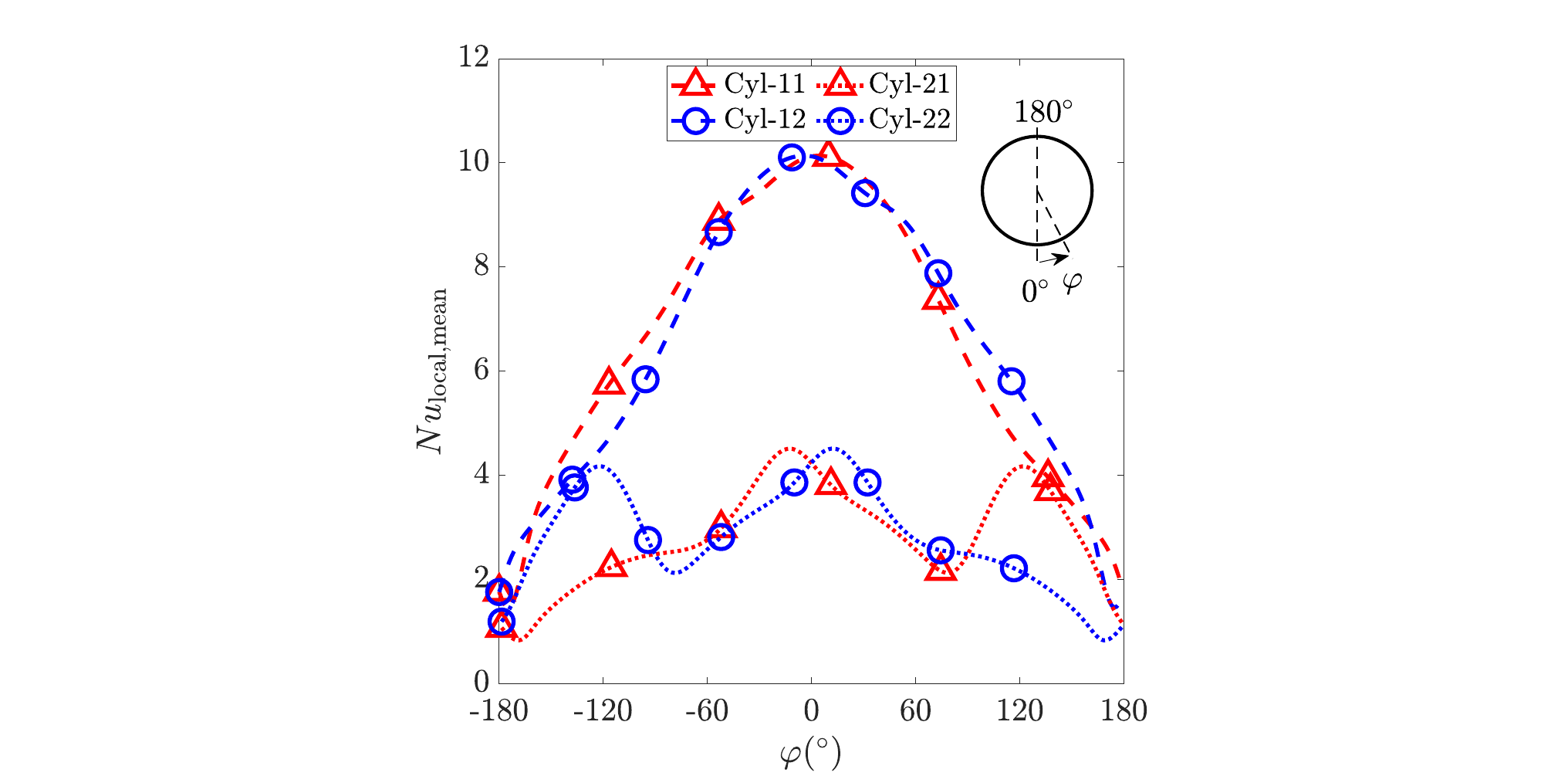}
        \caption{}
    \end{subfigure}%
    \begin{subfigure}{0.4\textwidth}
        \includegraphics[trim={9cm 0.1cm 8cm 1cm}, clip,width=\textwidth]{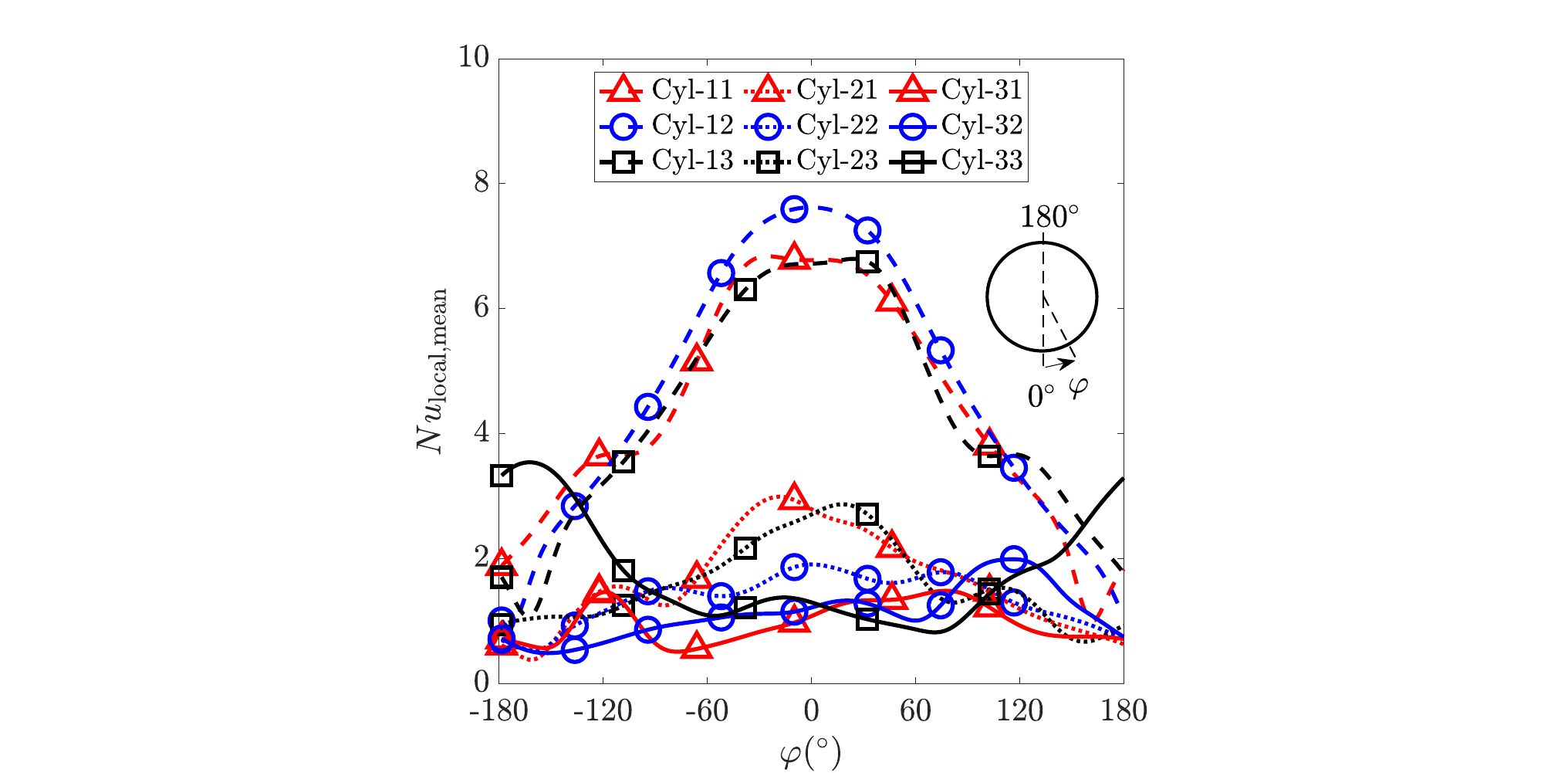}
        \caption{}
    \end{subfigure}
    
    \begin{subfigure}{\textwidth}
        \includegraphics[trim={8cm 0.1cm 9cm 1cm}, clip,width=0.25\textwidth]{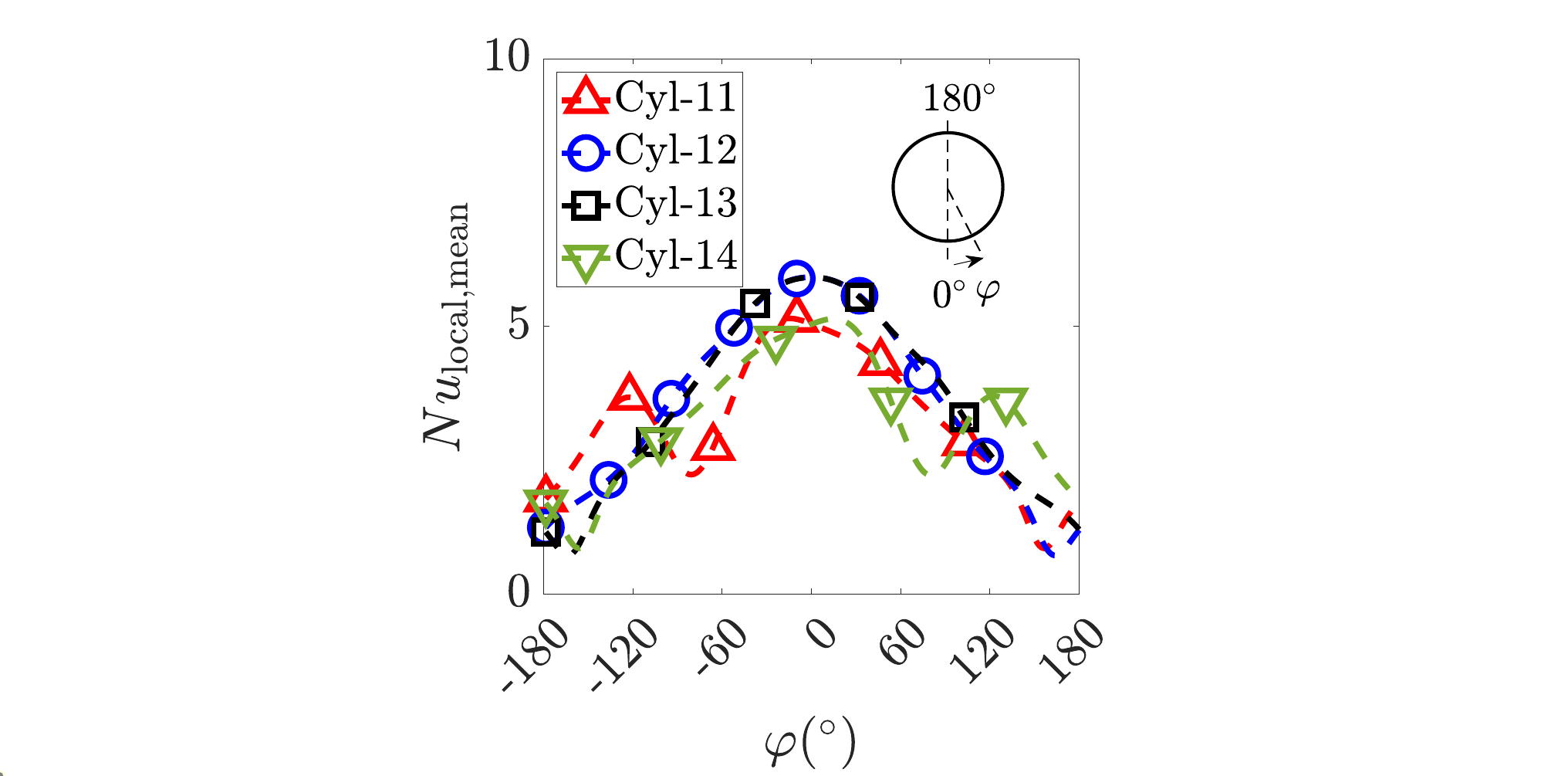}%
        \includegraphics[trim={8cm 0.1cm 9cm 1cm}, clip,width=0.25\textwidth]{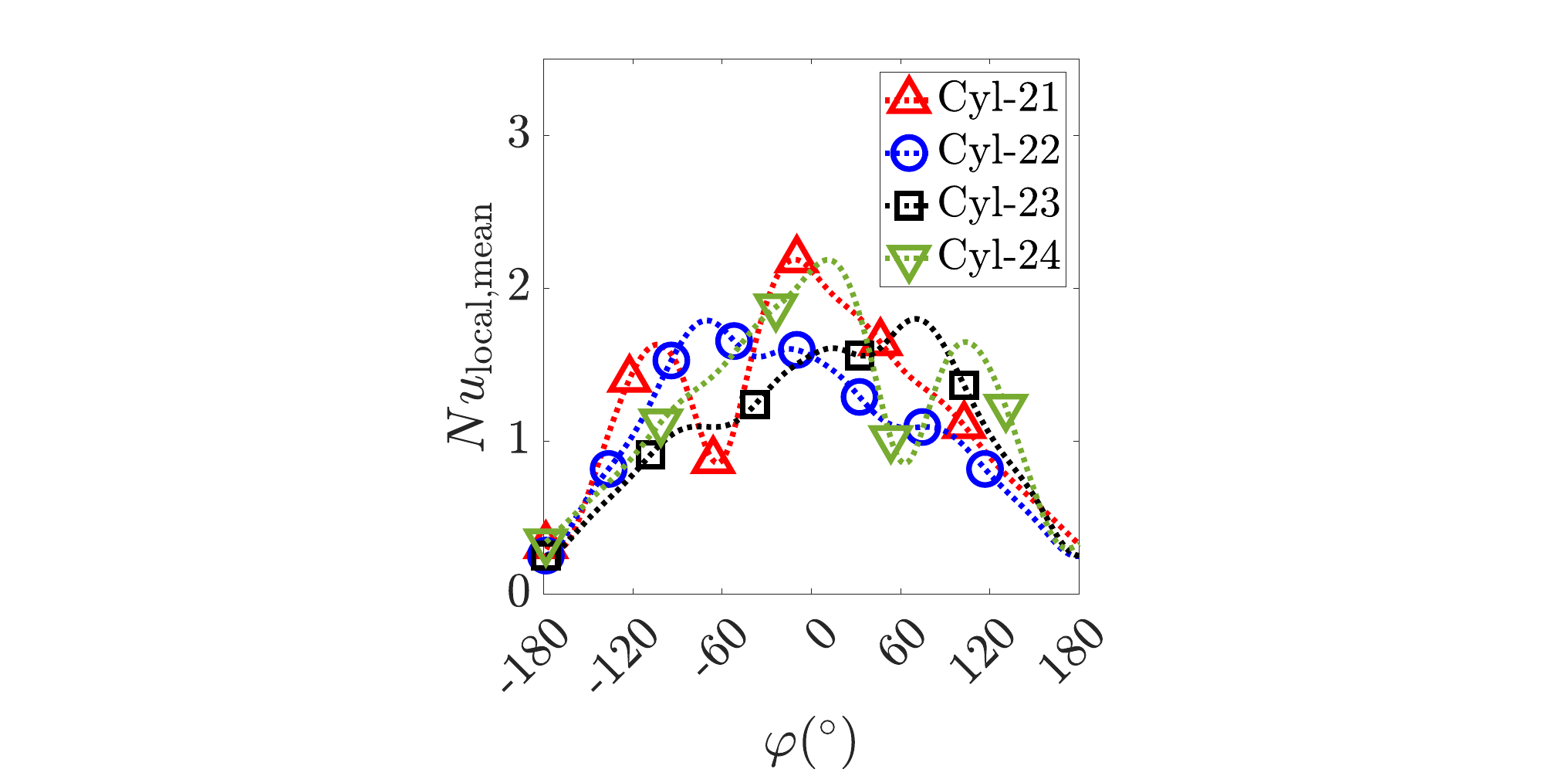}%
        \includegraphics[trim={8cm 0.1cm 9cm 1cm}, clip,width=0.25\textwidth]{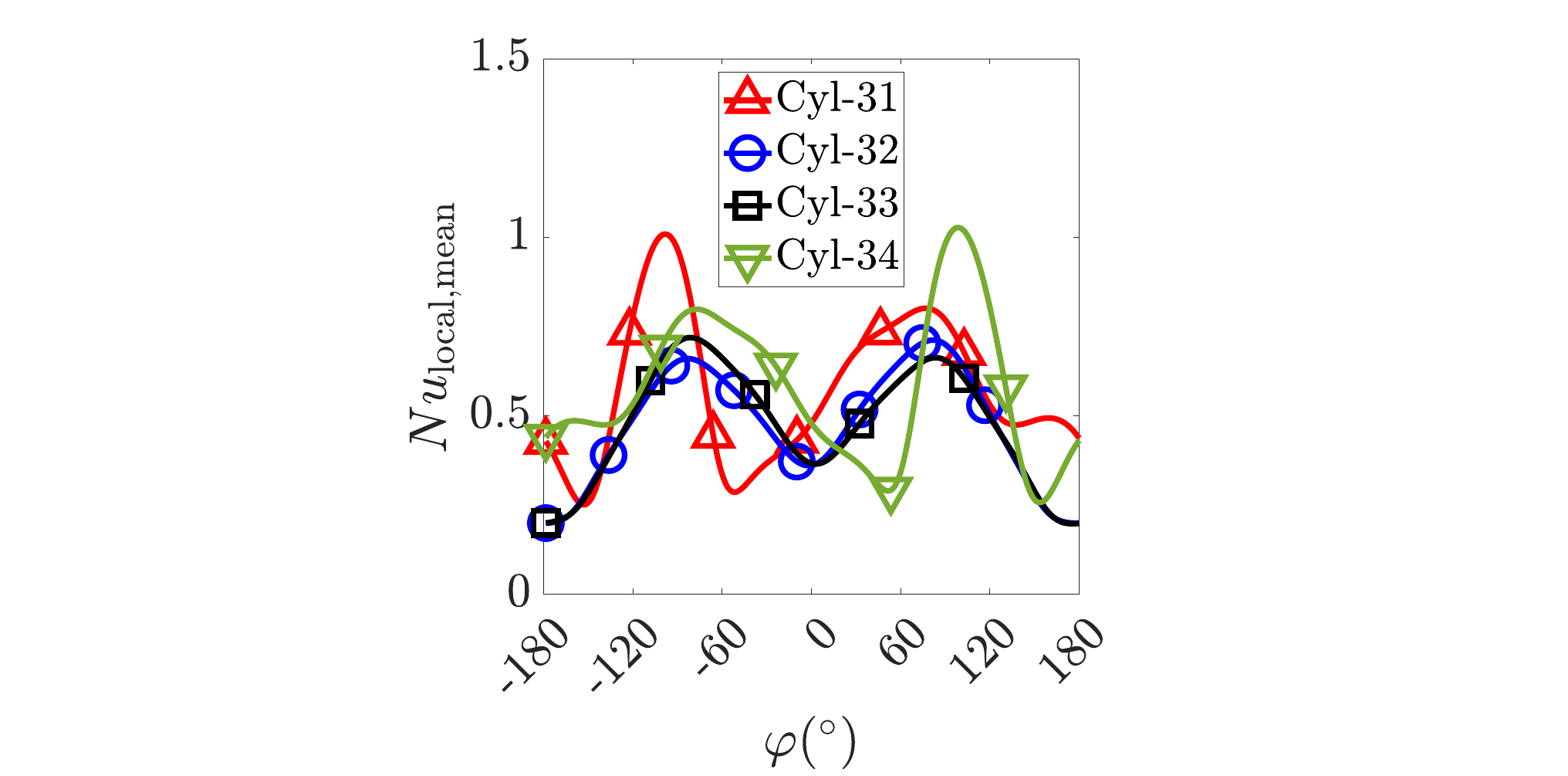}%
        \includegraphics[trim={8cm 0.1cm 9cm 1cm}, clip,width=0.25\textwidth]{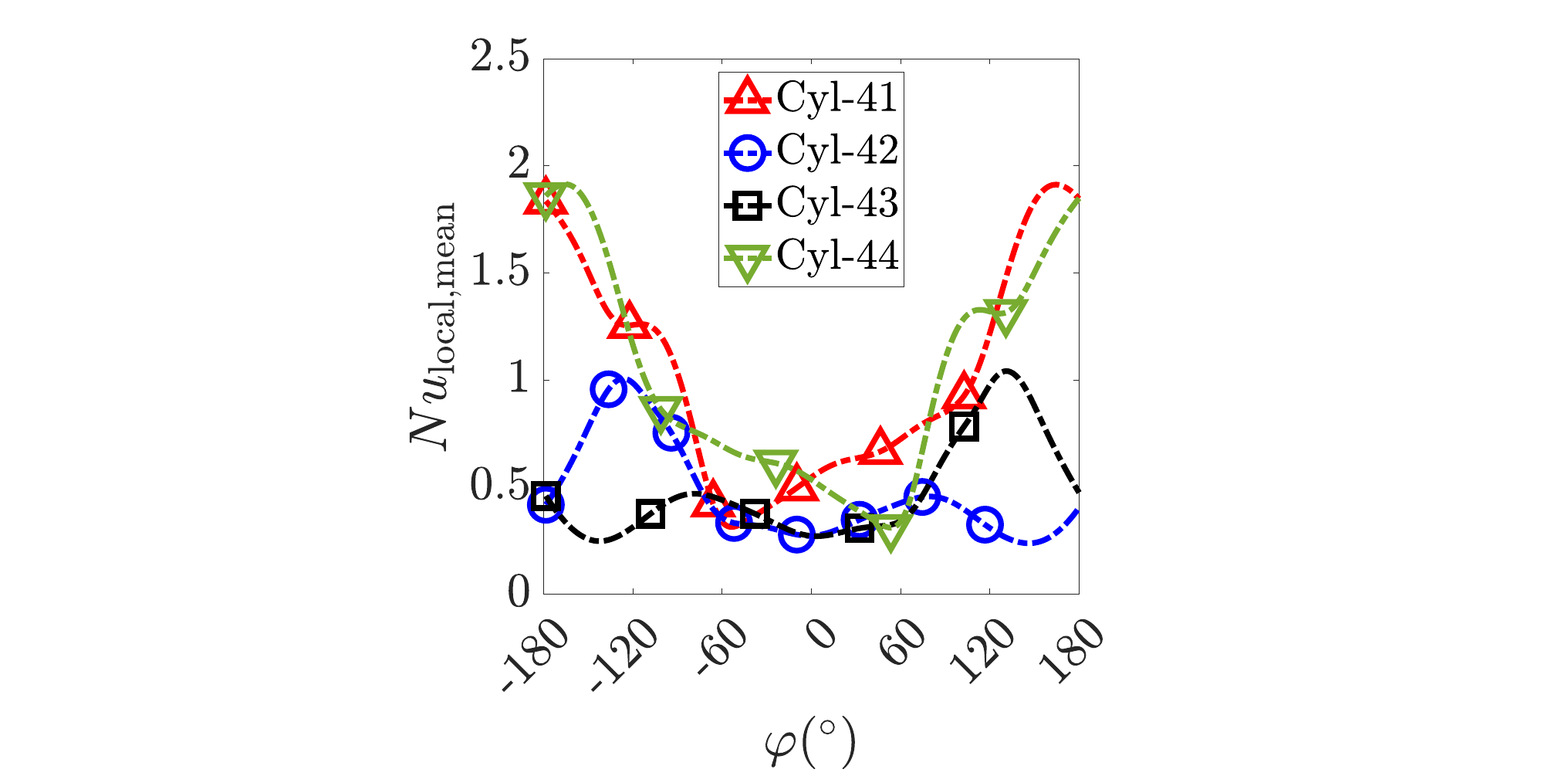}
        \caption{}
    \end{subfigure}
    \caption{Time-averaged Nusselt number variation on the surface of the cylinders in (a) $2\times 2$, (b) $3\times 3$, and (c) $4\times 4$ array.}
    \label{fig:Time_averaged_Nu}
\end{figure}

The symmetry in $2\times 2$ and $4\times 4$ configurations can also be observed in the time-averaged local Nusselt number distribution along the cylinders, shown in Fig. \ref{fig:Time_averaged_Nu}. The location around the circumference of the cylinder is defined based on the angle $\varphi$ with $\varphi = 0^\circ$ at the bottom of the cylinder. Corresponding cylinders showing mirror symmetry geometrically have similar local variation in the Nusselt number. For example, $Nu$ for Cyl-11 in the $4\times 4$ array has a peak at $\varphi \approx -120^\circ$ which is also observed for Cyl-14 at $\varphi \approx 120^\circ$. In contrast, Cyl-31 and Cyl-33 in $3\times 3$ array show the asymmetry in the top row where Cyl-33 has much larger $Nu_\mathrm{local, mean}$ at $\varphi \approx -180^\circ$ compared to Cyl-31. 

\begin{figure}
    \centering
    \begin{subfigure}{0.33\textwidth}
        \includegraphics[trim={7.5cm 0.1cm 9cm 1cm}, clip,width=\textwidth]{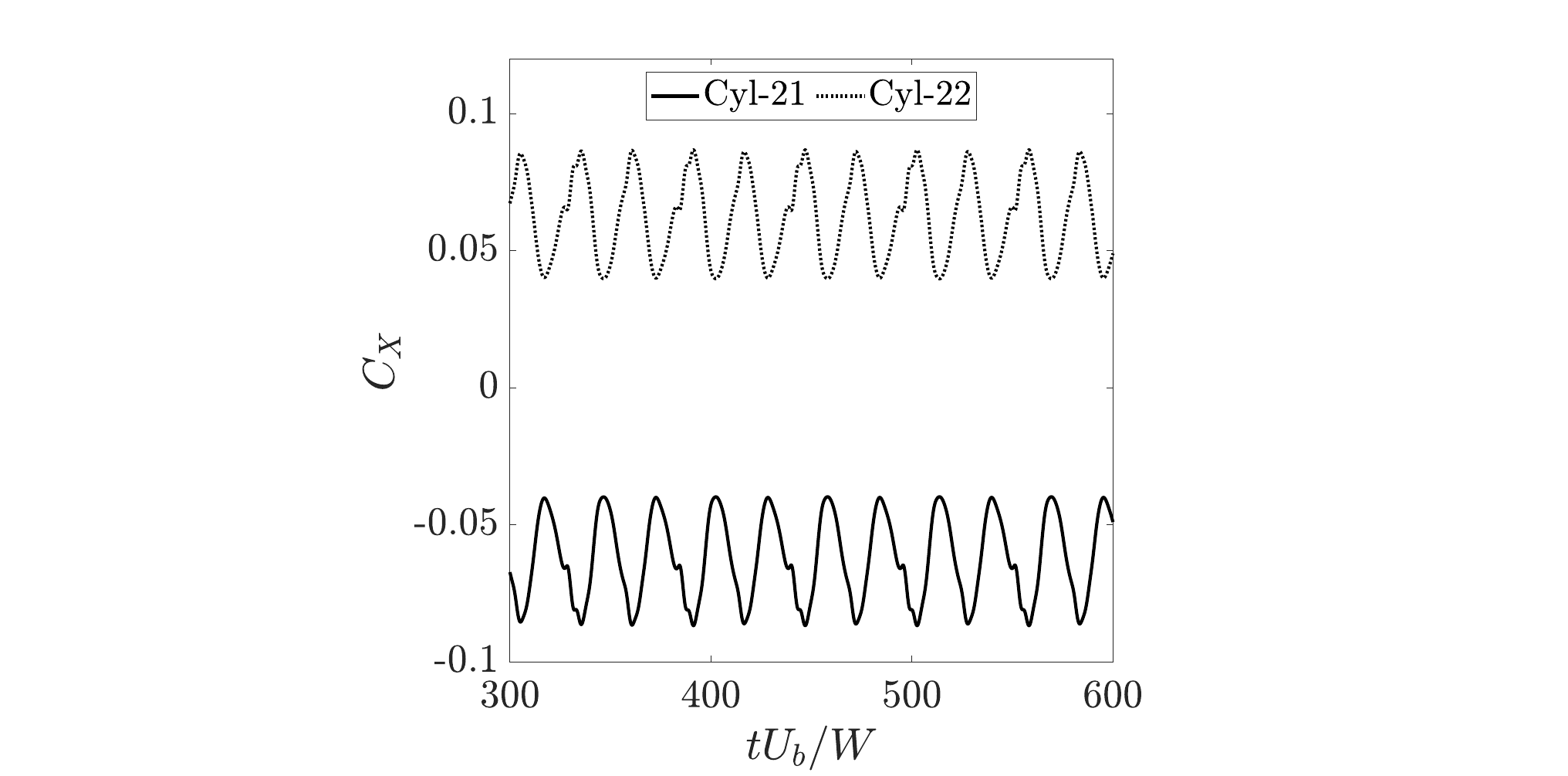}%
        \caption{}
    \end{subfigure}%
    \begin{subfigure}{0.33\textwidth}
        \includegraphics[trim={7.5cm 0.1cm 9cm 1cm}, clip,width=\textwidth]{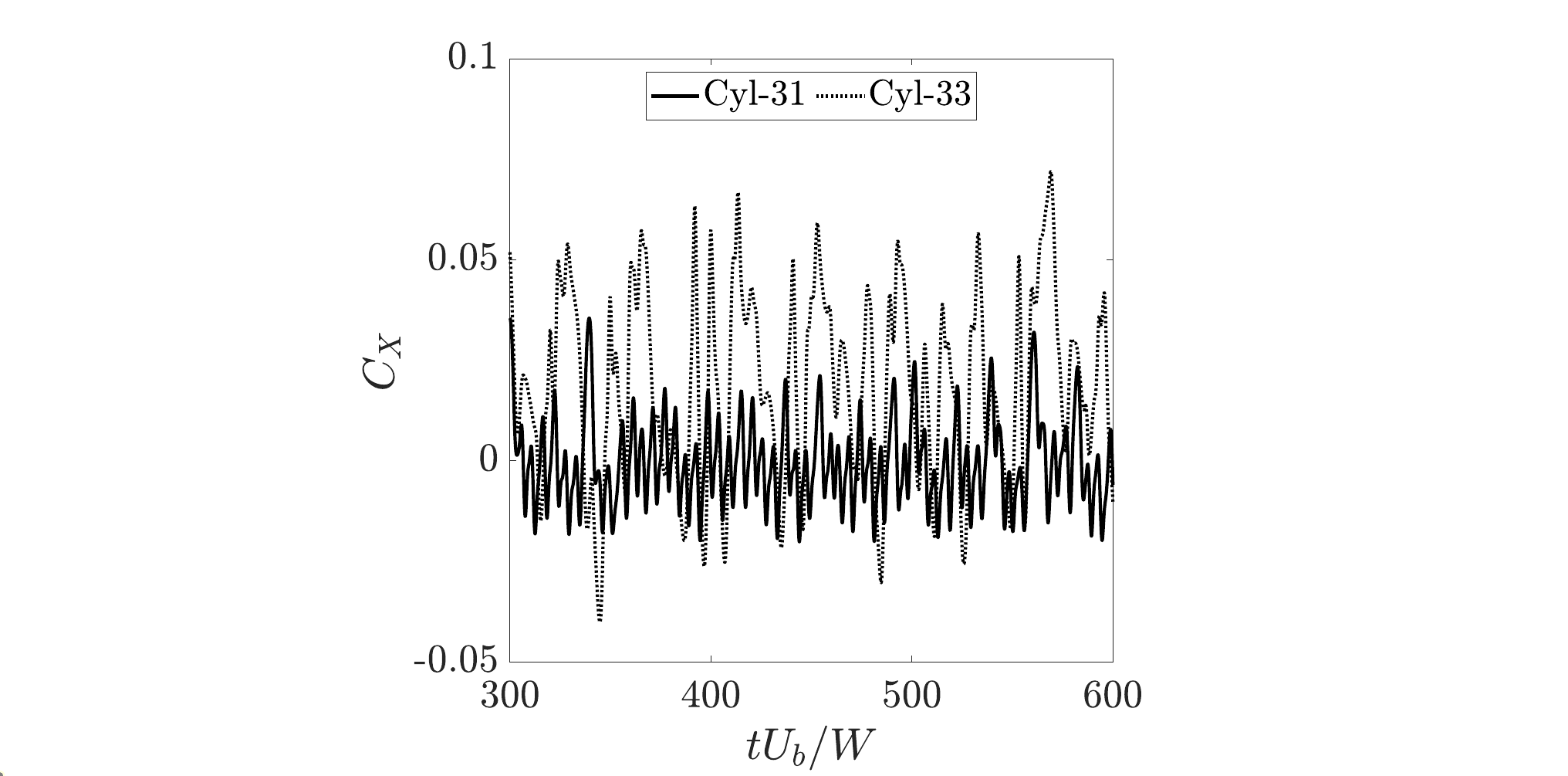}%
        \caption{}
    \end{subfigure}%
    \begin{subfigure}{0.33\textwidth}
        \includegraphics[trim={7.5cm 0.1cm 9cm 1cm}, clip,width=\textwidth]{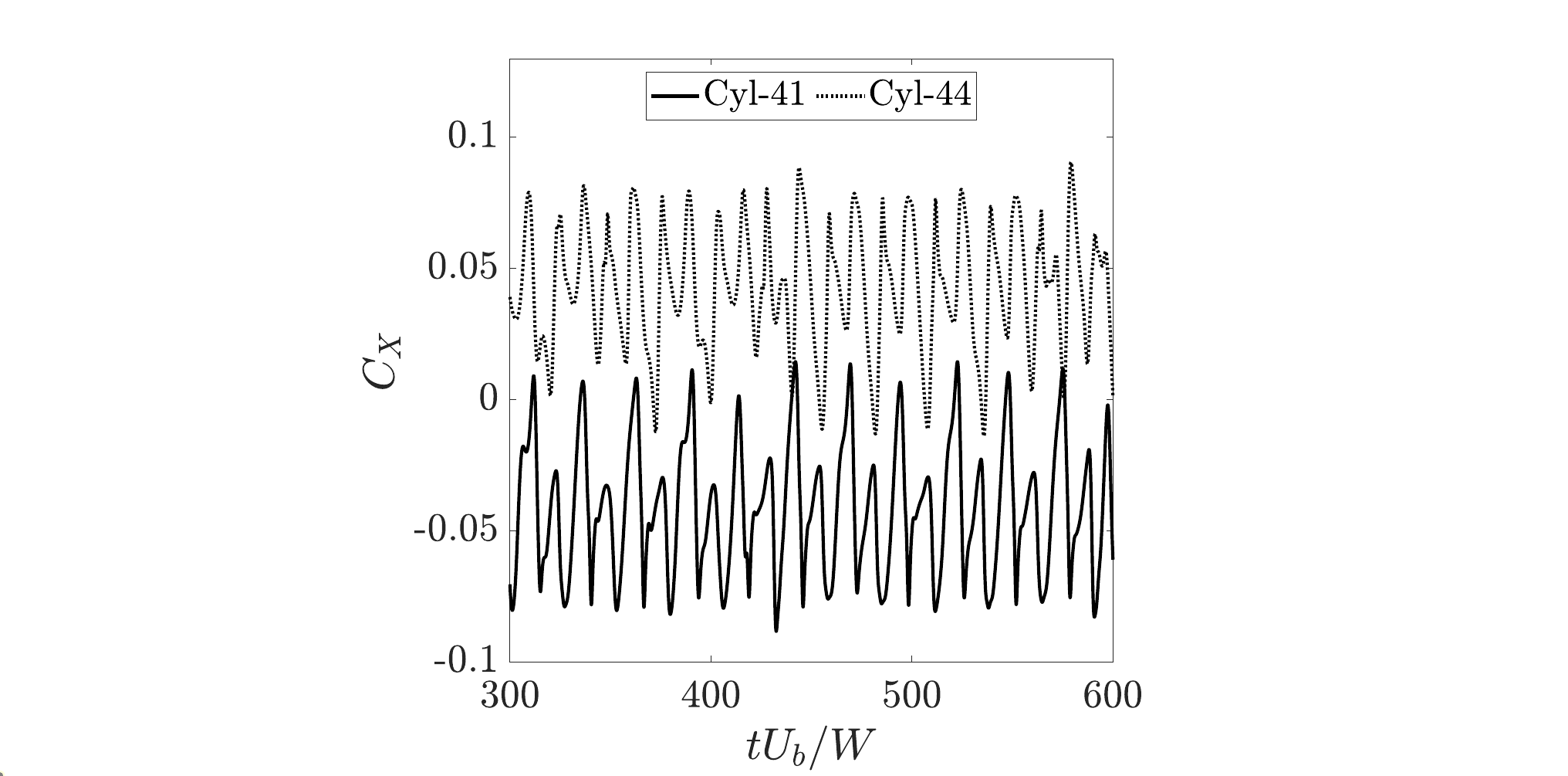}%
        \caption{}
    \end{subfigure}%
    \caption{Comparison of time histories of force coefficient $C_X$ for (a) $2\times 2$, (b) $3\times 3$, and (c) $4\times 4$ arrays.}
    \label{fig:Time_history_stationary_CX}
\end{figure}

\begin{figure}
    \centering
    \begin{subfigure}{0.33\textwidth}
        \includegraphics[trim={8cm 0.1cm 9cm 1cm}, clip,width=\textwidth]{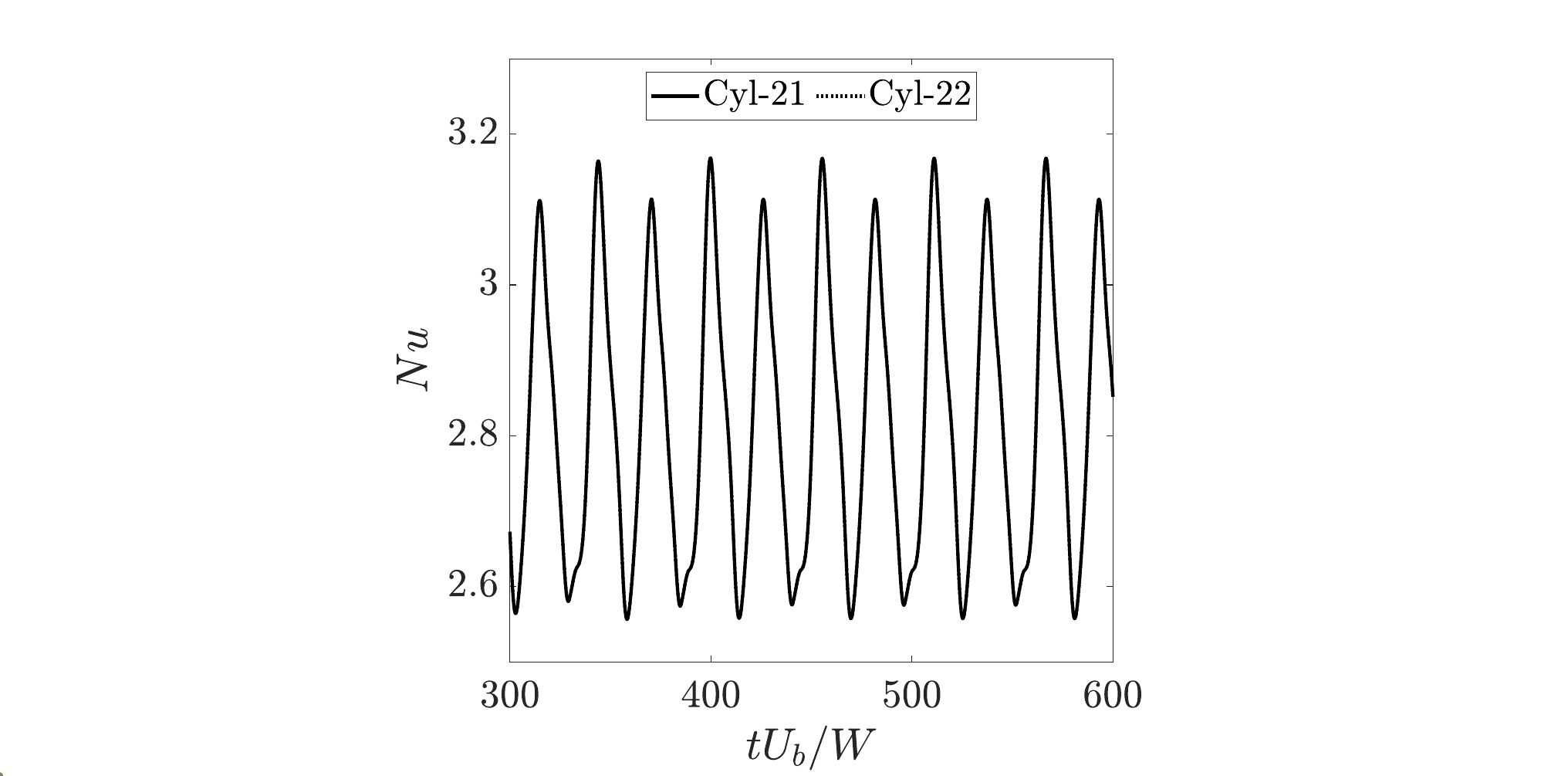}%
        \caption{}
    \end{subfigure}%
    \begin{subfigure}{0.33\textwidth}
        \includegraphics[trim={8cm 0.1cm 9cm 1cm}, clip,width=\textwidth]{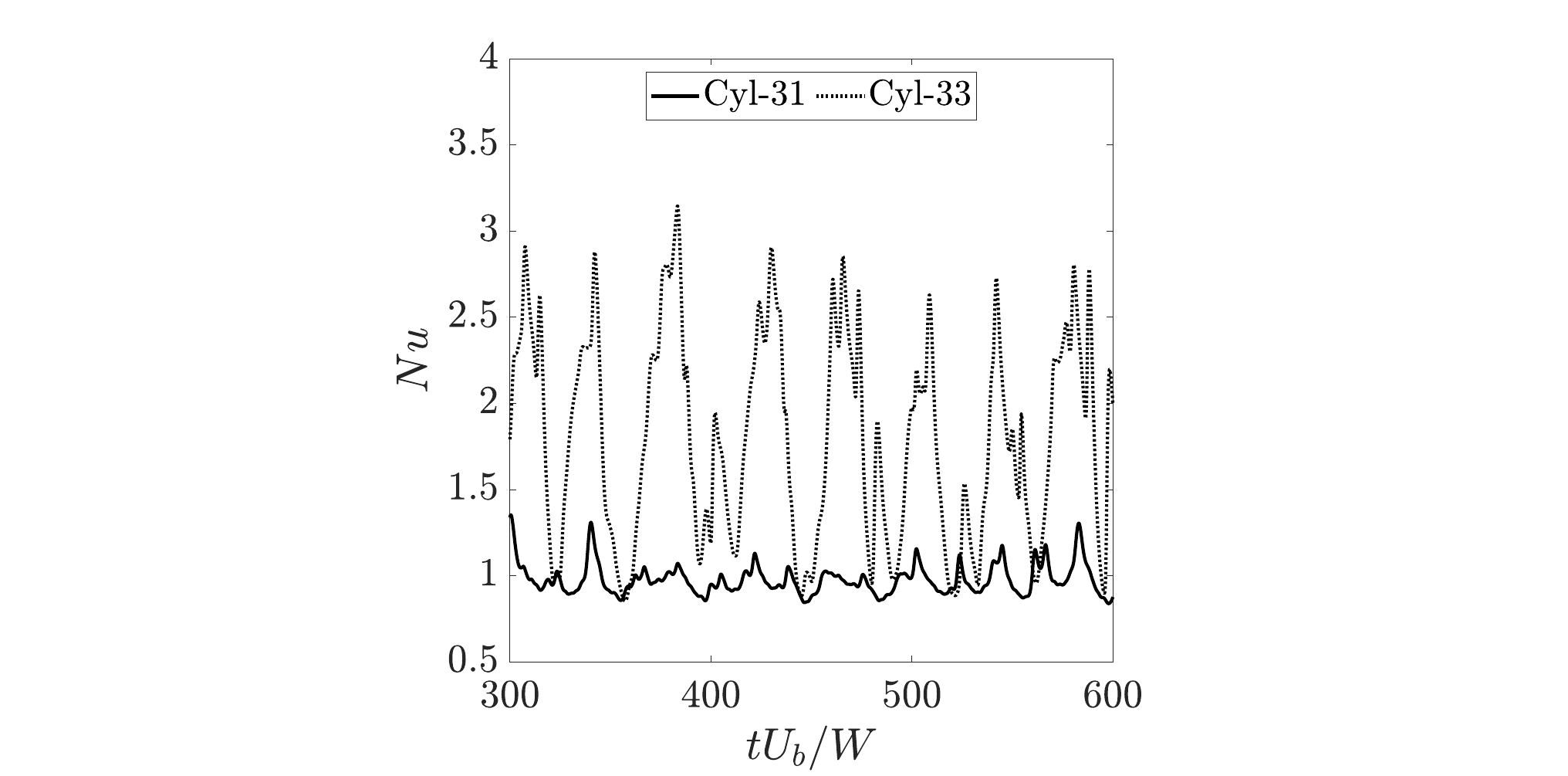}%
        \caption{}
    \end{subfigure}%
    \begin{subfigure}{0.33\textwidth}
        \includegraphics[trim={8cm 0.1cm 9cm 1cm}, clip,width=\textwidth]{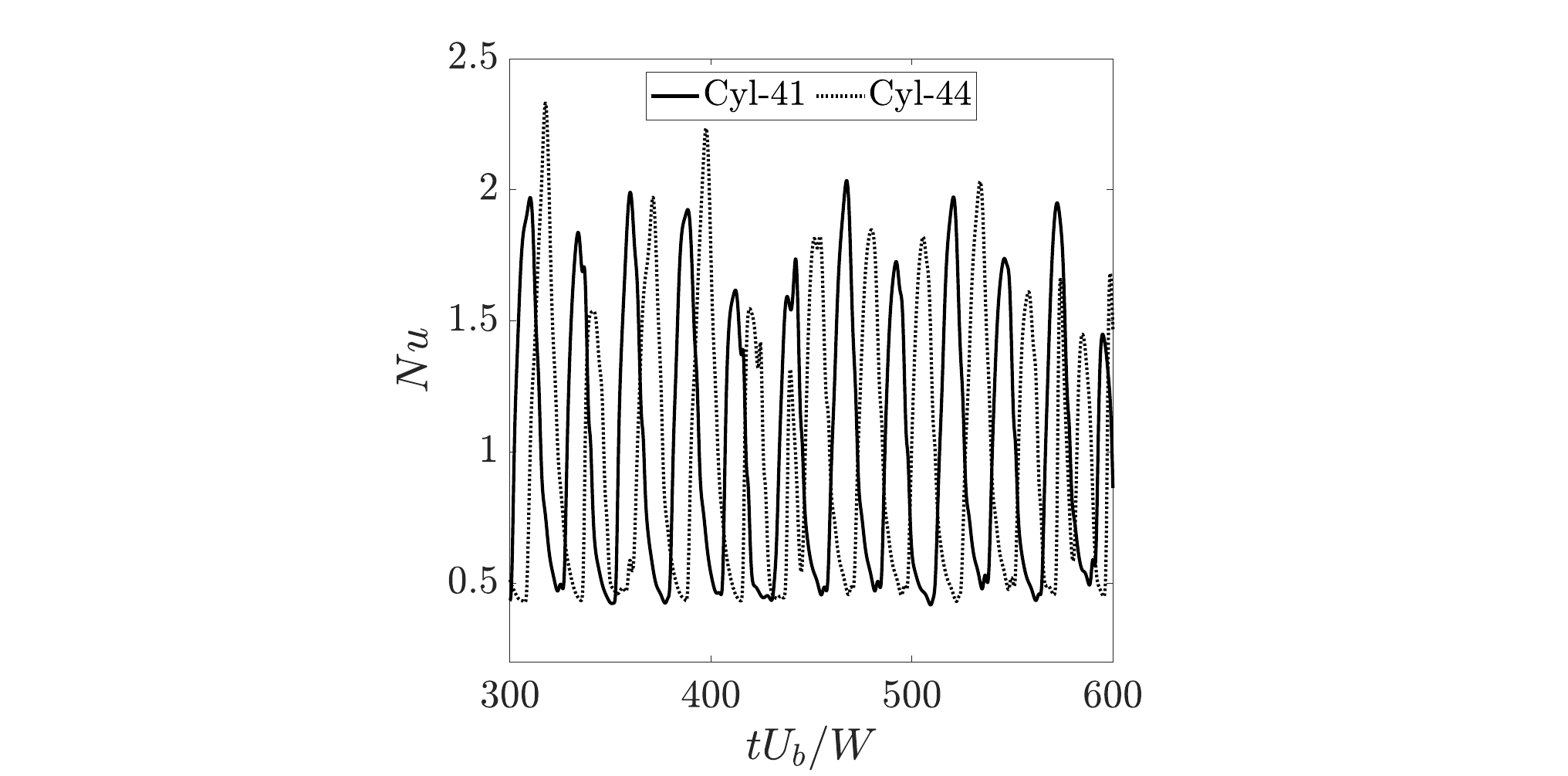}%
        \caption{}
    \end{subfigure}%
    \caption{Comparison of time histories of surface-averaged Nusselt number $Nu$ on the top corner cylinders for (a) $2\times 2$, (b) $3\times 3$, and (c) $4\times 4$ arrays.}
    \label{fig:Time_history_stationary_Nu}
\end{figure}

The time history of force coefficient $C_X$ and surface-averaged Nusselt number for the top row corner cylinders is depicted in Figs. \ref{fig:Time_history_stationary_CX} and \ref{fig:Time_history_stationary_Nu}, respectively. The instantaneous variation of $C_{X,\mathrm{mean}}$ for $2\times 2$ array is found to be perfectly mirror symmetric for Cyl-21 and Cyl-22, confirming the symmetry in the instantaneous flow dynamics. The Nusselt number also traces identically for both the cylinders. In the $3\times 3$ array, the force coefficient breaks symmetry. The shift in the plume direction results in a higher positive X direction force on Cyl-33 compared to Cyl-31. Furthermore, the Nusselt number is higher for Cyl-33 compared to Cyl-31 as the hot plume shifts towards Cyl-31. The transient shedding of the plume is evident from the transient variation in the force coefficient and Nusselt numbers. It is interesting to note that the $4\times 4$ array is symmetric only in averaged sense. The top corner cylinders Cyl-41 and 44 alternately produce peaks in the Nusselt number and thus heat transfer, observed in Fig. \ref{fig:Time_history_stationary_Nu}(c). 

\begin{figure}
    \centering
    \includegraphics[trim={4cm 0.1cm 18cm 1.0cm}, clip,width=0.4\textwidth]{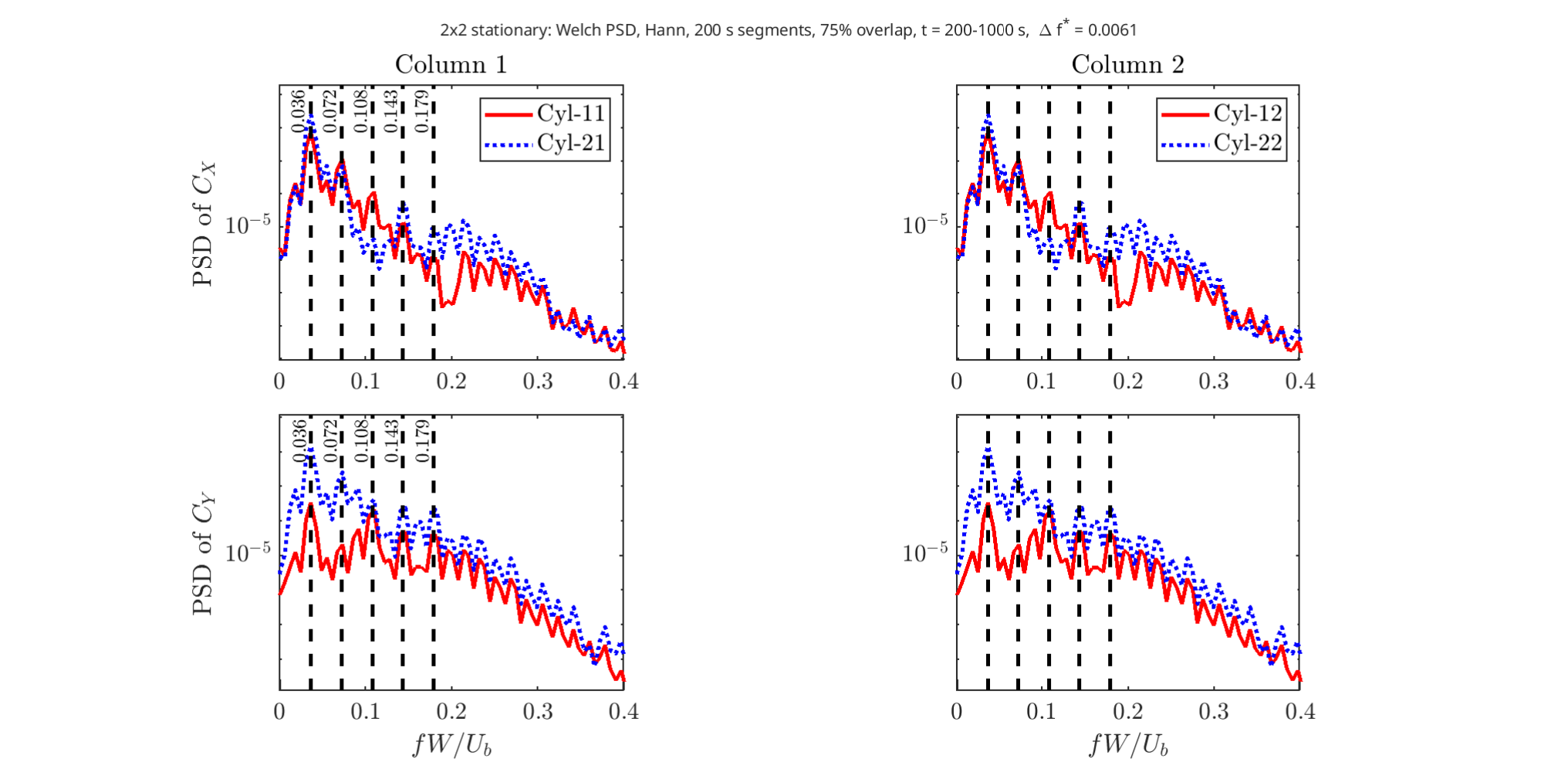}%
    \hspace{-1cm}
    \includegraphics[trim={18cm 0.1cm 4cm 1.0cm}, clip,width=0.4\textwidth]{Images/4Cyl_CXCYSpectrum-eps-converted-to.pdf}
    \caption{Welch power spectra of the force coefficients for the cylinders in $2\times 2$ array system.}
    \label{powerSpectra_2by2_stationary}
\end{figure}

\subsection{Spectra of force coefficients}

The instantaneous variation of the force coefficients is observed to be broadband and partly periodic for each cylinder, resulting in a noisy depiction using fast Fourier transform. Therefore, to isolate the dominant frequencies from the force signal, the Welch's power spectral density (PSD) method is utilized. Here, the signal is averaged over overlapping segments (with 75\% overlap) which reduces the variance, thus resulting in easier identification of the dominant peak frequencies. Each segment is tapered with a Hann window with a frequency resolution of $\Delta f W/U_b \approx 0.006$. The Hann taper suppresses spectral leakage from the strong low-frequency content into neighboring frequency bands. Because the PSD is variance-normalised, integrating it over a frequency band gives the fraction of the force fluctuation carried by that band. This is utilized to quantify the narrowband forcing near the dominant frequencies. 

\begin{figure}
    \centering
    \includegraphics[trim={1.5cm 0.1cm 2cm 1.1cm}, clip,width=\textwidth]{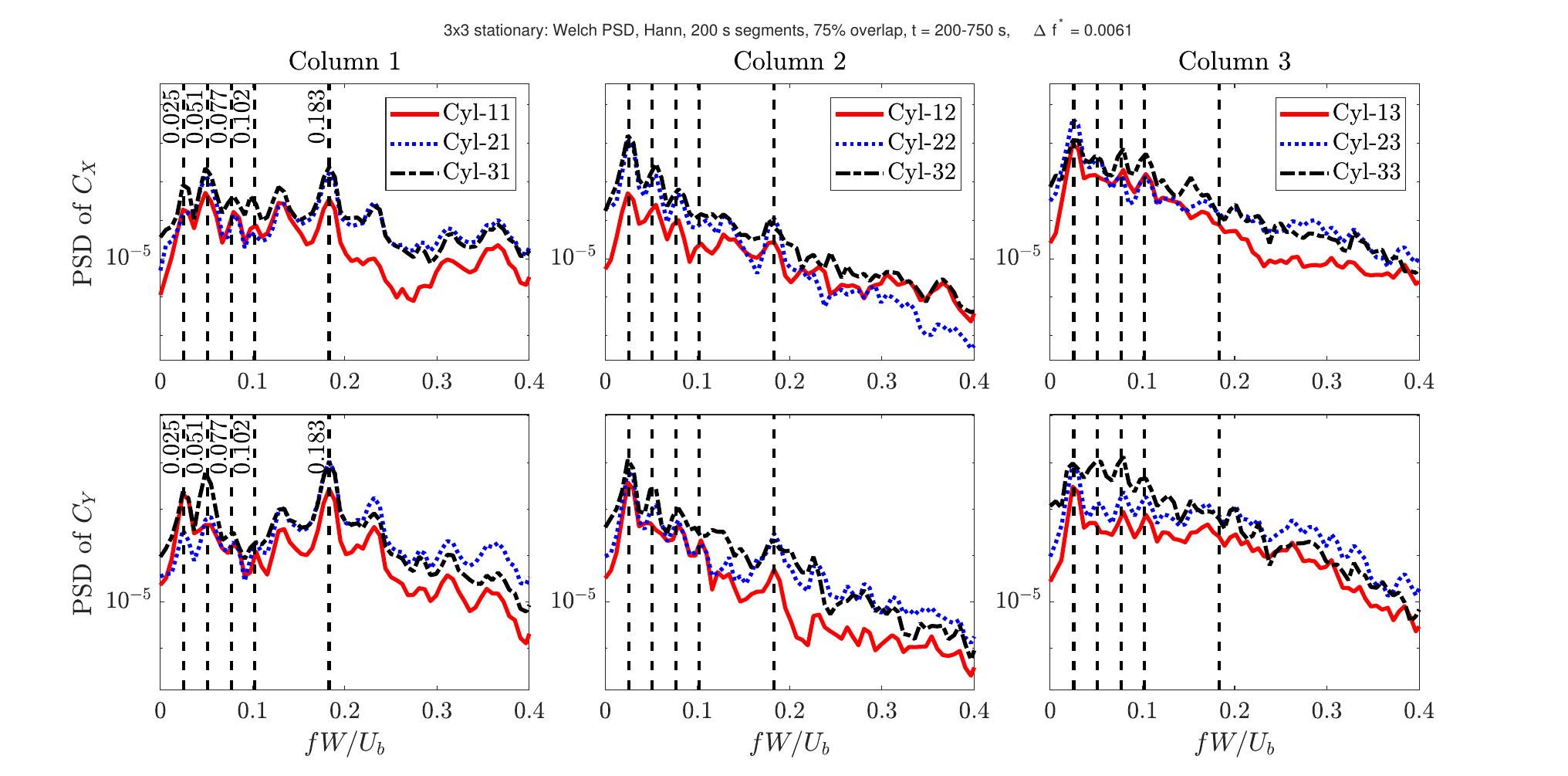}%
    \caption{Welch power spectra of the force coefficients for the cylinders in $3\times 3$ array system.}
    \label{powerSpectra_3by3_stationary}
\end{figure}

\begin{figure}
    \centering
    \includegraphics[trim={1.5cm 0.1cm 2.5cm 1.2cm}, clip,width=\textwidth]{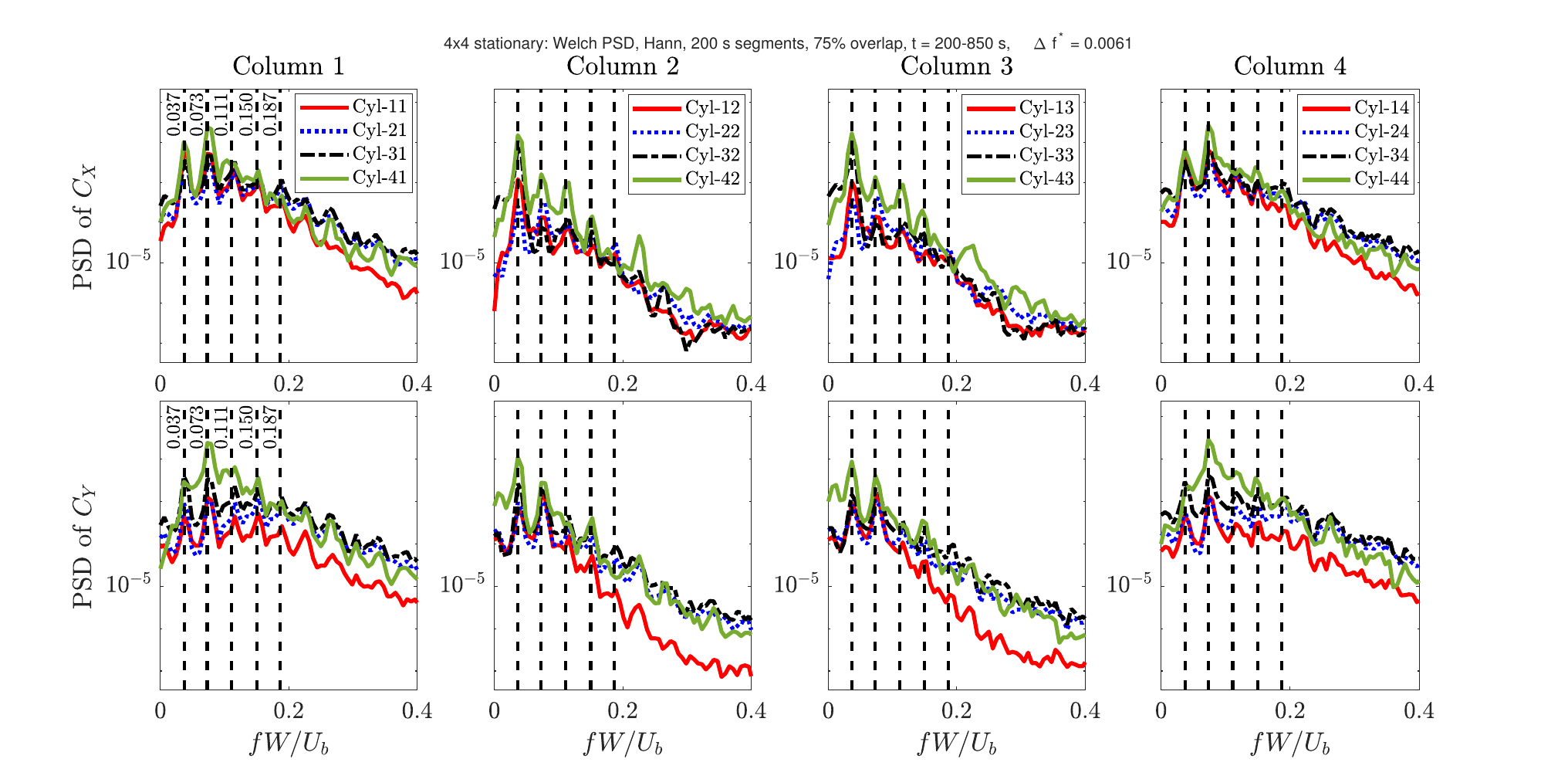}%
    \caption{Welch power spectra of the force coefficients for the cylinders in $4\times 4$ array system.}
    \label{powerSpectra_4by4_stationary}
\end{figure}

The PSD of the force coefficients for $2\times 2$, $3\times 3$, and $4\times 4$ array is shown in Figs. \ref{powerSpectra_2by2_stationary}, \ref{powerSpectra_3by3_stationary}, and \ref{powerSpectra_4by4_stationary}, respectively. The various frequencies of the force coefficients are also highlighted in the plot by dashed lines. Symmetry in the spectrum can be observed for $2\times 2$ array system in Fig. \ref{powerSpectra_2by2_stationary}. The dominant frequency for $C_X$ is $fW/U_b = 0.036$ for all the cylinders with its higher modes at $fW/U_b = 0.072, 0.108, 0.143,$ and $0.179$. The bottom row seems to have multiple dominant frequencies of $0.036$ and $0.108$ (third harmonic) for $C_Y$ while the frequency of $0.036$ dominates for the top row. For $3\times 3$ array in Fig. \ref{powerSpectra_3by3_stationary}, the first column (Cyl-11, 21, 31) has a dominant frequency for both the force coefficients at $fW/U_b=0.183$, which is not observed for any other cylinders. The low frequency content which is dominant is around $0.025$ for almost all the cylinders except the top-left corner cylinder (Cyl-31) where it is the second mode ($0.051$). For the $4\times 4$ array in Fig. \ref{powerSpectra_4by4_stationary}, harmonics of the frequency $fW/U_b = 0.037$ are observed. 

The arrays of cylinders have very low frequency plume sway representing the circulation of the fluid due to natural convection inside the cavity. However, a higher frequency of $fW/U_b = 0.183$ is observed for the first column of cylinders in the $3\times 3$ array configuration. Identification of these forcing frequencies is imperative for the subsequent sections where flow-induced oscillations of the elastically mounted cylinders will be investigated.

\subsection{Origin of $fW/U_b = 0.183$ in the $3\times 3$ array}
\label{vortex_core_origin}
\begin{figure}
    \centering
    \begin{subfigure}{\textwidth}
        \includegraphics[trim={0.1cm 0.1cm 0.1cm 0.1cm}, clip,width=\textwidth]{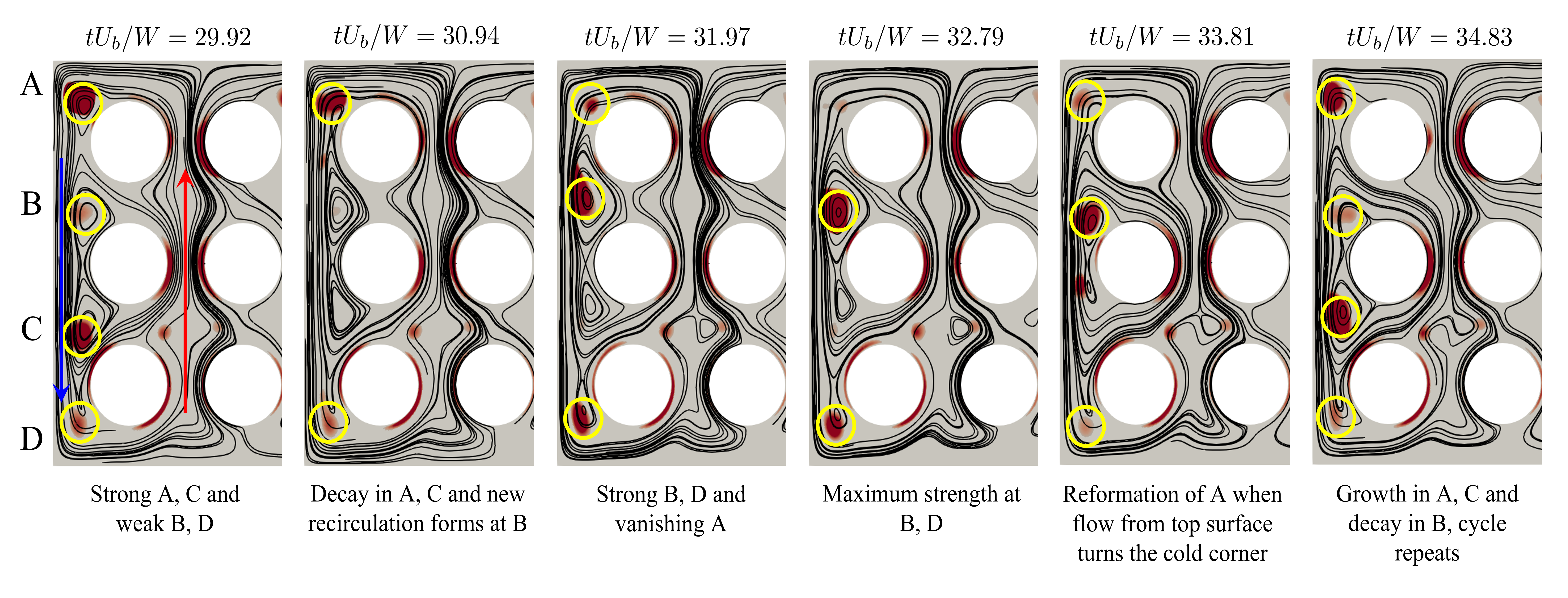}
        \caption{}
    \end{subfigure}
    
    \begin{subfigure}{\textwidth}
        \includegraphics[trim={0.1cm 5cm 0.1cm 0.1cm}, clip,width=\textwidth]{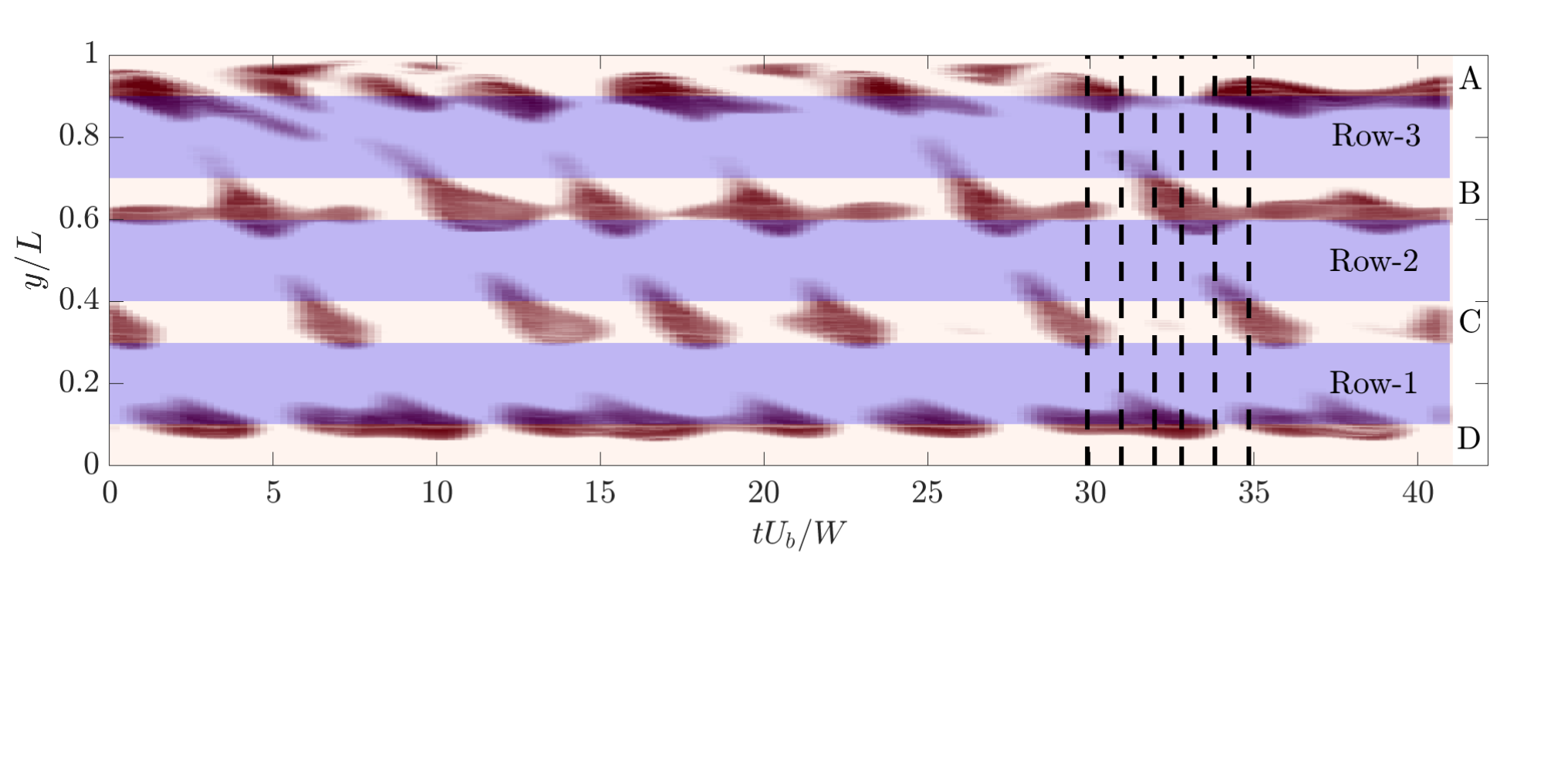}
        \caption{}
    \end{subfigure}
    \caption{Origin of $f_{wgp}W/U_b = 0.183$ using flow visualization in $3\times 3$ array: (a) Transition in the vortex core (visualized by $Q$-criterion along with streamlines) near the left wall of the array, and (b) Space-time map of $Q>0$ vortex cores near the left wall.}
    \label{fig:vortex_Cores_stationary}
\end{figure}

To understand the origin of the $fW/U_b=0.183$ for the first column cylinders (Cyl-11, 21, 31) in the $3\times 3$ array configuration, the flow is visualized using $Q$-criterion and streamlines in Fig. \ref{fig:vortex_Cores_stationary}(a). The figure shows the zoomed view of the left side of the enclosure focusing on Cyl-11, 21, and 31. The vortex cores are identified by the red patches and are encircled. Vortices are observed at four locations, A being the top-left corner, B and C representing the gaps between Cyl-21/Cyl-31 and Cyl-11/Cyl-21, respectively, and the bottom-left corner vortex located at D. The gaps lying between the cylinders and the wall (throat) are almost devoid of vorticity. The overall plume circulation direction is also shown by arrows where the hot plume rises between the first and second column and the colder fluid comes down from the gap between the first column and the left wall of the enclosure. The travelling of the vortex core from the top-left corner ($y/L = 1$) to the bottom-left corner ($y/L = 0$) within $0.012 < x/L < 0.15$ as time progresses is visualized in Fig. \ref{fig:vortex_Cores_stationary}(b). The maroon patches depict the vortex cores ($Q>0$) and the blue-colored rectangular patch identifies the throat region (between cylinders and wall). The instantaneous vortex cores with streamlines shown in Fig. \ref{fig:vortex_Cores_stationary}(a) refer to the time instances shown by dashed lines in Fig. \ref{fig:vortex_Cores_stationary}(b). Tracking the vortex cores shows that all the vortices at the four locations (A, B, C, and D) move with the same frequency of $fW/U_b \approx 0.183$.  

The flow carries hot fluid at the top surface of the enclosure into the top-left corner where the flow separates and forms the source vortex at A. The vortex then slides down around $0.08L$ before it is squeezed at the throat region (shown as slanted streaks in Fig. \ref{fig:vortex_Cores_stationary}(b)). One complete cycle is shown in the Fig. \ref{fig:vortex_Cores_stationary}(a). It starts with a strong vortex at A which travels down via the throat region near Cyl-31. As the vortex travels down, the vorticity at location A vanishes ($tU_b/W = 31.97$). Between each gap between the cylinders, the warm fluid passing through this gap interacts with the cold fluid coming down the left-wall side, thus forming a vortex.   During the same time duration, the strong vortex at C travels down the throat region near Cyl-11 to reach D. Therefore, at an instant ($tU_b/W=32.79$), the vorticity is at its maximum at locations B and D. As time progresses, vortex at A is formed again as vortex at B travels down the throat near Cyl-21 to reach C and vortex at D diffuses, and the cycle repeats. Therefore, vortex formation at neighboring gaps alternate because it takes about half a period for a disturbance to cross each throat region. For traversing the given row pitch of the array (half of total length from top to bottom) $p=D+g_c=0.2L+0.1L = 0.3L$ ($g_c$ being the gap between the cylinder surfaces), the phase velocity of the vortex core is about $c\approx 0.2U_b$. Thus, the total non-dimensional frequency of this vortex core oscillation would be $fW/U_b = (c/2p)W/U_b \approx 0.18$. This is called wall-gap plume oscillation frequency in the following sections and is denoted by $f_{wgp}W/U_b$.

\section{Quantification of added mass and fluid damping by free decay test}
\label{sec:added_mass_damping}
Since the cylinder array system is submerged in a fluid, there will be added mass and fluid damping effects, especially due to the compact arrangement of the cylinders. To quantify the same, free decay tests are performed where the cylinder of interest is given an initial displacement of $\delta/D = 0.034, 0.5, $ and $0.067$ for $2\times 2$, $3\times 3$, and $4\times 4$ arrays, respectively while keeping other cylinders stationary or non-moving, and the decay response is recorded. The initial displacement is given in the Y direction. The extracted added mass coefficient for each cylinder is defined as $C_M = m_a/\rho^\text{f}V_\text{disp}$, where $V_\text{disp}=(\pi/4)D^2$ is the volume of the fluid displaced by the cylinder. The fluid damping ratio $\zeta_\text{f} = c_\text{f}/(2\omega_{n,wet}(m+m_a))$, where $c_\text{f}$ is the fluid-induced damping and $\omega_{n,wet} = 2\pi f_{n,wet}$. These values are tabulated in Table \ref{tab:CM_Zeta_individual}.

Since the $2\times 2$ array is symmetric, all the cylinders (Cyl-11, 12, 21, 22) give the same decay response resulting in identical fluid damping and added mass. For the $3\times 3$ array, the corner cylinders (Cyl-11/13/31/33) are more confined than the other cylinders (with walls on two sides and 3 cylinders on the other sides), thus giving $\zeta_\text{f}$ and $C_M$ highest in the group. The least confinement is observed for Cyl-22 with 8 cylinders having curved walls surrounding it. The damping and added mass of the other cylinders Cyl-21/23 and Cyl-12/32 fall in between the values of corner and middle cylinders, as expected. The symmetric nature of the arrangement returns in the $4\times 4$ array for which Cyl-11,14,41,44 lie at the extreme corners, Cyl-12,13,42,43 lie in the inner side of first and last rows, Cyl-21,24,31,34 lie at the outer sides of the inner rows and Cyl-22,23,32,33 are the inner cylinders of the inner rows. Note that there will be some variation in the added mass coefficient when applied to the cylinder moving in X direction, since the decay test was only performed with initial displacement in Y direction. However, Table \ref{tab:CM_Zeta_individual} shows that the variation in the added mass coefficient is not that significant across the configurations and can be assumed to vary between $C_M \approx 1.5-1.7$.
\begin{table}
\centering 
\caption{Fluid damping ratio and added mass coefficient from the free decay test performed for each cylinder in the array.} 
\label{tab:CM_Zeta_individual} 
\begin{tabular}{llcc}
\hline
\textbf{Array} & \textbf{Cylinder} & $\zeta_\text{f}$ & $C_M$\\ \hline
$2\times2$ & All cylinders & 0.033 &  1.714 \\ \hline
\multirow{4}{*}{$3\times3$} & Cyl-11, 13, 31, 33 & 0.048 &  1.714 \\ 
& Cyl-12, 32 & 0.045 &  1.604 \\ 
& Cyl-21, 23 & 0.046 &  1.656 \\ 
& Cyl-22 & 0.044 & 1.547  \\ \hline
\multirow{4}{*}{$4\times4$} & Cyl-11, 14, 41, 44 & 0.063 & 1.764  \\ 
& Cyl-12, 13, 42, 43 & 0.059 &  1.646 \\ 
& Cyl-21, 24, 31, 34 & 0.061 &  1.695 \\ 
& Cyl-22, 23, 32, 33 & 0.057 &  1.578 \\ \hline
\end{tabular}
\end{table}

After the identification of the flow and thermal behavior for stationary arrays and quantification of the added mass for the cylinders in the array configuration, the response of the elastically mounted cylinders under the action of natural convection inside the enclosed cavity is investigated for $3\times 3$ and $4\times 4$ arrays. The $2\times 2$ array has not been considered for further analyses as due to the large diameter of the cylinders, the motion resulted in large deformation of the surrounding mesh leading to mesh failure.  

\section{Response of elastically mounted array subjected to natural convection for $3\times 3$ cylinder array}
\label{sec:FSI-3by3}

From the power spectra of the stationary cylinders in the $3\times 3$ array configuration shown in Fig. \ref{powerSpectra_3by3_stationary}, it is observed that only the left column cylinders (Cyl-11, 21, and 31) feel the wall-gap plume oscillation frequency at $f_{wgp}W/U_b = 0.183$. All the other cylinders experience slow fluid forcing of small frequencies due to slow enclosure circulation (or plume sway) at $f_{ps}W/U_b \approx 0.025, 0.051,$ and $0.077$, while no peak is observed at $fW/U_b = 0.183$. Therefore, the left column cylinders have a potential frequency to lock-in when the cylinders are elastically mounted.

To analyze the response of the cylinders subjected to the fluid forcing, each cylinder is elastically mounted on a spring system. Since the fluid forcing is fixed at $Ra=1.5\times 10^6$, the reduced velocity of each cylinder is varied by varying the stiffness of the mounted spring such that the wet natural frequency of each cylinder reaches the wall-gap plume oscillation forcing frequency of $f_{wgp}W/U_b = 0.183$. Thus, the reduced velocity range vary as $U_r \in [10-23]$ with reduced velocity for each cylinder defined based on the wet natural frequency calculated from the added mass obtained from the free decay test. The domain with elastically mounted cylinders is initialized with the stationary cylinder results to bypass the initial transient developing flow to reduce the computational cost. 

Furthermore, as the forcing on the cylinders has a mean component due to buoyancy (See Tables \ref{tab:CXCYNu_Mean_3by3} and \ref{tab:CXCY_Mean_4by4}), the elastic springs of each cylinder is pre-loaded by the mean forces such that only the amplitude fluctuations about mean are recorded. Otherwise, static drift and sag are observed for the cylinders which changes the mean positions and thus the underlying geometry, influencing the coupled thermal-fluid-structure interactions. This is carried out by subjecting the elastically mounted rigid body to a preload ($\boldsymbol{P}^\mathrm{s}$ in Eq. (\ref{MBD})) depending on the time-averaged force coefficients.   

\subsection{Amplitude and frequency response}
\label{sec:amp_freq_3by3}
\begin{figure}
\centering
\begin{subfigure}{0.5\textwidth}
    \includegraphics[trim={8cm 0.1cm 9cm 0.1cm}, clip,width=\textwidth]{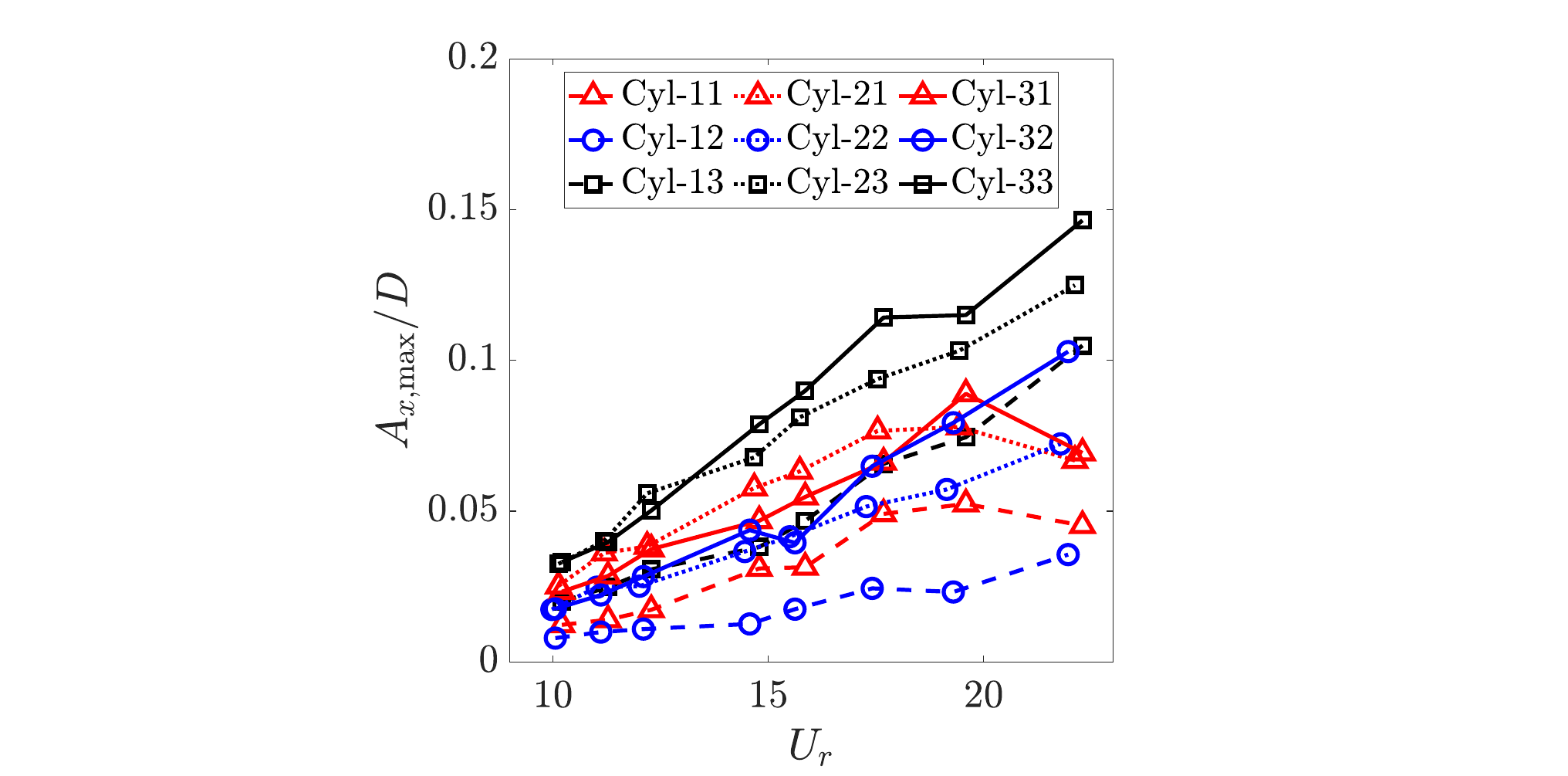}
    \caption{}
\end{subfigure}%
\begin{subfigure}{0.5\textwidth}
    \includegraphics[trim={8cm 0.1cm 9cm 0.1cm}, clip,width=\textwidth]{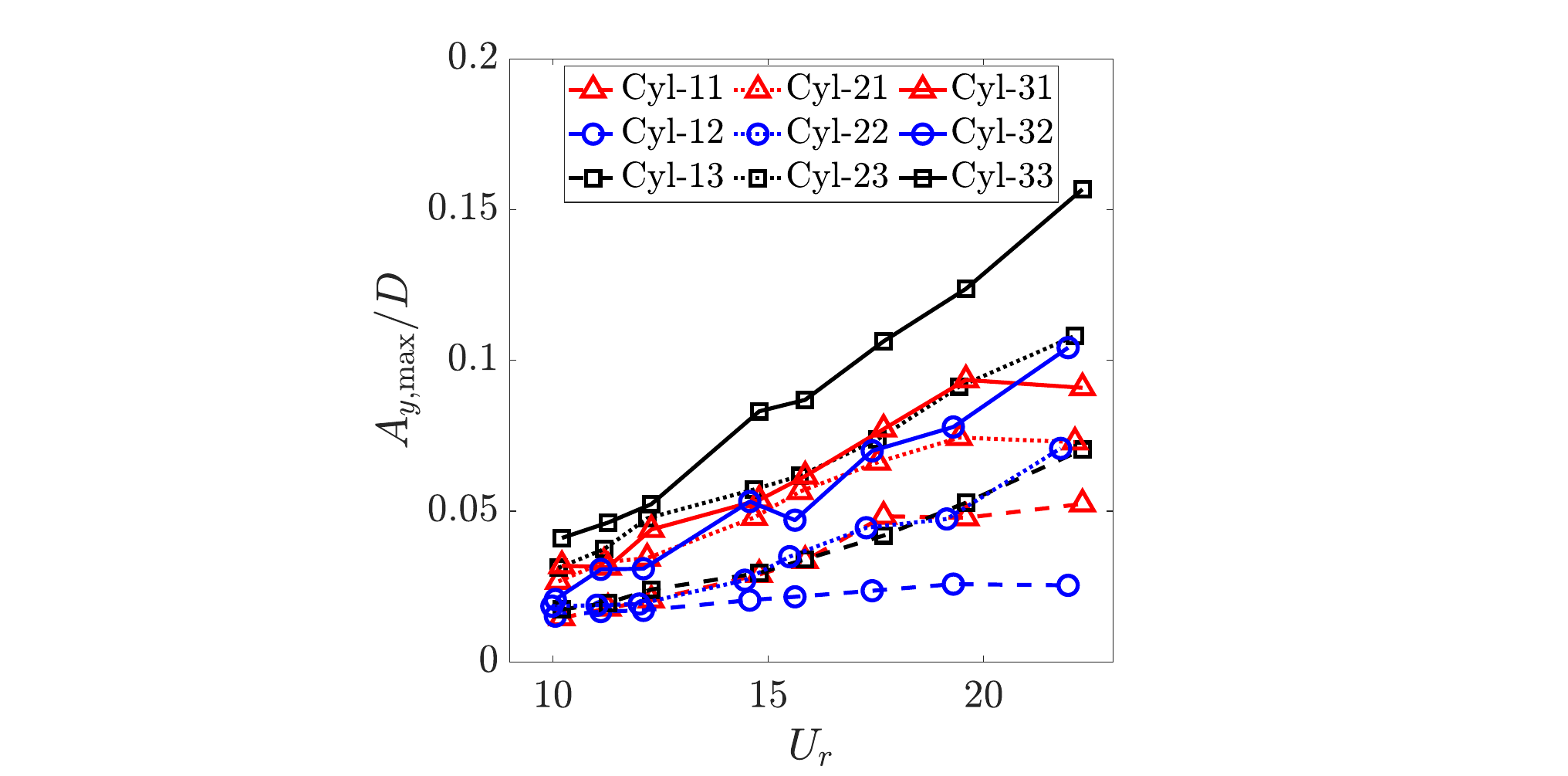}
    \caption{}
\end{subfigure}%
\caption{Maximum amplitude of the response of the cylinders in $3\times 3$ array with the reduced velocity $U_r$: (a) $A_{x,\mathrm{max}}/D$, and (b) $A_{y,\mathrm{max}}/D$.}
\label{fig:Amp-9Cyl-FSI-MAX}
\end{figure}

The non-dimensional maximum amplitude response for X and Y directions for each cylinder with corresponding reduced velocity is shown in Fig. \ref{fig:Amp-9Cyl-FSI-MAX}. It can be observed that the amplitudes of Cyl-11, 21, and 31 (1st column) reach the maximum amplitude around $U_r = 17.5-20$ and then decrease with further increase in $U_r$. The amplitudes of all other cylinders increase monotonically with $U_r$. The behavior remains almost identical in both the directions. In Y direction, apart from the cylinders in the first column, Cyl-12 also shows a maximum around $U_r \approx 19$ although the amplitude is very small. 

\begin{figure}
\centering
\begin{subfigure}{0.5\textwidth}
    \includegraphics[trim={8cm 0.1cm 9cm 0.1cm}, clip,width=\textwidth]{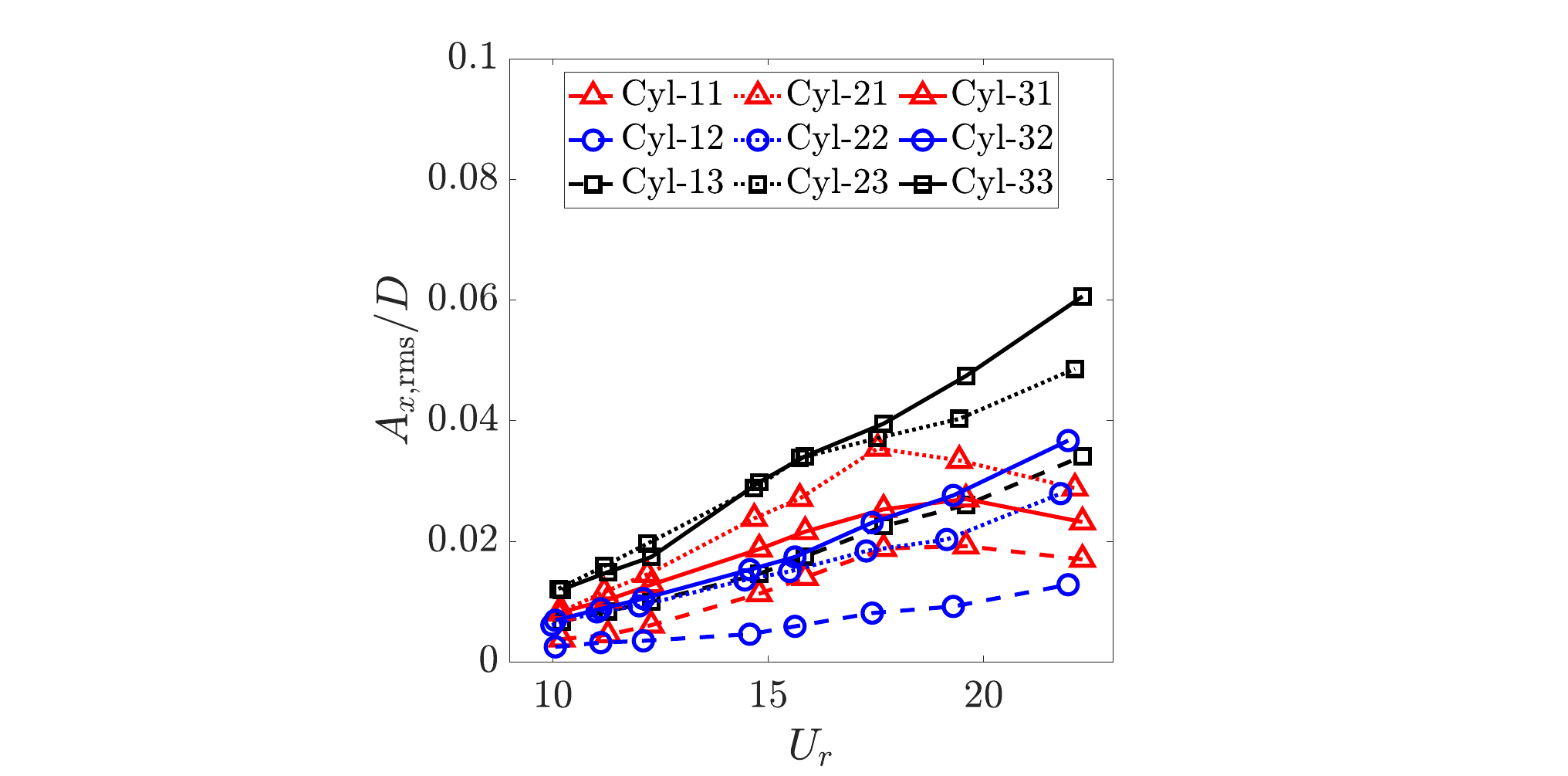}
    \caption{}
\end{subfigure}%
\begin{subfigure}{0.5\textwidth}
    \includegraphics[trim={8cm 0.1cm 9cm 0.1cm}, clip,width=\textwidth]{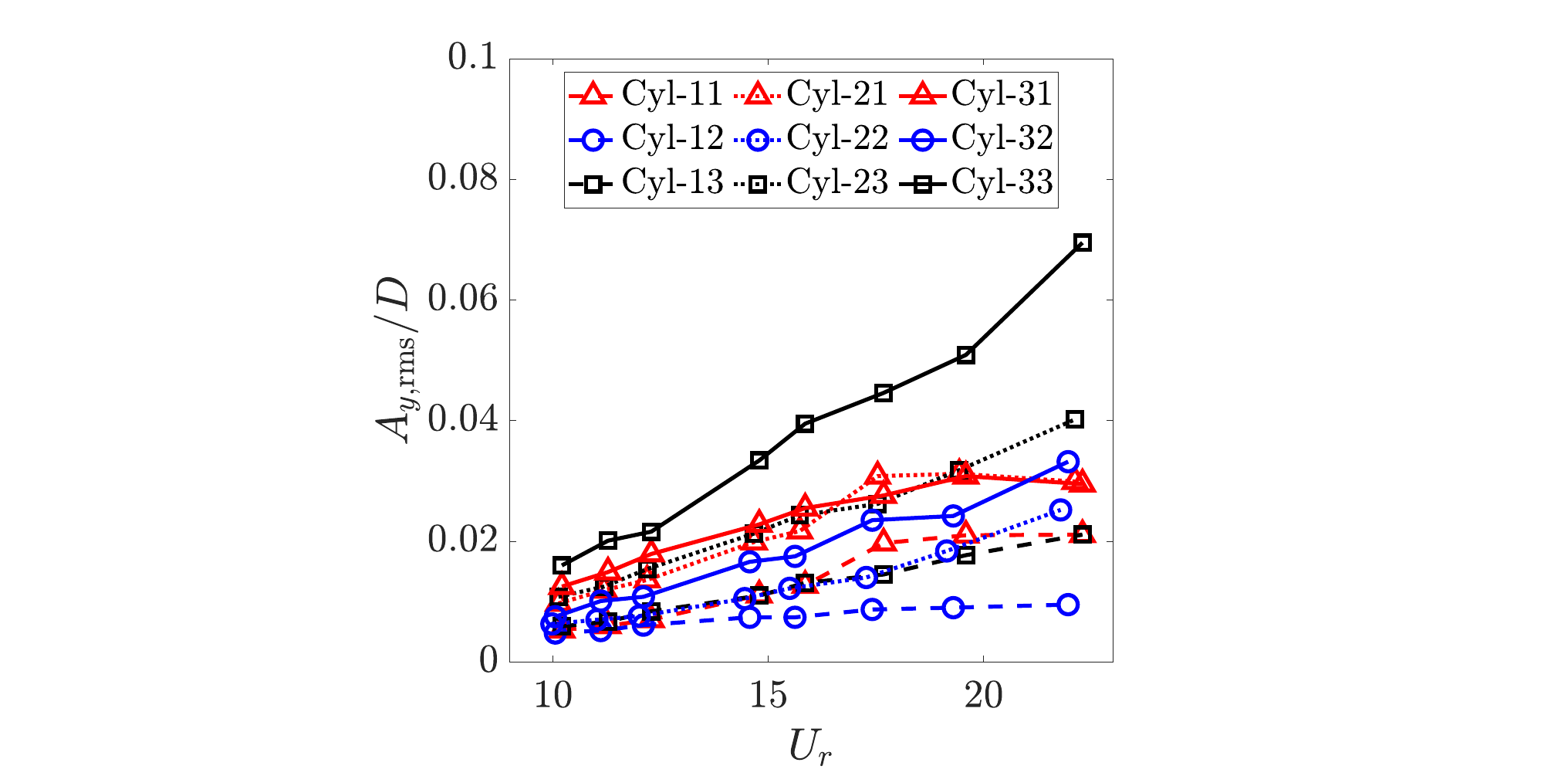}
    \caption{}
\end{subfigure}%
\caption{Root-mean-square amplitude of the response of the cylinders in $3\times 3$ array with the reduced velocity $U_r$: (a) $A_{x,\mathrm{rms}}/D$, and (b) $A_{y,\mathrm{rms}}/D$.}
\label{fig:Amp-9Cyl-FSI-RMS}
\end{figure}

The root-mean-square (RMS) amplitude of the cylinders is shown in Fig. \ref{fig:Amp-9Cyl-FSI-RMS}. The trends remain similar to that of maximum amplitude where maximum is observed for Cyl-11, 21, and 31, while the response keeps increasing with $U_r$ for other cylinders and is the largest for Cyl-33. The large RMS amplitude represents the slow sway motion of the cylinders and are a result of the slow fluid forcing at low frequencies. Therefore, the response of the cylinders may be categorized into two types: a bounded, synchronized response observed for the first column (Cyl-11, 21, 31) and a growing forced buffeting response as a result of circulation in the enclosure for all the remaining cylinders. 

The dominant response frequency of each displacement component is obtained from the periodogram of the displacement fluctuation over the analysis window $tU_b/W \ge t_0 = 41$ (initial transient excluded) consisting of 8300-11000 sample data points. Let $u^\mathrm{s}_{j,n} = u^\mathrm{s}_j(t_0 + n\Delta tU_b/W)$, $n = 0,\dots,N-1$, denote the $N$ samples of the cylinder displacement in direction $j \in \{x, y\}$, with sampling interval $\Delta t$ equal to the time step, and let $u^\mathrm{s}_{j,\mathrm{mean}}$ be its mean over the sampling window. To reduce spectral leakage, the fluctuation is multiplied by the Hann window $w_n = \tfrac12\left[1-\cos\left(\tfrac{2\pi n}{N-1}\right)\right]$ and zero-padded to $N_p = 8N$ points before computing its discrete Fourier transform,
\begin{align}
  X_{p,j}(f_k) &= \sum_{n=0}^{N-1} w_n \left(u^\mathrm{s}_{j,n} - u^\mathrm{s}_{j,\mathrm{mean}}\right)
                 e^{-\mathrm{i} 2\pi k n / N_p},
  \qquad f_k = \frac{k}{N_p\,\Delta t},\quad k = 0,\dots,\frac{N_p}{2},\\
  f_j &= \underset{f_k \in \mathcal{B}}{\arg\max}\; \left|X_{p,j}(f_k)\right|^2 ,
\end{align}
where $\mathrm{i}=\sqrt{-1}$. Zero-padding interpolates the spectrum and the frequency resolution is $\Delta f = 1/(N\Delta t)$, i.e. $\Delta f\,W/U_b \approx 0.002$. Two search bands $\mathcal{B}$ are used, viz., (i) the overall band $\mathcal{B}_\mathrm{tot} = \{f_k : 0.015 < f_k W/U_b < 0.70\}$, which gives the overall dominant frequency including the low-frequency enclosure-scale sway, and (ii) the band around the natural frequency, $\mathcal{B}_\mathrm{res} = \{f_k : \max(0.10\,U_b/W,\; 0.75 f_{n,wet}) \le f_k \le 2.0 f_{n,wet}\}$, which gives the response frequency near the natural frequency of the cylinder. The lower limit $0.10\,U_b/W$ excludes the sway components at the lowest stiffness, where $0.75 f_{n,wet}$ falls below it.

\begin{figure}
\centering
\begin{subfigure}{0.5\textwidth}
    \includegraphics[trim={8cm 0.1cm 9cm 0.1cm}, clip,width=\textwidth]{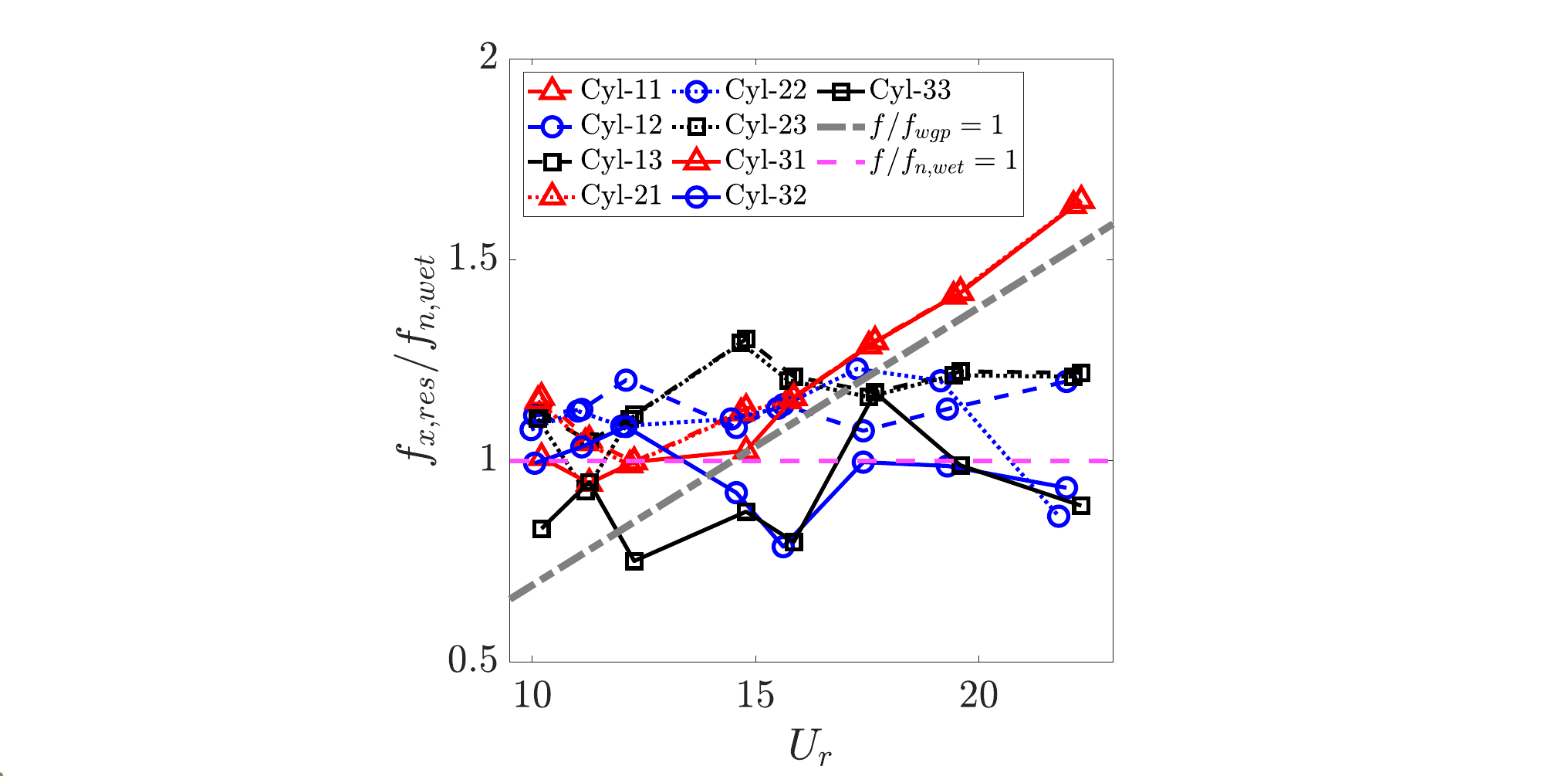}
    \caption{}
\end{subfigure}%
\begin{subfigure}{0.5\textwidth}
    \includegraphics[trim={8cm 0.1cm 9cm 0.1cm}, clip,width=\textwidth]{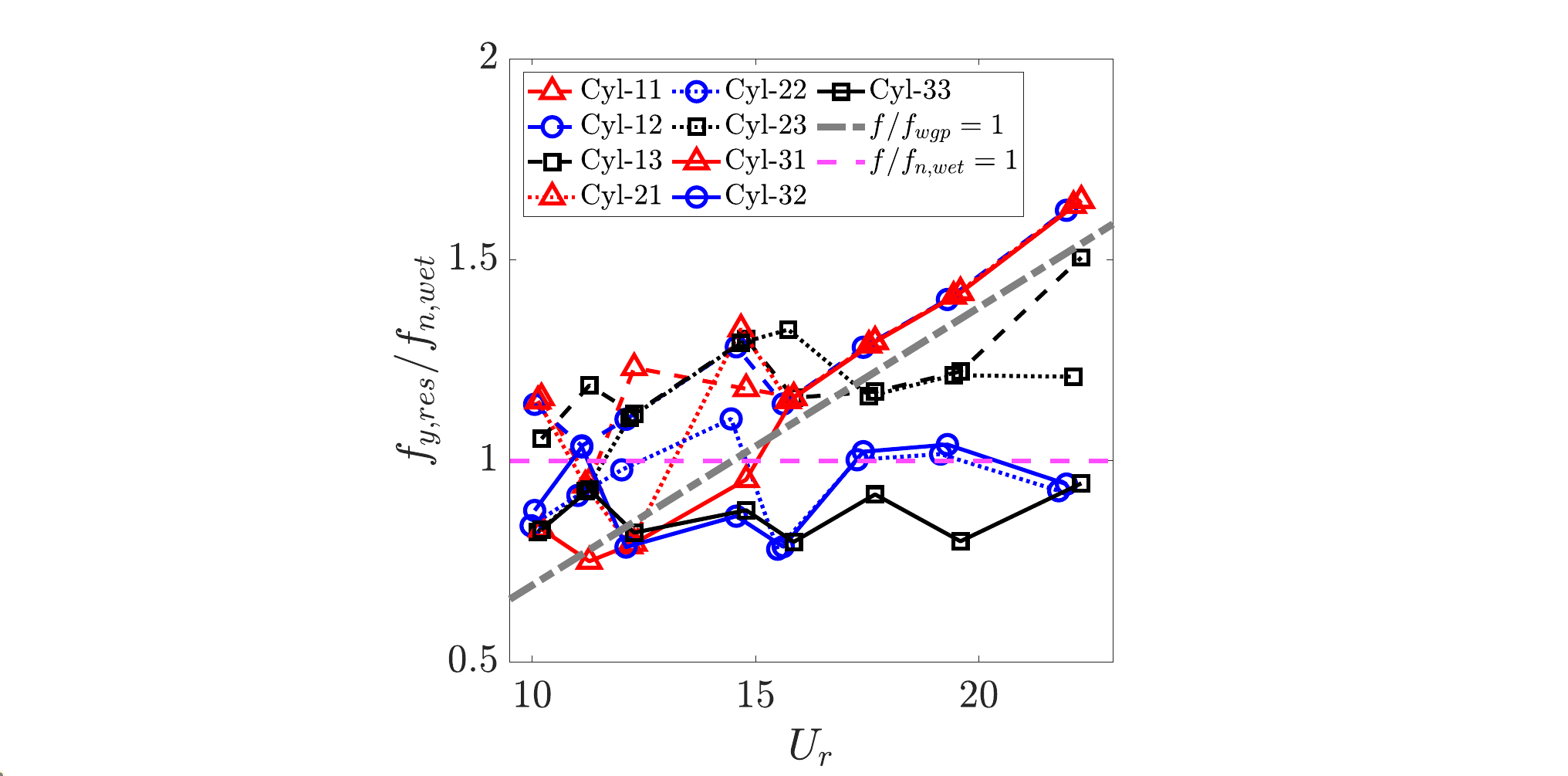}
    \caption{}
\end{subfigure}%
\caption{Non-dimensional frequency response of the cylinders in $3\times 3$ array with the reduced velocity $U_r$: (a) $f_{x,res}/f_{n,wet}$, and (b) $f_{y,res}/f_{n,wet}$.}
\label{fig:Freq-9Cyl-FSI}
\end{figure}

The dominant frequency near the natural frequency $(\mathcal{B}_\mathrm{res})$ is shown in Fig. \ref{fig:Freq-9Cyl-FSI} for each cylinder with their corresponding reduced velocity range. The synchronization response of the first column of cylinders is evident from the frequency ratio plots. The horizontal dashed line denotes the resonance where $f_{res}/f_{n,wet} = 1$ and the dash-dotted inclined line represents the higher plume frequency ratio ($f/f_{wgp} = 1$) where the response is synchronized with the wall-gap plume oscillation frequency. It is noticed that the left column synchronizes with the plume frequency around $U_r \approx 15.7$ in both X and Y directions. Cyl-12 also seems to be synchronized with the plume frequency in Y direction. This onset of synchronization also happens when the frequency ratio reaches 1, i.e., when the natural frequency of the cylinders approach the plume frequency.  Furthermore, the motion of the cylinders changes the wall-gap plume oscillation frequency from $f_{wgp}W/U_b = 0.183$ to $f_{wgp}W/U_b = 0.195$ which is shown as the shifted line for those cylinders. The disordered frequencies for all the other cylinders come from the fact that the frequency spectrum is broadband. 

\begin{figure}
\centering
\includegraphics[trim={1cm 0.1cm 1cm 0.1cm}, clip,width=\textwidth]{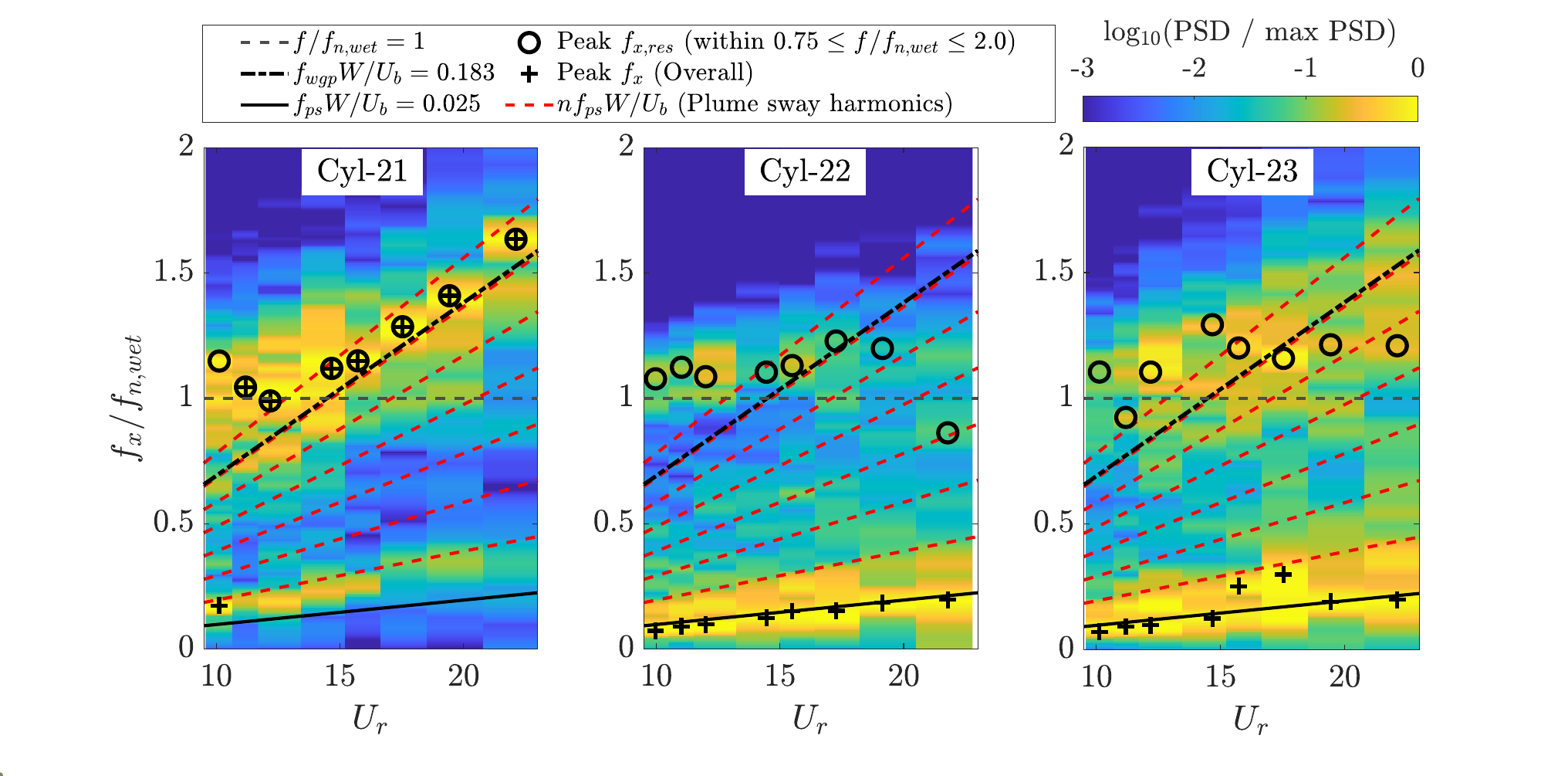}
\caption{Welch PSD of the displacement in X, normalised by its maximum over $fW/U_b > 0.015$ at each $U_r$ overlaid with the response frequency for the middle row (Cyl-21, 22, and 23).}
\label{fig:Freq-map-X-9Cyl-FSI}
\end{figure}

\begin{figure}
\centering
\includegraphics[trim={1cm 0.1cm 1cm 0.1cm}, clip,width=\textwidth]{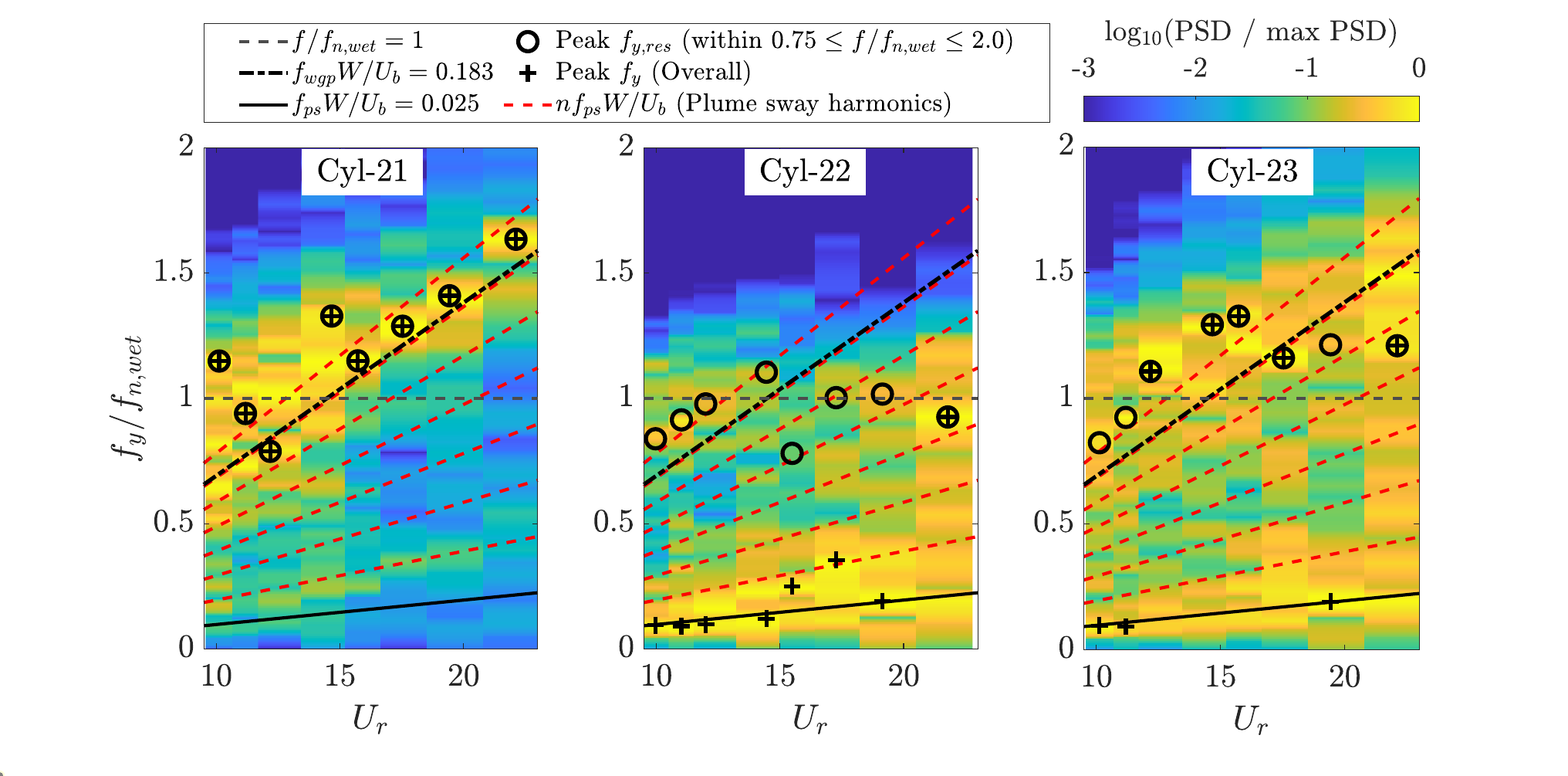}
\caption{Welch PSD of the displacement in Y, normalised by its maximum over $fW/U_b > 0.015$ at each $U_r$ overlaid with the response frequency for the middle row (Cyl-21, 22, and 23).}
\label{fig:Freq-map-Y-9Cyl-FSI}
\end{figure}

To further understand the variation in the dominant frequencies, frequency maps in X and Y directions are shown in Figs. \ref{fig:Freq-map-X-9Cyl-FSI} and \ref{fig:Freq-map-Y-9Cyl-FSI}, respectively for the middle row consisting of Cyl-21, 22, and 23. The figure consists of the Welch PSD spectrum of the displacement in X and Y directions, normalised by its maximum over $fW/U_b > 0.015$ at each reduced velocity $U_r$. The dominant frequencies considering the two bands stated above have been overlaid on the Welch spectrum. The solid line depicts the first sway mode at $f_{ps}W/U_b = 0.025$ along with its harmonics by dashed red-colored lines. The higher wall-gap plume oscillation frequency of 0.183 is also shown by dash-dotted line lying near the 7th mode. The overall peak response frequency is depicted by \enquote*{$+$} symbol while the dominant frequency in the range of $f/f_{n,wet} \in [0.75, 2]$ is shown by \enquote*{$\medcirc$} for each $U_r$. This is carried out to isolate the response due to the slow plume sway. The synchronization for the left column cylinder (Cyl-21) can be clearly observed from the frequency map in both the directions. In the X direction, the overall dominant frequency starts with the weak sway at $U_r = 10.1$ (Fig. \ref{fig:Freq-map-X-9Cyl-FSI}), but the response stays near $1.0-1.15 f_{n,wet}$ before the synchronization begins. At $U_r \approx 15.7$, when $f/f_{n,wet}=1$, all the frequencies coincide. The response after $U_r = 15.7$ deviates from $f_{n,wet}$ line and follows $f_{wgp}$, signifying the lock-in phenomenon. This is also observed for the response in Y direction in Fig. \ref{fig:Freq-map-Y-9Cyl-FSI}. The responses of Cyl-22 and Cyl-23 never follow the wall-gap plume oscillation frequency. The overall response is dominated by the slow plume sway closer to the first mode while some peaks are observed near to the wet natural frequency of the cylinders. 


In the synchronized regime $(U_r\geq 15.7)$, the left column responds at $fW/U_b = 0.193–0.196$, i.e. 5–7\% above the wall-gap plume oscillation frequency of the stationary array ($f_{wgp}W/U_b = 0.183$). The synchronized frequency is independent of the spring stiffness, remaining fixed while $f_{n,wet}$ decreases by 29\% across this range. The frequency is therefore set by the wall-gap plume oscillation and is modified by the motion of the cylinders it interacts with. This modification is consistent with a mutual synchronization between a self-sustaining flow oscillation and the structure, analogous to vortex-induced vibration, where the synchronized frequency differs from the stationary body Strouhal frequency \cite{KHALAK1999813, ANNREV_Govardhan2004}. Note that in the considered range for the reduced velocity, the end of the synchronization was not captured as the stiffness of the mounted cylinders tends to be very low at high $U_r$ resulting in large distortion of the mesh.

\begin{figure}
\centering
\begin{subfigure}{0.5\textwidth}
    \includegraphics[trim={8cm 0.1cm 9cm 0.1cm}, clip,width=\textwidth]{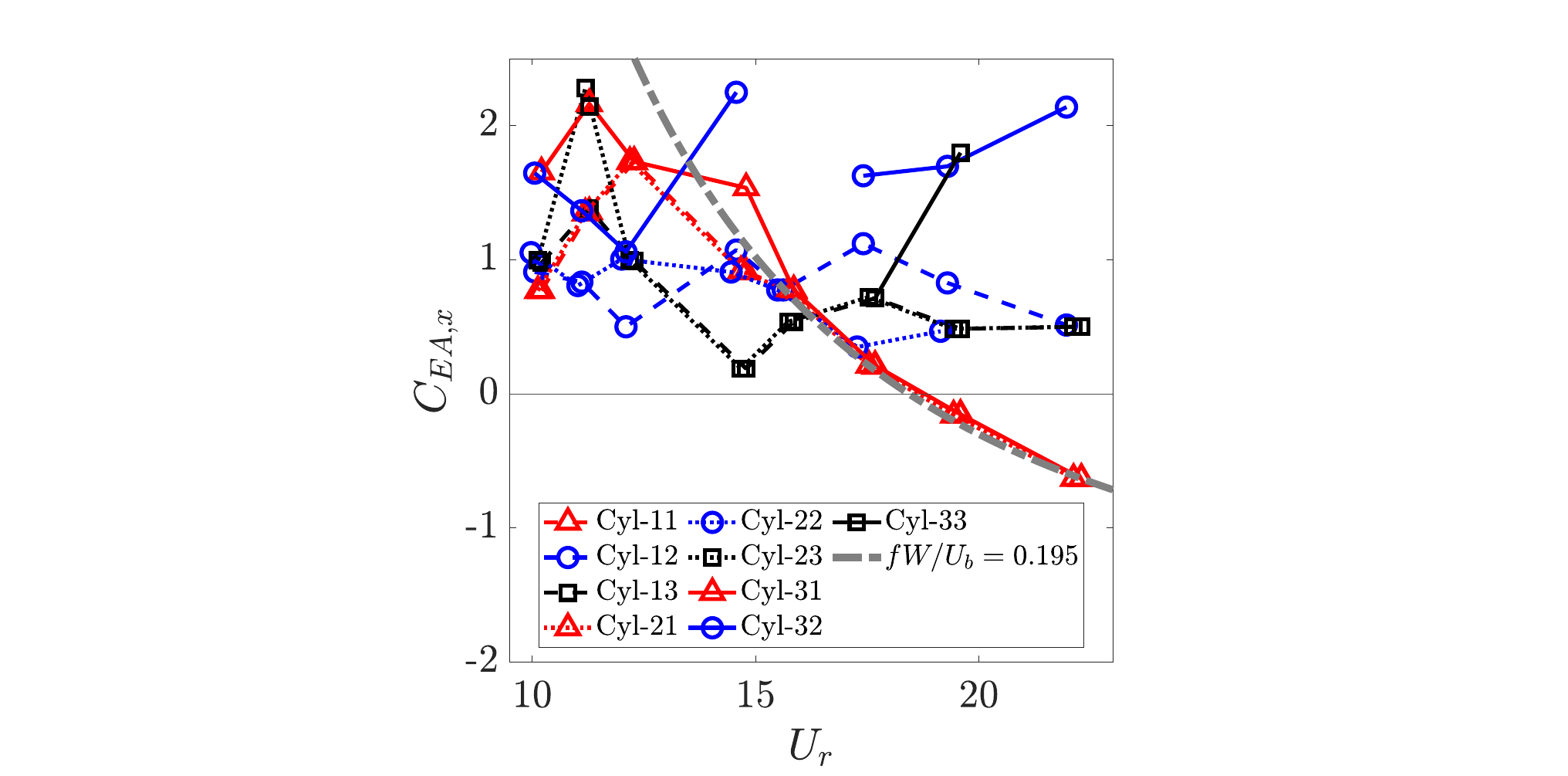}
    \caption{}
\end{subfigure}%
\begin{subfigure}{0.5\textwidth}
    \includegraphics[trim={8cm 0.1cm 9cm 0.1cm}, clip,width=\textwidth]{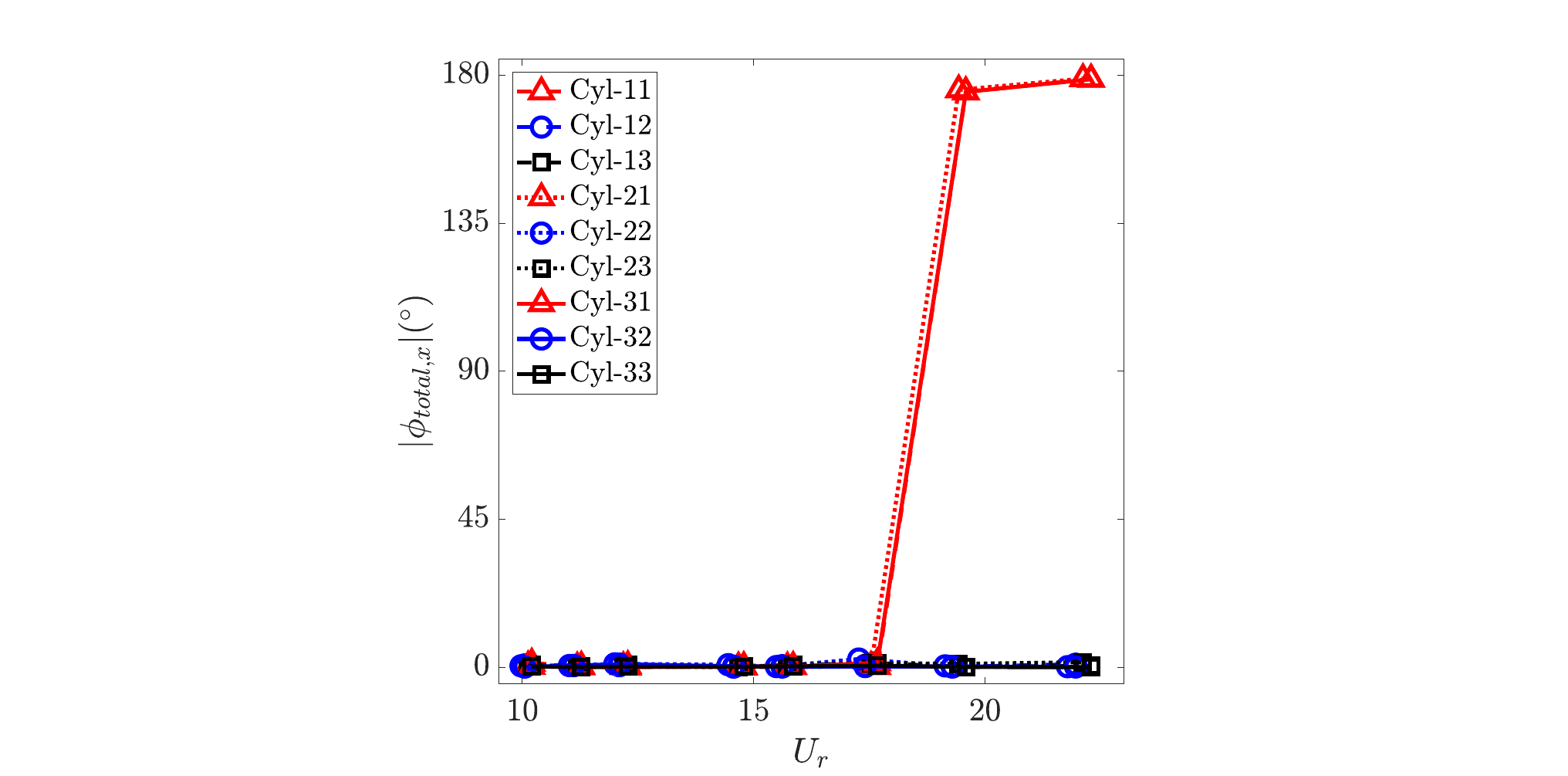}
    \caption{}
\end{subfigure}%
\caption{Variation of the (a) effective added mass coefficient, and (b) total phase between fluid forces and the displacement, with reduced velocity $U_r$ for the $3\times 3$ array system.}
\label{fig:EffectiveAddedMass-Phase-X-9Cyl-FSI}
\end{figure}

\begin{figure}
\centering
\begin{subfigure}{0.25\textwidth}
    \includegraphics[trim={8cm 0.1cm 9cm 0.1cm}, clip,width=\textwidth]{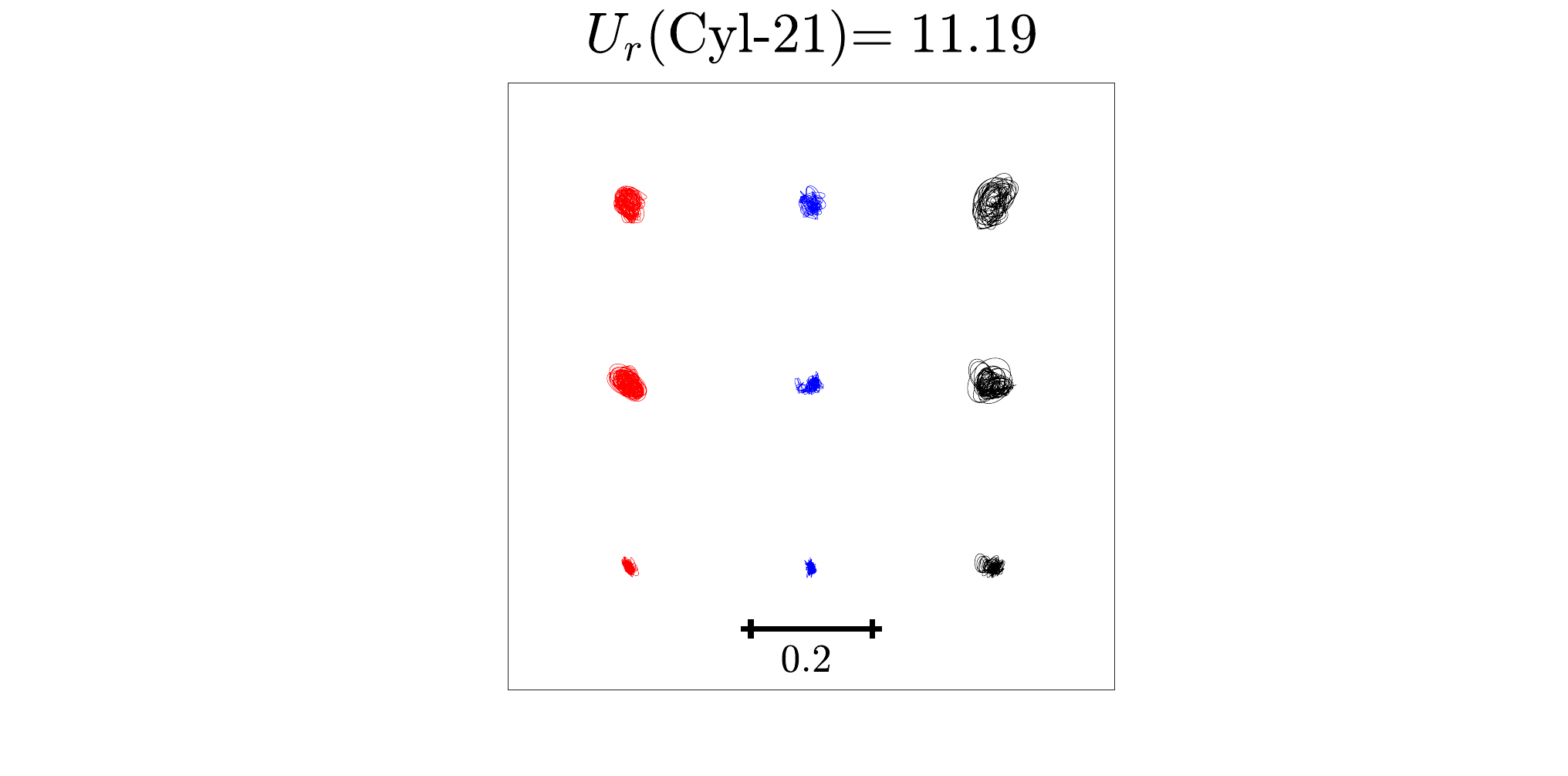}
    \caption{}
\end{subfigure}%
\begin{subfigure}{0.25\textwidth}
    \includegraphics[trim={8cm 0.1cm 9cm 0.1cm}, clip,width=\textwidth]{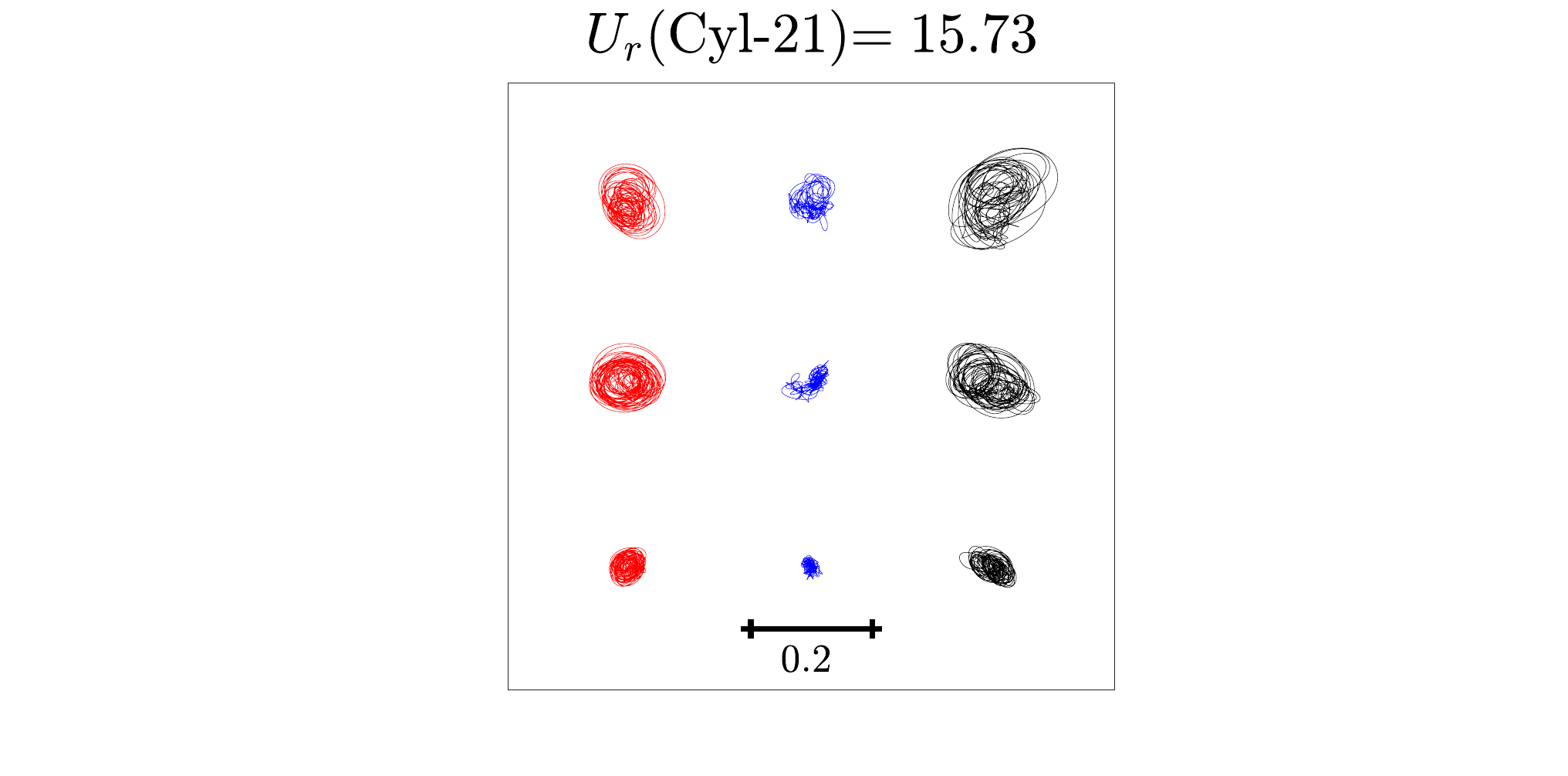}
    \caption{}
\end{subfigure}%
\begin{subfigure}{0.25\textwidth}
    \includegraphics[trim={8cm 0.1cm 9cm 0.1cm}, clip,width=\textwidth]{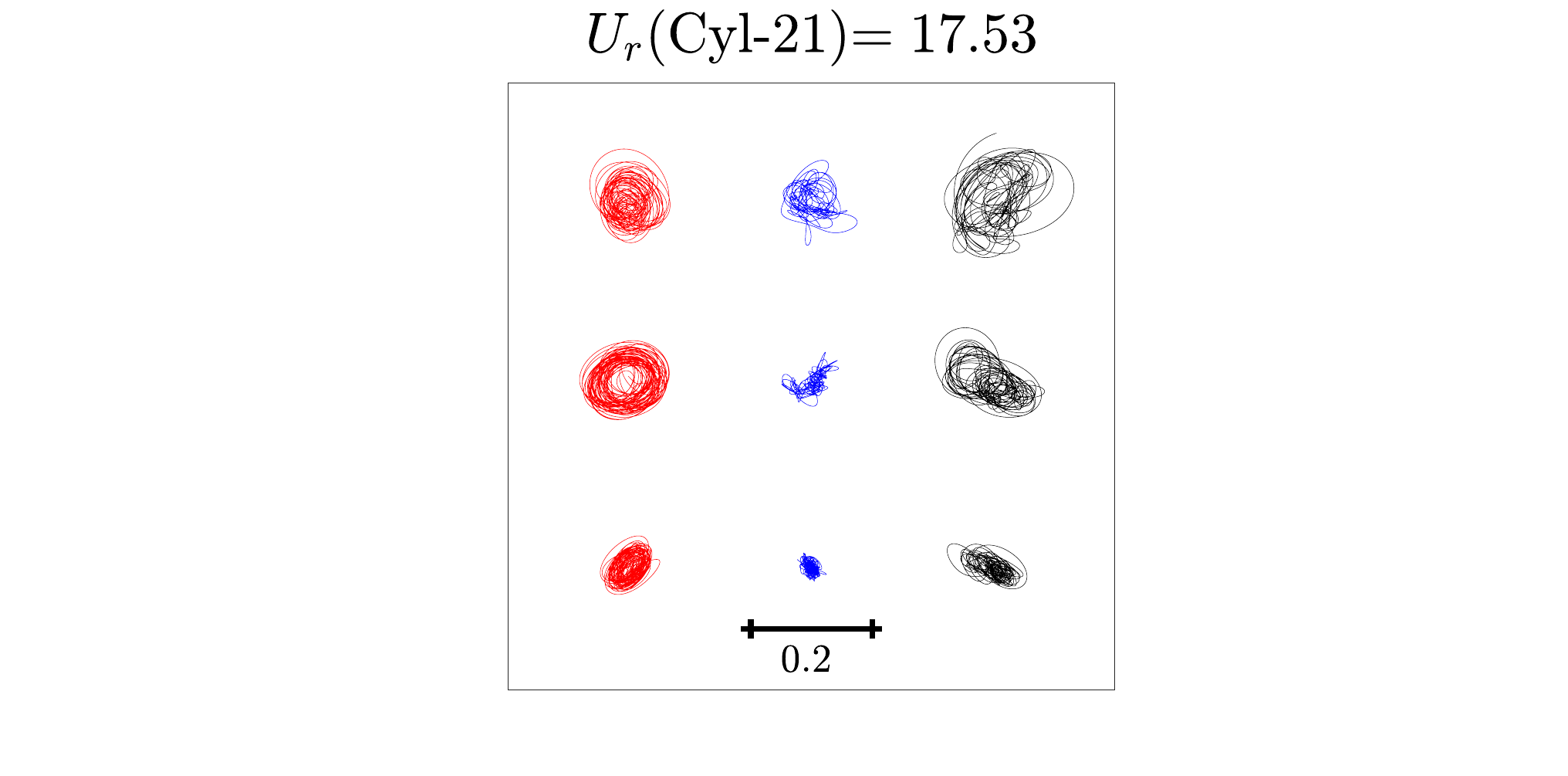}
    \caption{}
\end{subfigure}%
\begin{subfigure}{0.25\textwidth}
    \includegraphics[trim={8cm 0.1cm 9cm 0.1cm}, clip,width=\textwidth]{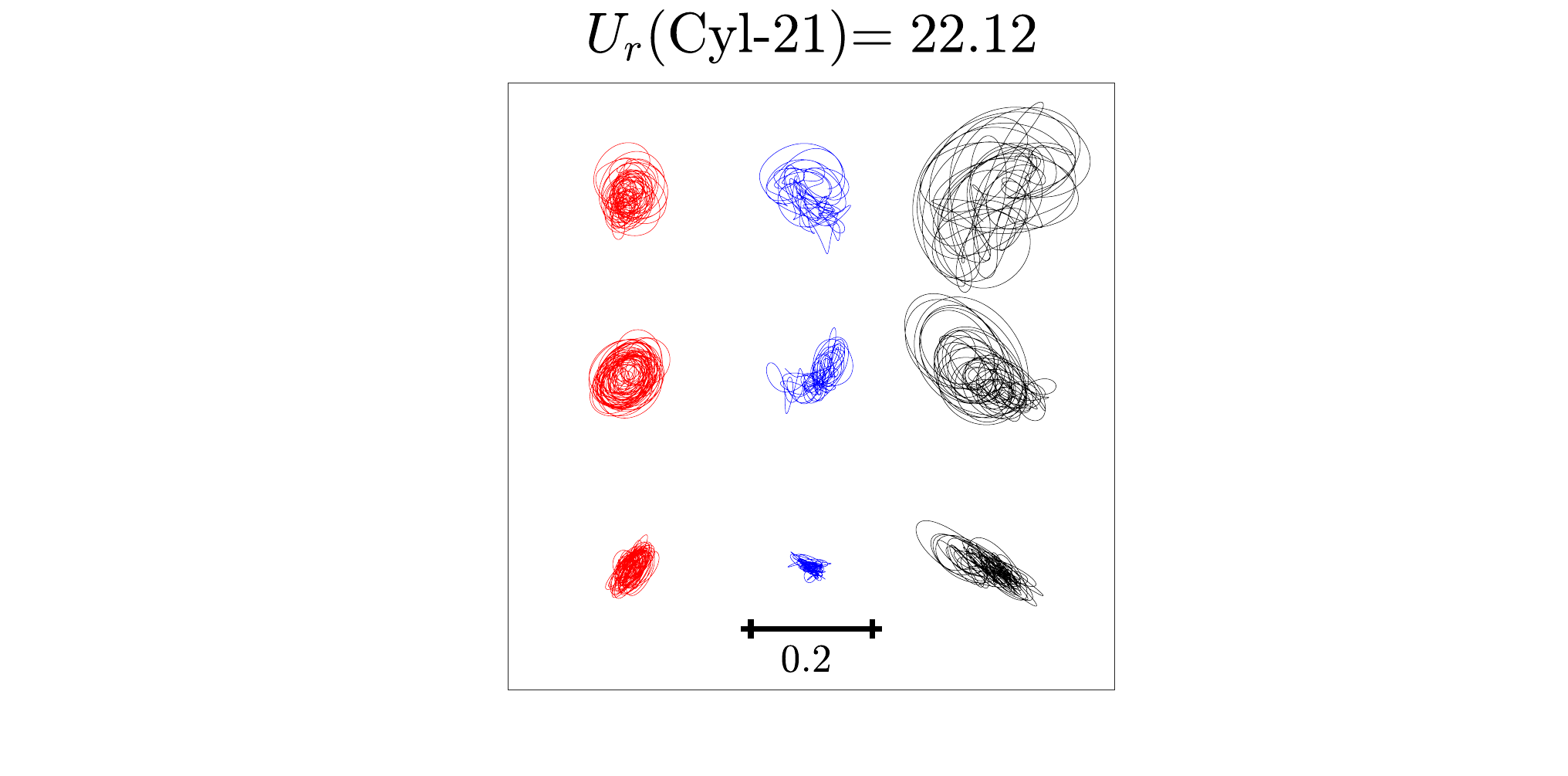}
    \caption{}
\end{subfigure}%
\caption{Trajectory of cylinders in the $3\times 3$ array system at reduced velocities of $U_r$ (evaluated for Cyl-21): (a) 11.2, (b) 15.7,  (c) 17.5 and (d) 22.1. The horizontal line shows the measure of the non-dimensional amplitude (normalized by $D$) for scale.}
\label{fig:Trajectories-9Cyl-FSI}
\end{figure}

The variation in the effective added mass and the phase difference between forces and displacement is also shown in Figs. \ref{fig:EffectiveAddedMass-Phase-X-9Cyl-FSI}(a) and \ref{fig:EffectiveAddedMass-Phase-X-9Cyl-FSI}(b), respectively, for the response in the X direction. The effective added mass is defined as 
\begin{align}
    C_{EA,x} &= (m^* + C_M)\bigg( \frac{f_{n,wet}}{f_{x,res}} \bigg)^2 - m^*,
\end{align}
where $f_{x,res}$ is the response frequency in the X direction. Before synchronization or lock-in, the effective added mass coefficient for the first column is $0.9-1.6$ (except Cyl-31), which is at or below the still-fluid value ($1.55-1.71$ in Table \ref{tab:CM_Zeta_individual}). From $U_r = 15.7$, $C_{EA,x}$ of Cyl-11, 21, and 31 cylinders converge to the curve depicting $fW/U_b = 0.195$ and cross zero around $U_r \approx 19$. This behavior is similar to the lower branch of vortex-induced vibrations for low mass ratio bodies. The third column (Cyl-13, 23, and 33) does not follow the lock-in curve. Note that the data points pertaining to frequencies $< 0.9f_{n,wet}$ are not shown in the figure. They correspond to slow sway peaks for which added mass has no meaning. The change in the total phase between force and displacement in the X direction (Fig. \ref{fig:EffectiveAddedMass-Phase-X-9Cyl-FSI}(b)) also shows the shift in phase from $0^\circ$ to $180^\circ$ between $U_r = 17.5-19.4$. This is the same location where $C_{EA,x}$ changes sign and represents the lock-in regime. The trajectory of the cylinders at different reduced velocities is shown in Fig. \ref{fig:Trajectories-9Cyl-FSI}. The first column which comes under lock-in with increase in $U_r$ manifests an ordered elliptical-trajectory while other cylinders exhibit irregular trajectories. Note that the end of the lock-in or synchronization regime for the the array is not reached due to large distortion of the mesh at higher $U_r$. Therefore, it remains to be determined when the lock-in will end for the configuration, which is a limitation of the study.  

The synchronized amplitudes ($A_\mathrm{max}/D \approx 0.06–0.09$) are around an order of magnitude smaller than the upper-branch amplitudes of vortex-induced vibration (VIV) at comparable mass ratio ($A_\mathrm{max}/D \approx 1$). This could be due to the following reasons: (i) the plume-induced fluid force is weaker compared to the forces due to vortex shedding in VIV, and (ii) the array configuration results in confinement of the cylinders, i.e., the small inter-cylinder and wall gaps restrict large motion, similar to a cylinder near wall. The amplitudes are, however, consistent with flow-induced vibration of tube bundles below the fluid-elastic threshold, and the response retains the qualitative features of synchronization. Moreover, the frequency response differs from the classical VIV behavior, in which the response follows the Strouhal frequency outside the lock-in range while it remains near the natural frequency otherwise. Outside of the synchronization range, the plume excitation is weak due to a weak fluctuating force and is superposed on the broadband forcing of the enclosure-scale circulation. In the absence of structural damping, the response is dominated by the resonant amplification near $f_{n,wet}$.
Within the synchronization range, the buoyancy-driven wall-gap plume oscillation, generated by the array as a whole and only weakly modified by the motion ($\approx$ 6\%), imposes its frequency on the light structure ($m^* = 2$), which adapts through its effective added mass (decreasing $C_{EA}$). The synchronized frequency thus equals the effective natural frequency of the structure, including its motion-dependent added mass. This is the limiting case of the low mass ratio trend in VIV, where the lock-in frequency departs increasingly from $f_{n,wet}$ as $m^*$ decreases \cite{GOVARDHAN_WILLIAMSON_2000}.

\subsection{Flow visualization of vortex cores and influence of lock-in}

\begin{figure}
    \centering
    \begin{subfigure}{\textwidth}
        \includegraphics[trim={0.1cm 0.1cm 0.1cm 0.1cm}, clip,width=\textwidth]{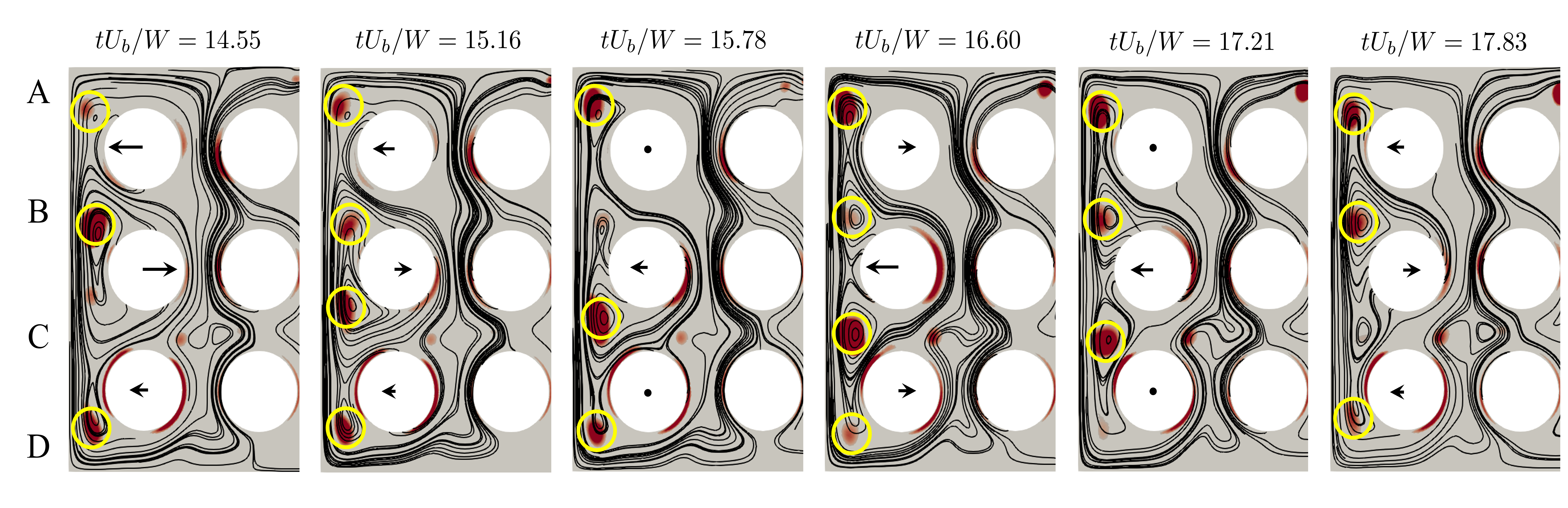}
        \caption{}
    \end{subfigure}
    
    \begin{subfigure}{\textwidth}
        \includegraphics[trim={0.1cm 4.5cm 0.1cm 0.1cm}, clip,width=\textwidth]{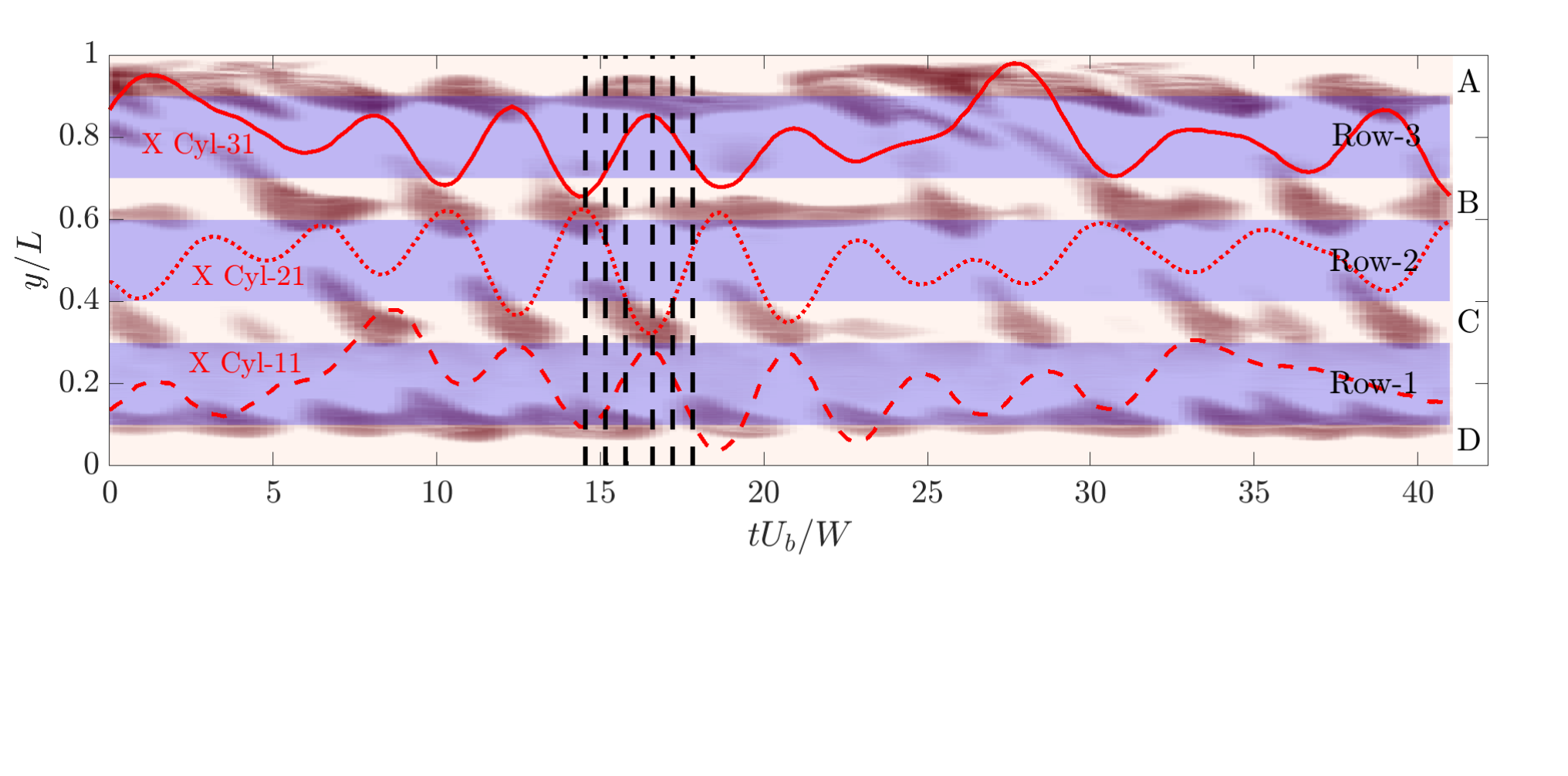}
        \caption{}
    \end{subfigure}
    \caption{Flow visualization in $3\times 3$ array at $U_r \approx 12.2$: (a) Transition in the vortex core (visualized by $Q$-criterion along with streamlines) near the left wall of the array, and (b) Space-time map of $Q>0$ vortex cores near the left wall. The time history of the X displacement of the cylinder is overlaid on the space-time map. The vertical dashed lines are the time instances shown in (a).}
    \label{fig:vortex_Cores_Ur12}
\end{figure}

\begin{figure}
    \centering
    \begin{subfigure}{\textwidth}
        \includegraphics[trim={0.1cm 0.1cm 0.1cm 0.1cm}, clip,width=\textwidth]{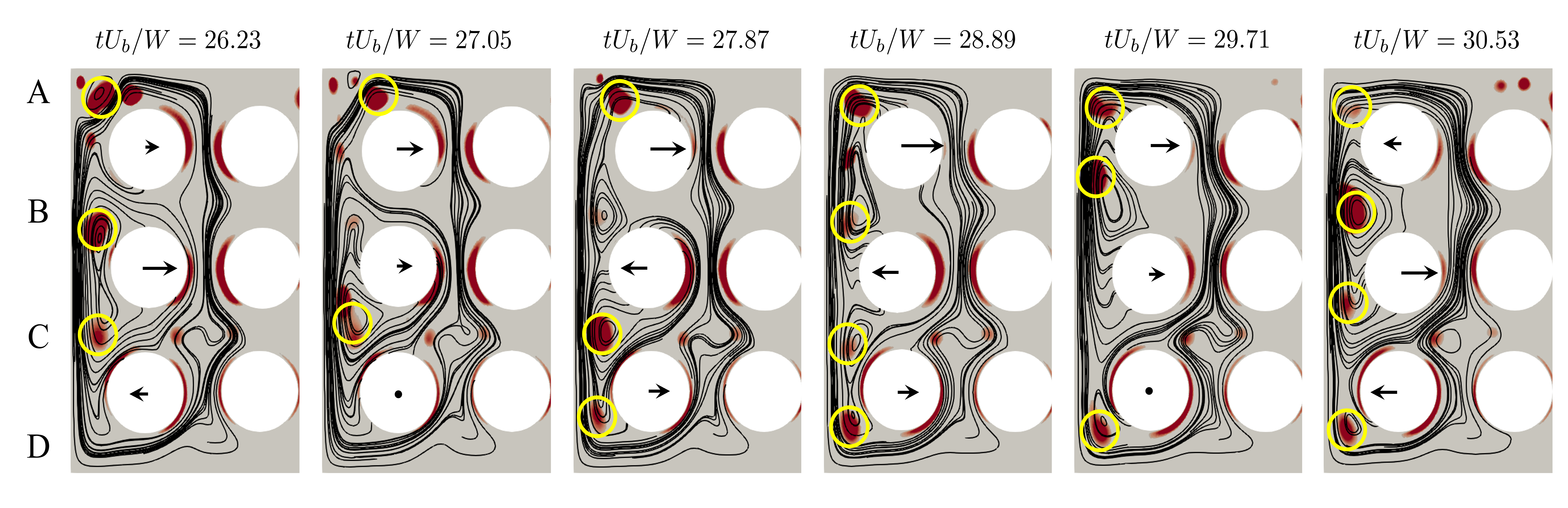}
        \caption{}
    \end{subfigure}
    
    \begin{subfigure}{\textwidth}
        \includegraphics[trim={0.1cm 4.5cm 0.1cm 0.1cm}, clip,width=\textwidth]{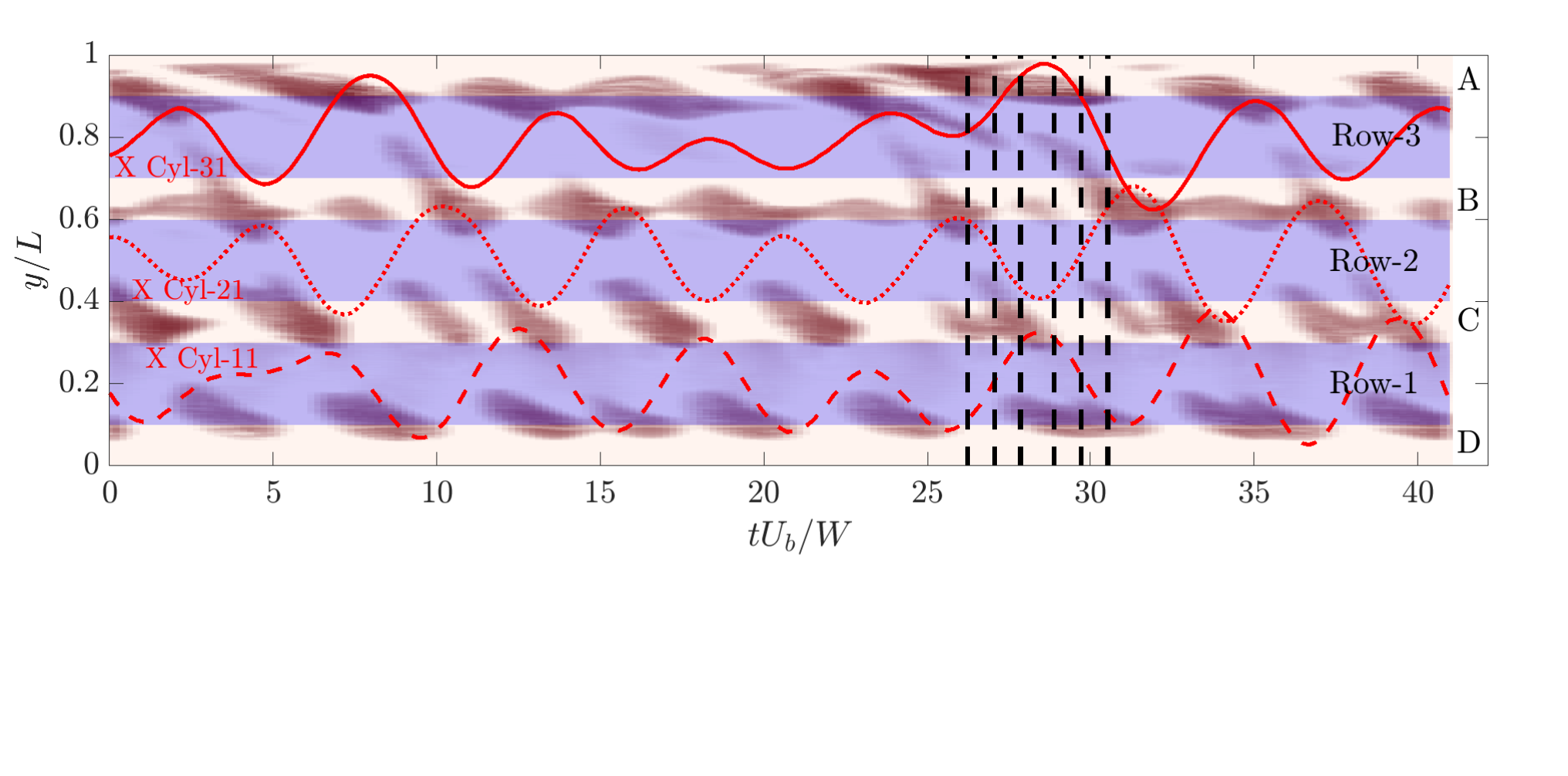}
        \caption{}
    \end{subfigure}
    \caption{Flow visualization in $3\times 3$ array at $U_r \approx 17.5$: (a) Transition in the vortex core (visualized by $Q$-criterion along with streamlines) near the left wall of the array, and (b) Space-time map of $Q>0$ vortex cores near the left wall. The time history of the X displacement of the cylinder is overlaid on the space-time map. The vertical dashed lines are the time instances shown in (a).}
    \label{fig:vortex_Cores_Ur17}
\end{figure}

When the cylinders are elastically mounted, their oscillatory motion periodically narrows and widens the gaps between the cylinders and the wall, and this directly affects the development of the vortex cores and its transition from the top corner to the bottom. To understand the effect of lock-in further, the vortex cores at different time instances have been visualized by $Q$-criterion for two cases: $U_r \approx 12.2$ (before lock-in) and $U_r \approx 17.5$ (after lock-in) in Figs. \ref{fig:vortex_Cores_Ur12} and \ref{fig:vortex_Cores_Ur17} respectively. For the elastically mounted cylinders, Cyl-11 and Cyl-31 oscillate anti-phase to Cyl-21, which is highlighted by the arrows inside the cylinders depicting their motion. The methodology to construct this vortex core map is similar to that of the stationary array configuration in Section \ref{vortex_core_origin}. In addition, the X displacement of the cylinders Cyl-11, 21, and 31 is overlaid on the space-time map of $Q>0$ vortex cores in Figs. \ref{fig:vortex_Cores_Ur12}(b), and \ref{fig:vortex_Cores_Ur17}(b).

Before lock-in at $U_r \approx 12.2$, the lower gap locations at C and D respond to the motion of the adjacent cylinders (Cyl-21 and Cyl-11, respectively). As observed in Fig. \ref{fig:vortex_Cores_Ur12}(b), the vortex cores at these locations form roughly at the same time instance of the cylinder (Cyl-11 and Cyl-21) oscillation. However, the corner vortex at A and that at location B keep forming at their own frequency. The vortex cores at A and B drift in time compared to the time instances of Cyl-31 motion. In Fig. \ref{fig:vortex_Cores_Ur12}(a), the corner vortex at A grows and decays almost independently of the Cyl-31 motion. Therefore, the flow oscillates with its own frequency $f_{wgp}W/U_b \approx 0.183$ which dominates the fluid forces, the corner vortex A and the Y direction motion of Cyl-31. The lower cylinders of the first column (Cyl-11 and Cyl-21) oscillate near their natural frequency and the gaps near these cylinders follow them. The X direction motion of Cyl-31 consists of superposition of both the plume frequency and the natural frequency of the cylinder, with no fixed phase relation. Therefore, only the lower openings (C and D) follow the cylinders while the upper-corner vortex at A and that in the gap between Cyl-21 and 31 (at B) oscillate at the wall-gap plume oscillation frequency. 

Within lock-in ($U_r \approx 17.5$), the motion of the cylinders becomes more regular and alternating. As observed from the X displacements shown in Fig. \ref{fig:vortex_Cores_Ur17}(b), Cyl-21 moves opposite to the motion of Cyl-11 and Cyl-31. The gaps between the cylinders and the wall (throat regions) thus narrow and widen alternately. All the frequencies of the vortex core movement at each location (A, B, C, and D) and the cylinder movements, coincide with a fixed phase. A vortex core grows in the gap as the cylinder above it swings back towards the wall and closes the throat region (Fig. \ref{fig:vortex_Cores_Ur17}(a)). The core weakens as the throat region opens with the movement of the cylinder opposite to the wall. The corner vortex at A is also captured by the cylinder movement and is released once every cylinder cycle. The movement of cylinders work together to push the vortex core down in unison. The vortex cores at each location appear at the same time instant relative to the motion of the cylinders, as seen in Fig. \ref{fig:vortex_Cores_Ur17}(b). The streaks that are observed for $U_r \approx 12.2$ case transform to regular pattern under lock-in.  


To summarize, the $3\times 3$ array exhibits two distinct mechanisms. The first column (Cyl-11, 21, 31), which lies in the wall-gap oscillating buoyant plume of the heated array (stationary-array frequency $f_{wgp}W/U_b = 0.183$), undergoes instability-induced excitation with lock-in. For $15.7 \leq U_r \leq 22.1$, the dominant response frequency remains fixed at $f/f_{wgp} = 1.05–1.07$ while $f/f_{n,wet}$ increases from 1.15 to 1.64, accommodated by a decrease of the effective added mass from $0.77$ to $-0.63$. Below $U_r \approx 15$, the same cylinders respond at their natural frequency. The remaining cylinders are not synchronised, apart from a weak entrainment of Cyl-12 in Y direction. Their response is buffeting by the enclosure-scale slow circulation, whose forcing ($fW/U_b \approx 0.025, 0.05$ and $0.08$) is broadband and lies well below the natural frequencies in the tested range. It produces a resonant response near $1.2 f_{n,wet}$ together with a low-frequency sway that dominates the RMS amplitude and grows monotonically with $U_r$. The largest vibration amplitude in the array come from the unsynchronized buffeting response and not from synchronization. Furthermore, under synchronization, the motion on the left column cylinders is such that they work together to push the top corner vortex core in unison. This unison also modifies the plume frequency for the stationary array, which was the original source for the synchronization.


\section{Response of elastically mounted array subjected to natural convection for $4\times 4$ cylinder array}
\label{sec:FSI-4by4}
The power spectra of the stationary cylinders in the $4\times 4$ array configuration shown in Fig. \ref{powerSpectra_4by4_stationary} shows that the dominant frequency is the slow plume sway motion in the enclosure with the first mode at $f_{ps}W/U_b = 0.0376$ and its harmonics. The results are observed to be mirror symmetric in time-averaged sense. To investigate the effect of elastically mounting the cylinders in the $4\times 4$ configuration, the reduced velocity is varied in the range $U_r \in [10-25]$ by varying the stiffness of the mounted spring for each cylinder. 

\subsection{Amplitude and frequency response}
\begin{figure}
\centering
\begin{subfigure}{0.5\textwidth}
    \includegraphics[trim={8cm 0.1cm 9cm 0.1cm}, clip,width=\textwidth]{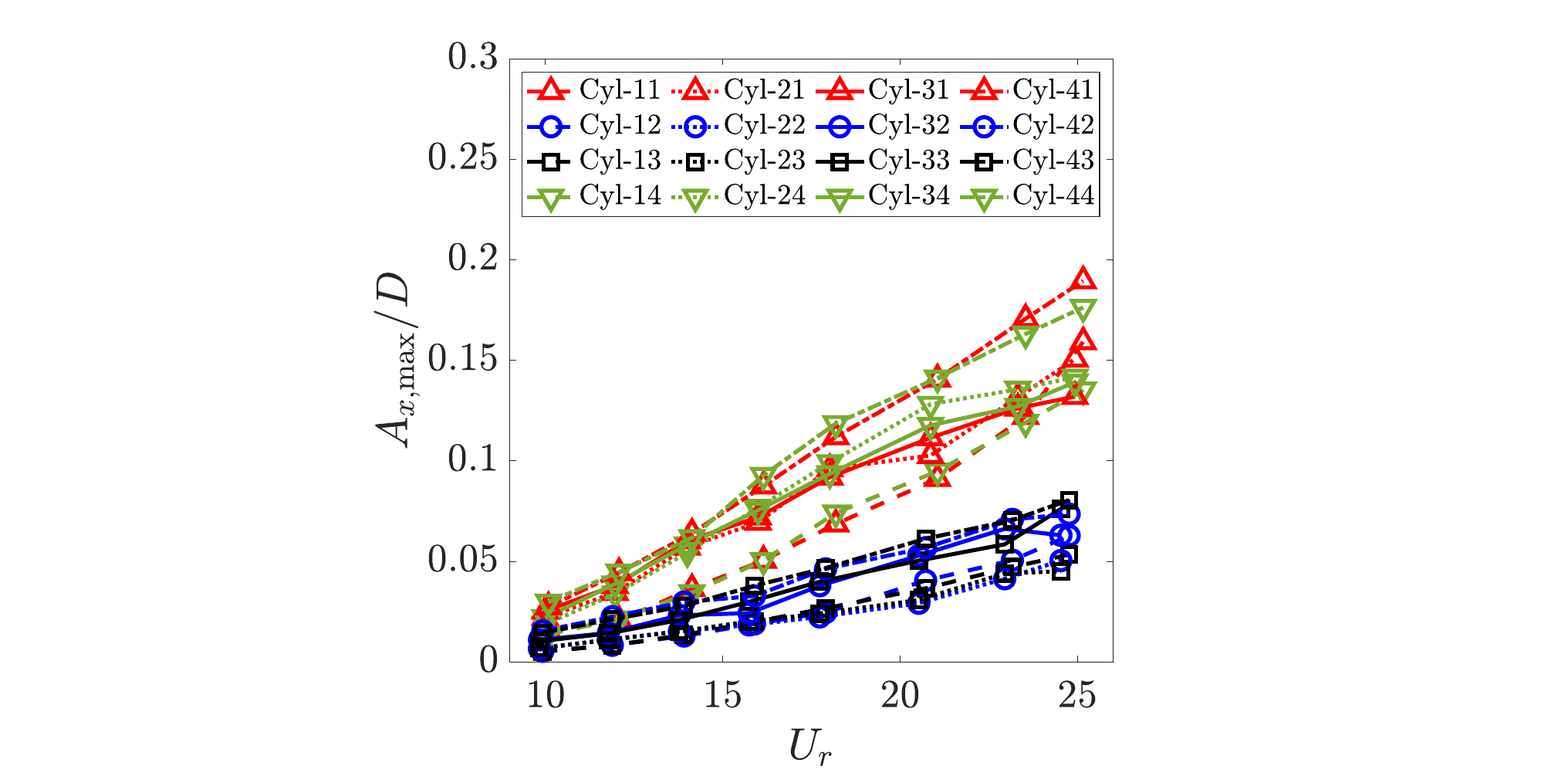}
    \caption{}
\end{subfigure}%
\begin{subfigure}{0.5\textwidth}
    \includegraphics[trim={8cm 0.1cm 9cm 0.1cm}, clip,width=\textwidth]{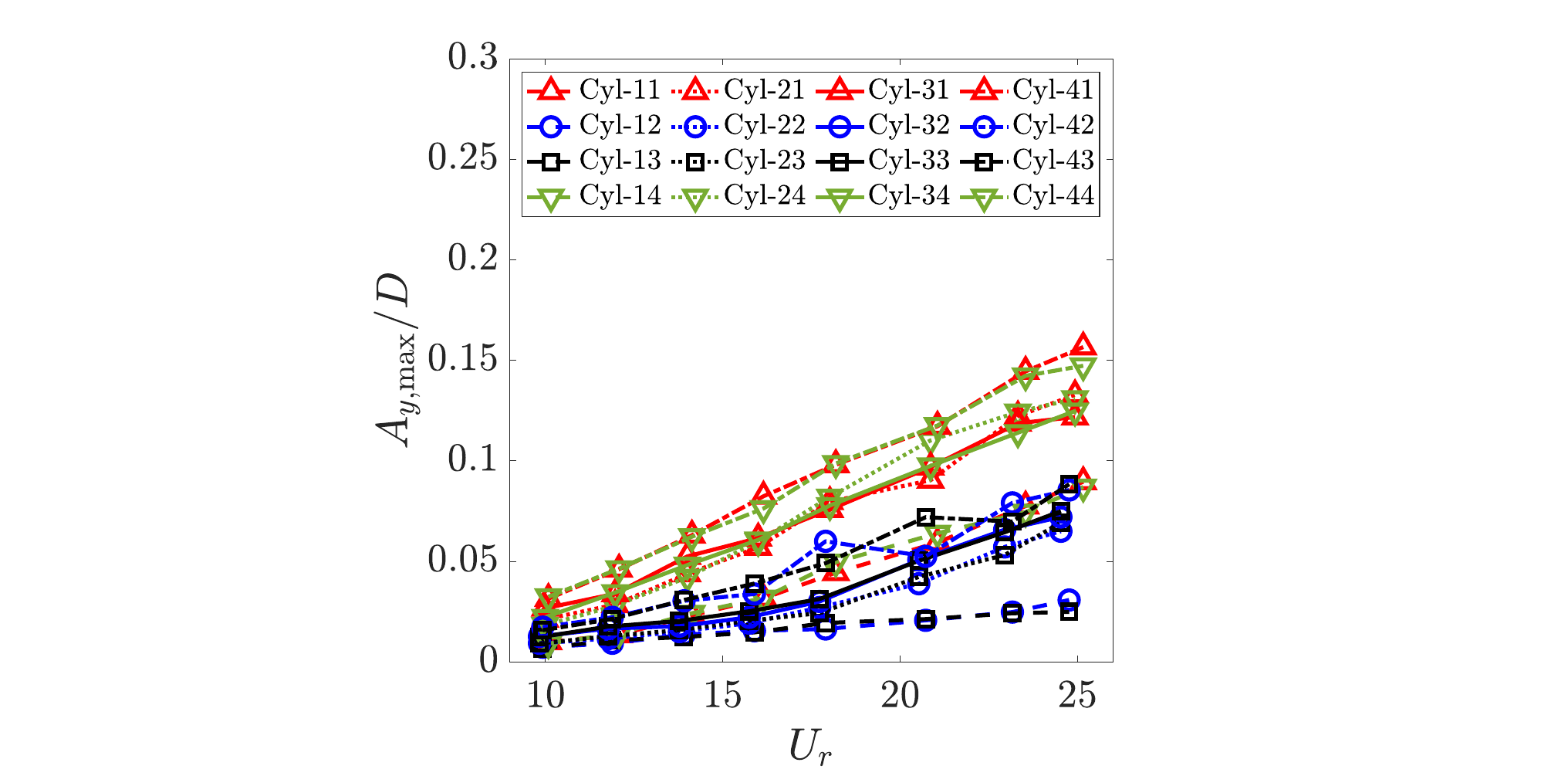}
    \caption{}
\end{subfigure}%
\caption{Maximum amplitude of the response of the cylinders in $4\times 4$ array with the reduced velocity $U_r$: (a) $A_{x,\mathrm{max}}/D$, and (b) $A_{y,\mathrm{max}}/D$.}
\label{fig:Amp-16Cyl-FSI-MAX}
\end{figure}

The non-dimensional maximum amplitude response of the cylinders in the $4\times 4$ configuration with reduced velocity is shown in Fig. \ref{fig:Amp-16Cyl-FSI-MAX}. It can be noticed that the cylinders of first and last column have amplitudes $3-5$ times higher than those in the inner columns, throughout the $U_r$ range. Furthermore, the amplitudes of symmetric cylinders (Cyl-11/14, Cyl-21/24, Cyl-31/34 and Cyl-41/44) are nearly the same in both the directions. The responses are statistically symmetric about the vertical centerline. 
With increase in $U_r$, the amplitudes grow till $A_{\mathrm{max}}/D\approx 0.2$ for the top corner cylinders. No amplitude peak or plateau similar to the $3\times 3$ array is observed in this configuration. Similar observations can be made for the root-mean-square amplitudes.

\begin{figure}
\centering
\includegraphics[trim={2cm 2cm 2.5cm 0.1cm}, clip,width=\textwidth]{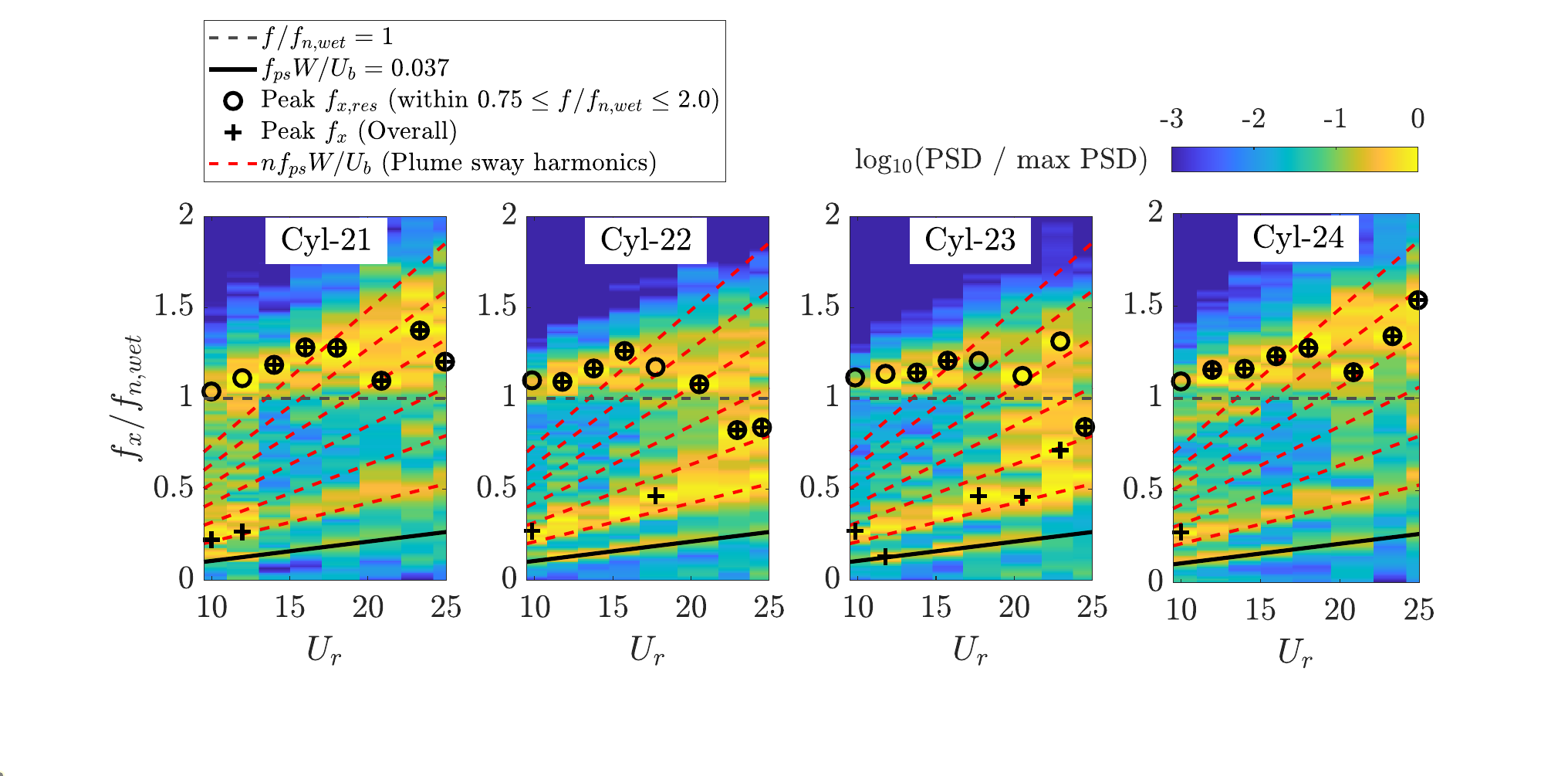}
\caption{Welch PSD of the displacement in X, normalised by its maximum over $fW/U_b > 0.015$ at each $U_r$ overlaid with the peak amplitude frequency for the second row (Cyl-21, 22, 23 and 24).}
\label{fig:Freq-map-X-16Cyl-FSI}
\end{figure}

\begin{figure}
\centering
\includegraphics[trim={2cm 2cm 2.5cm 0.1cm}, clip,width=\textwidth]{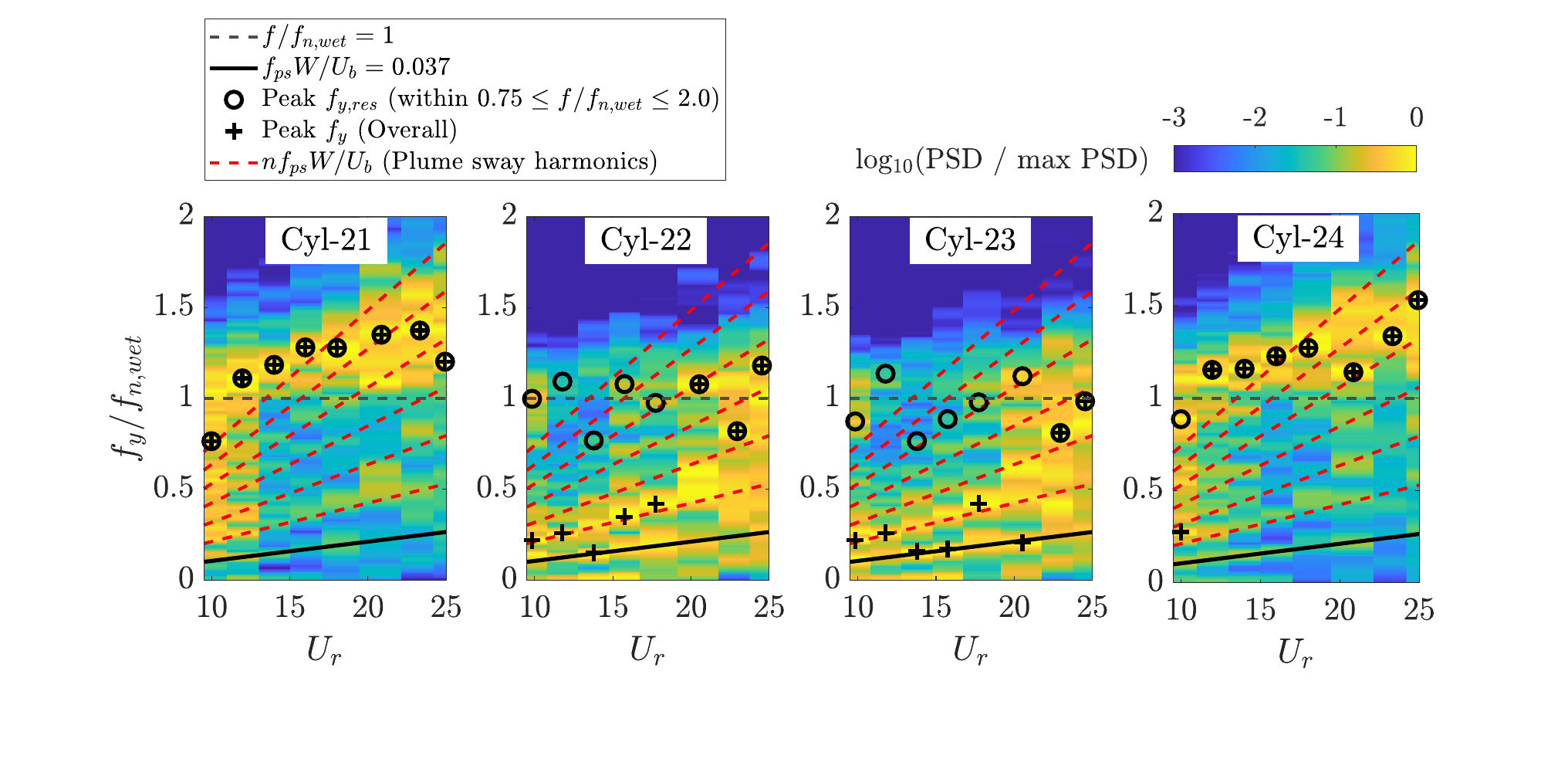}
\caption{Welch PSD of the displacement in Y, normalised by its maximum over $fW/U_b > 0.015$ at each $U_r$ overlaid with the peak amplitude frequency for the second row (Cyl-21, 22, 23 and 24).}
\label{fig:Freq-map-Y-16Cyl-FSI}
\end{figure}

Since the behavior of different rows of cylinders in the array is similar, the frequency map for the X and Y displacement for the second row of cylinder array is shown in Figs. \ref{fig:Freq-map-X-16Cyl-FSI} and \ref{fig:Freq-map-Y-16Cyl-FSI}, respectively. The figure consists of the Welch PSD spectrum of the displacement in X and Y directions, normalised by its maximum over $fW/U_b > 0.015$ at each reduced velocity $U_r$. The dominant frequencies considering the two bands stated above have been overlaid on the Welch spectrum. The solid line indicates the slow plume sway $f_{ps}W/U_b = 0.0376$ and all the red dashed slope lines are its harmonics. The overall peak response frequency is depicted by \enquote*{$+$} symbol while the dominant frequency in the range of $f/f_{n,wet} \in [0.75, 2]$ is shown by \enquote*{$\medcirc$} for each $U_r$, similar to the frequency maps in Section \ref{sec:amp_freq_3by3}. In X direction, Cyl-21 and 24 show the dominant amplitude response near the second harmonic of the plume sway frequency $f_{ps}W/U_b$ at $U_r \approx 10$. As the reduced velocity increases, the dominant response shifts to frequencies greater than $f_{n,wet}$ and the response becomes more broadband. In the Y direction, the response frequency is mostly dominated by frequencies greater than $f_{n,wet}$ except at $U_r = 10$. The frequency response also becomes broadband with increasing $U_r$. The response crosses the sway harmonics rather than following a particular harmonic. Therefore, no harmonic lock-in is observed for the $4\times 4$ array configuration. For the inner cylinders (Cyl-22 and 23), a strong first and second-harmonic response is noticed in both X and Y directions. Similar to outer cylinders, the response transitions to broadband response with $U_r$, with irregular dominant sway harmonics throughout the reduced velocity range. The slow sway fundamental frequency is very small compared to the natural frequency till $U_r = 25$, which prevents any kind of lock-in to occur. Further increase in $U_r$ results in interference between the motion of the cylinders leading to mesh failure.  

\subsection{Orbital trajectories}
\begin{figure}
\centering
\begin{subfigure}{0.25\textwidth}
    \includegraphics[trim={8cm 0.1cm 9cm 0.1cm}, clip,width=\textwidth]{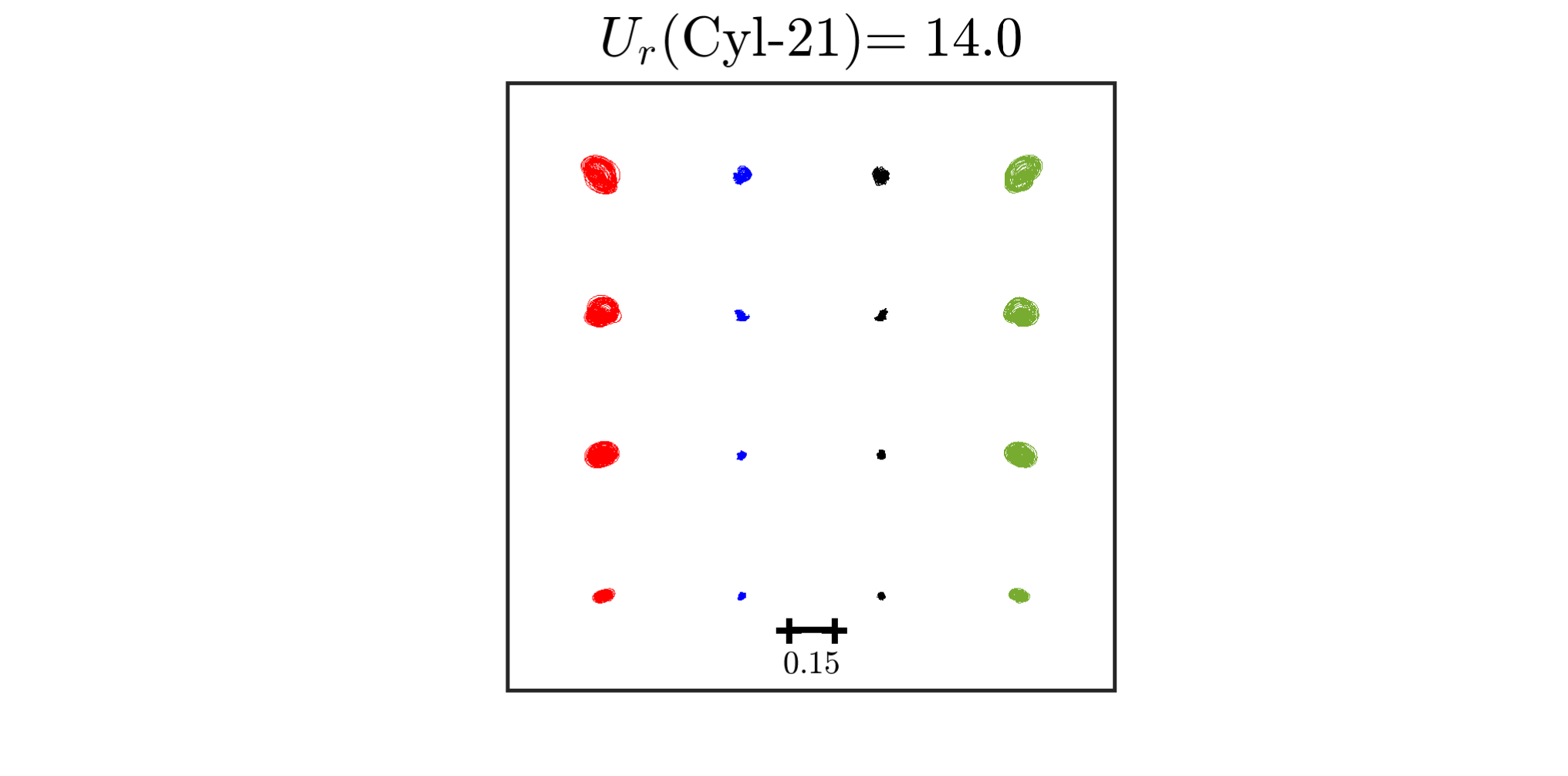}
    \caption{}
\end{subfigure}%
\begin{subfigure}{0.25\textwidth}
    \includegraphics[trim={8cm 0.1cm 9cm 0.1cm}, clip,width=\textwidth]{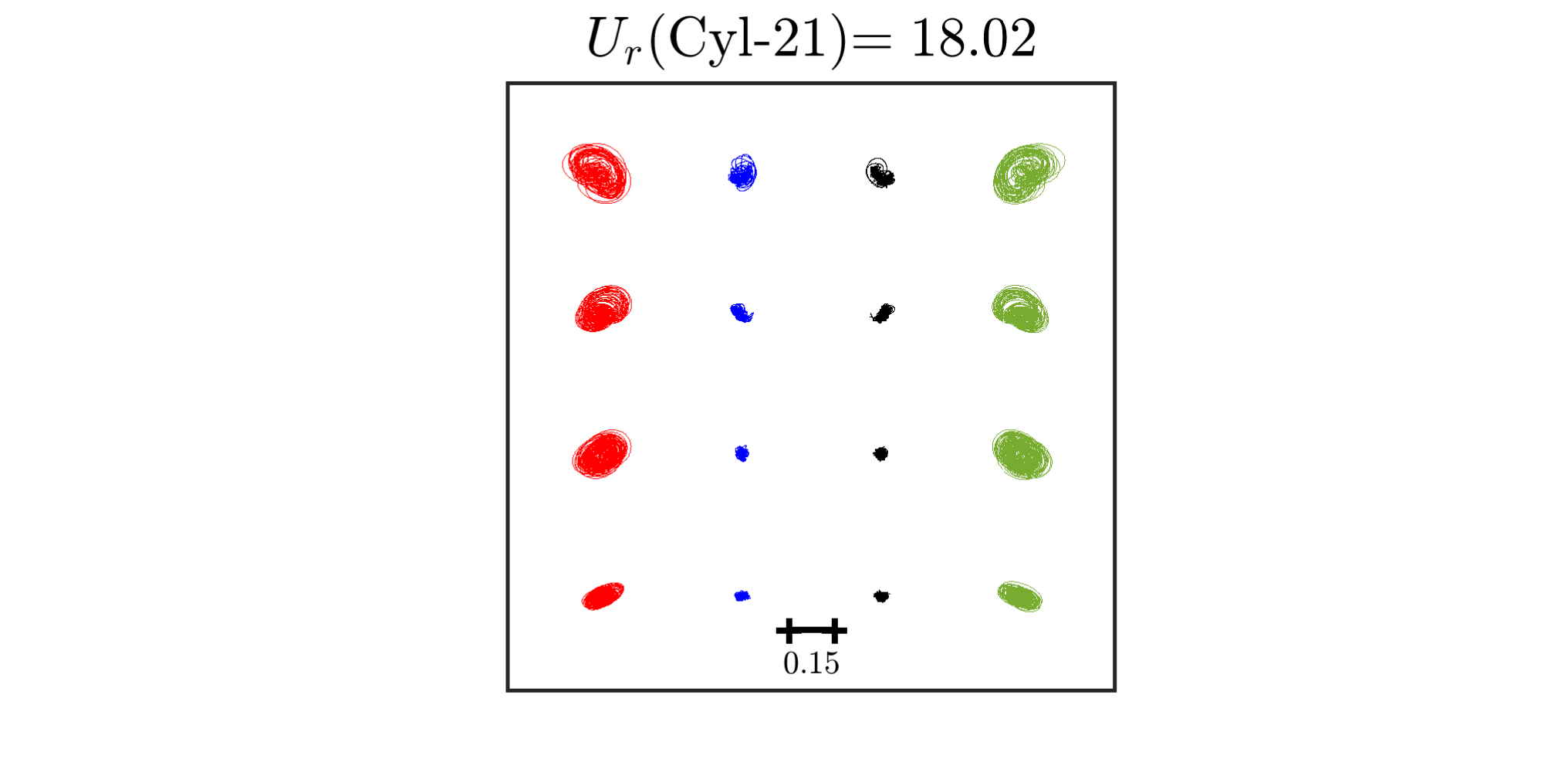}
    \caption{}
\end{subfigure}%
\begin{subfigure}{0.25\textwidth}
    \includegraphics[trim={8cm 0.1cm 9cm 0.1cm}, clip,width=\textwidth]{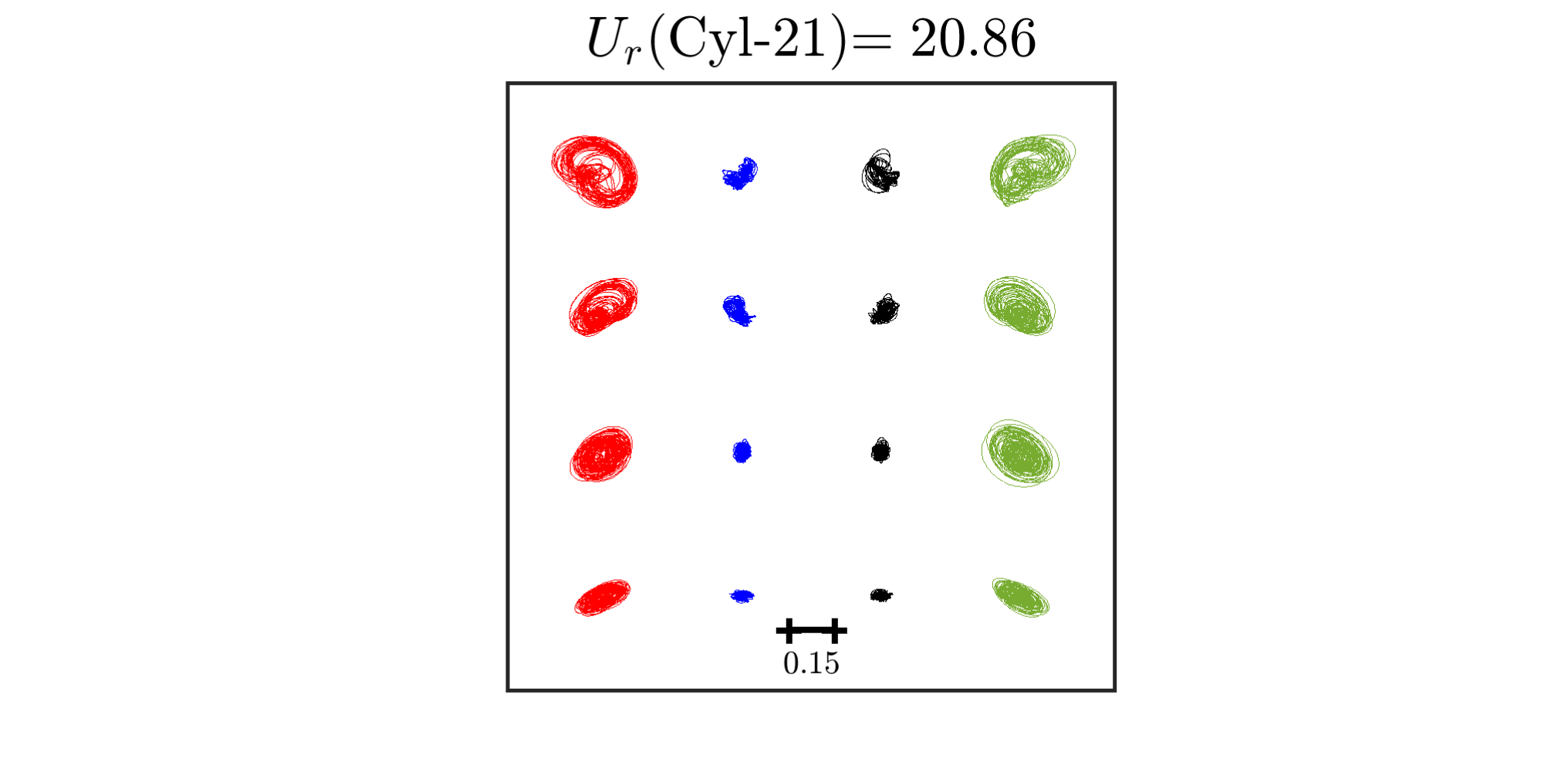}
    \caption{}
\end{subfigure}%
\begin{subfigure}{0.25\textwidth}
    \includegraphics[trim={8cm 0.1cm 9cm 0.1cm}, clip,width=\textwidth]{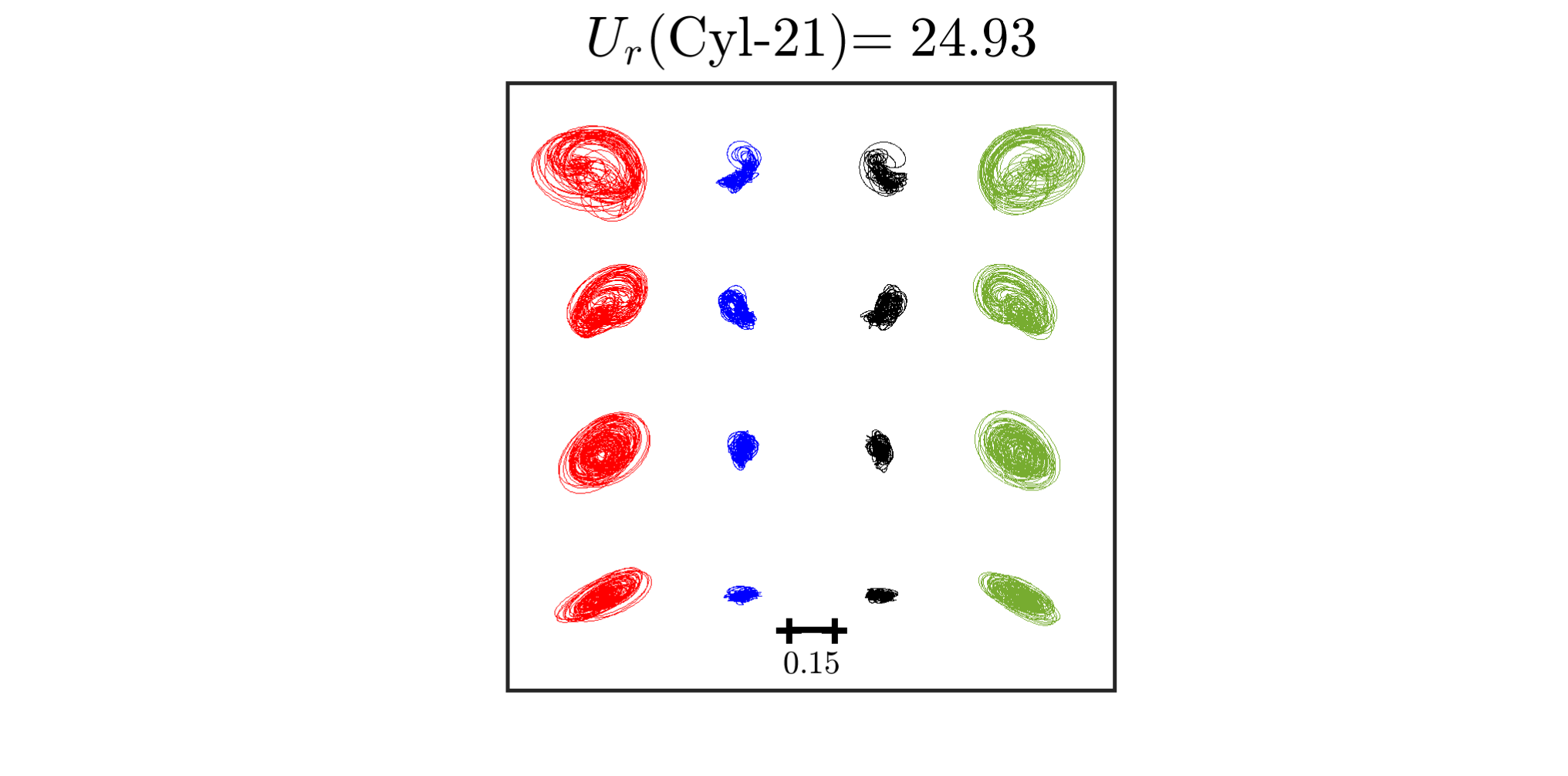}
    \caption{}
\end{subfigure}%
\caption{Trajectory of cylinders in the $4\times 4$ array system at reduced velocities of $U_r$ (evaluated for Cyl-21): (a) 14.0, (b) 18.02,  (c) 20.86 and (d) 24.93. The horizontal line shows the measure of the non-dimensional amplitude (normalized by $D$) for scale.}
\label{fig:Trajectories-16Cyl-FSI}
\end{figure}

The trajectories of each cylinder in the $4\times 4$ array is shown in Fig. \ref{fig:Trajectories-16Cyl-FSI} for four representative reduced velocities defined based on Cyl-21. At $U_r = 14$, the cylinder in the outer columns trace near-circular trajectories while the inner columns have negligible displacement. With increase in $U_r$, the size of the trajectories of outer cylinders grow. The motion is irregular with no closed limit cycle. The shapes of the trajectories are also noticed to be mirror-symmetric about the central vertical line. At $U_r = 24.93$, Cyl-11 and 14 follow flat ellipse-type of trajectory. The trajectories of Cyl-21, 24 (second row) and Cyl-31, 34 (third row) resemble tilted ellipse and Cyl-41, 44 become nearly circular. The inner cylinders show small and distorted shape of the trajectory as a result of slow plume sway and broadband response.  

To summarize, the $4\times 4$ array exhibits no synchronization for the considered reduced velocity range $U_r = 10–25$. The outer columns respond in a narrow band at 1.05–1.35 $f_{n,wet}$ where the frequency follows the structural stiffness. This indicates a stable buoyancy-driven buffeting with flow-induced stiffening. It becomes broadband for $U_r\geq 21$. The inner columns respond weakly, mainly through coupling with the outer columns in X direction, and quasi-statically to the global plume-sway mode ($f_{ps}W/U_b = 0.0376$) and its harmonics in Y direction. Amplitudes remain small and are observed to be mirror-symmetric about the vertical centreline. The amplitude for the outer column cylinders are the largest. Unlike the $3\times 3$ array, whose first column synchronizes with a wall-gap plume oscillation at $f_{wgp}W/U_b \approx 0.183$, the stationary $4\times 4$ buoyant plume forcing contains no discrete frequency near the natural frequencies of the cylinders. The low-frequency sway and its harmonics are only present, which the response crosses without coming in synchronization, for the considered reduced velocity range.

\section{Effect of cylinder motion on Nusselt number}
\label{sec:effectNu}
\begin{figure}
\centering
\begin{subfigure}{0.5\textwidth}
    \includegraphics[trim={8cm 0.1cm 9cm 0.1cm}, clip,width=\textwidth]{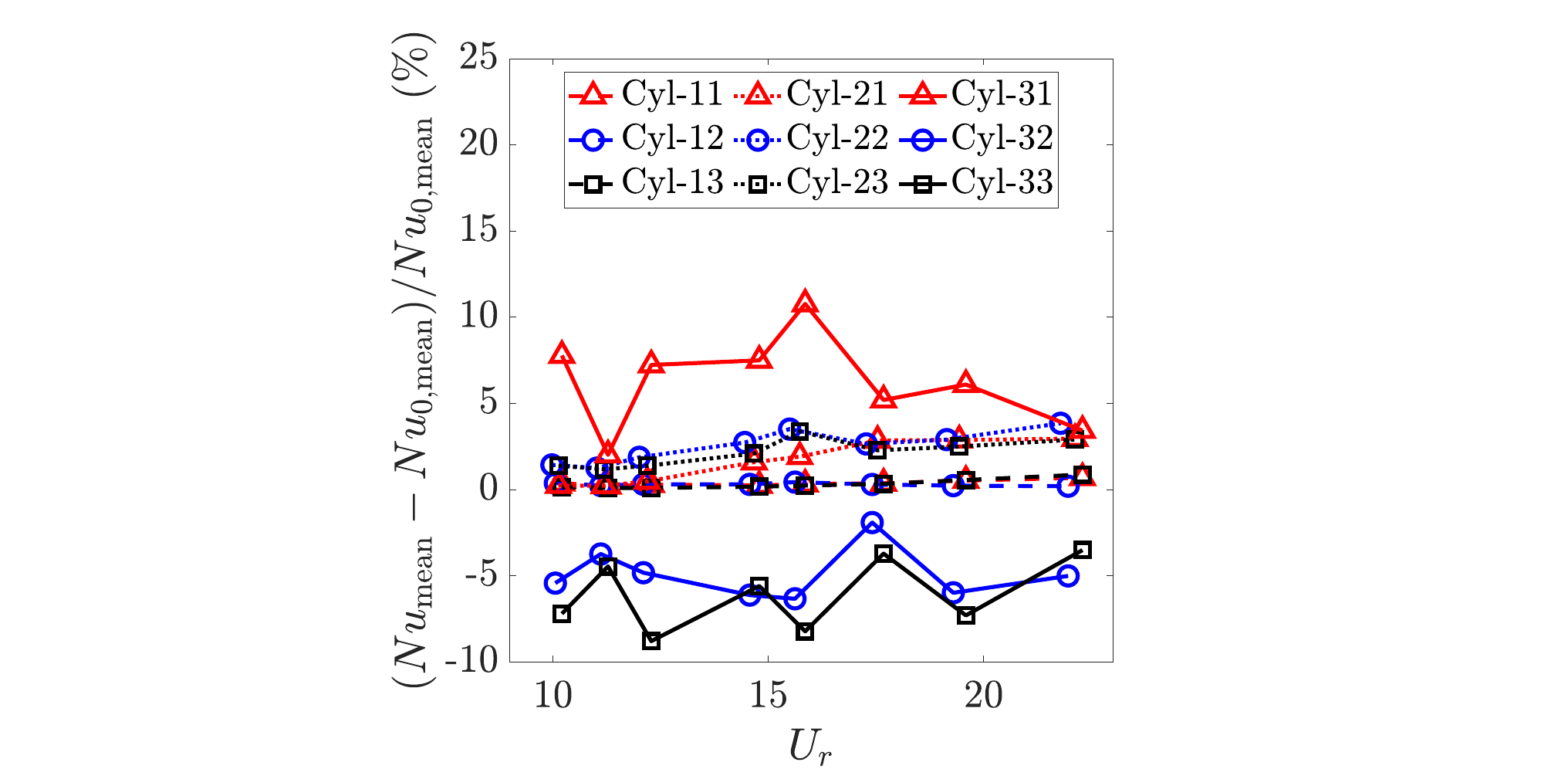}
    \caption{}
\end{subfigure}%
\begin{subfigure}{0.5\textwidth}
    \includegraphics[trim={8cm 0.1cm 9cm 0.1cm}, clip,width=\textwidth]{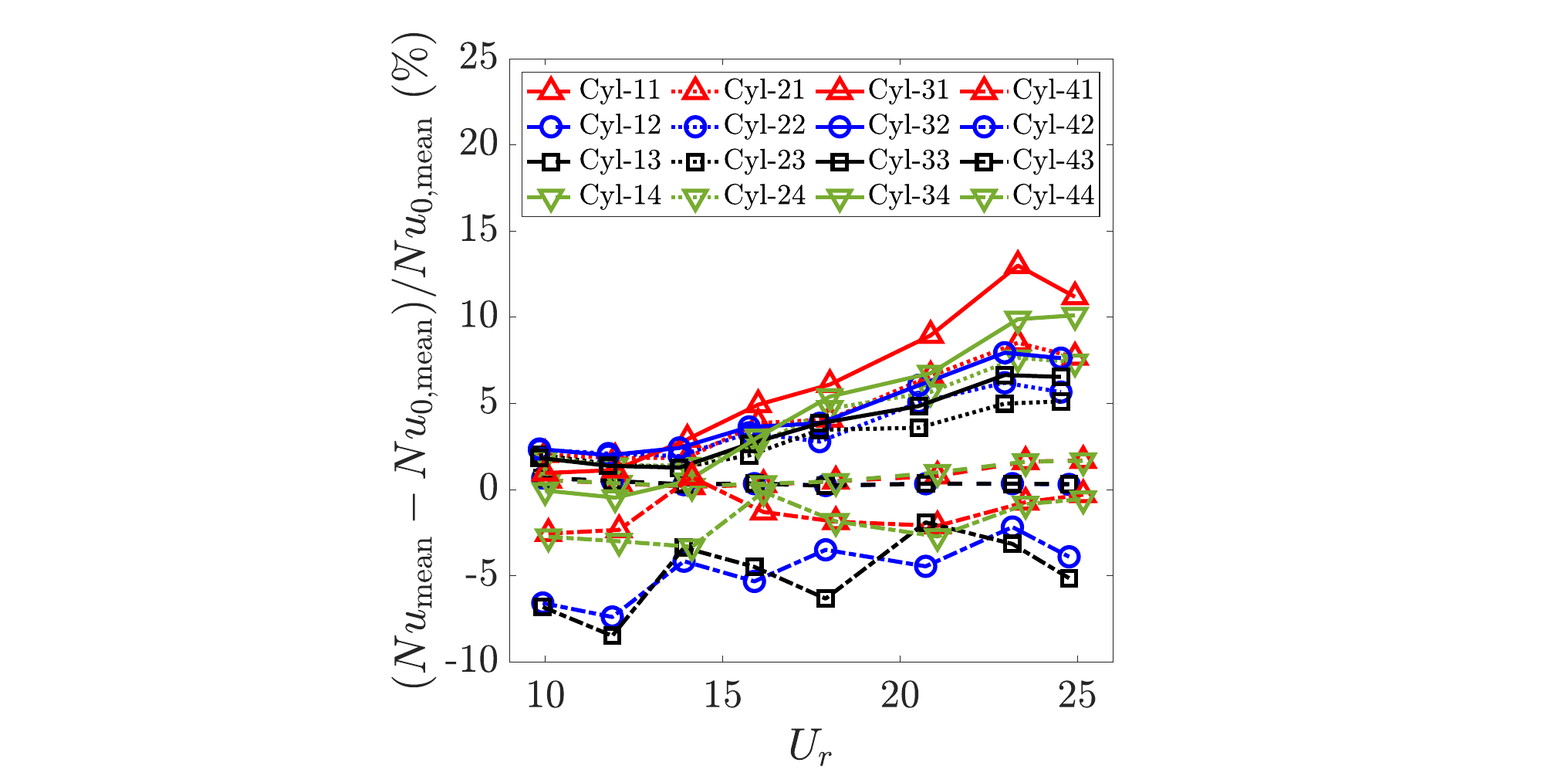}
    \caption{}
\end{subfigure}%
\caption{Variation of time-averaged Nusselt number on each cylinder compared to the stationary case with reduced velocity for: (a) $3\times 3$, and (b) $4\times 4$ array. Here, $Nu_{0,\mathrm{mean}}$ represents the time-averaged Nusselt number for the respective cylinder in the stationary configuration.}
\label{fig:NuMean-FSI}
\end{figure}

\begin{figure}
\centering
\begin{subfigure}{0.33\textwidth}
    \includegraphics[trim={12cm 0.5cm 12cm 1cm}, clip,width=\textwidth]{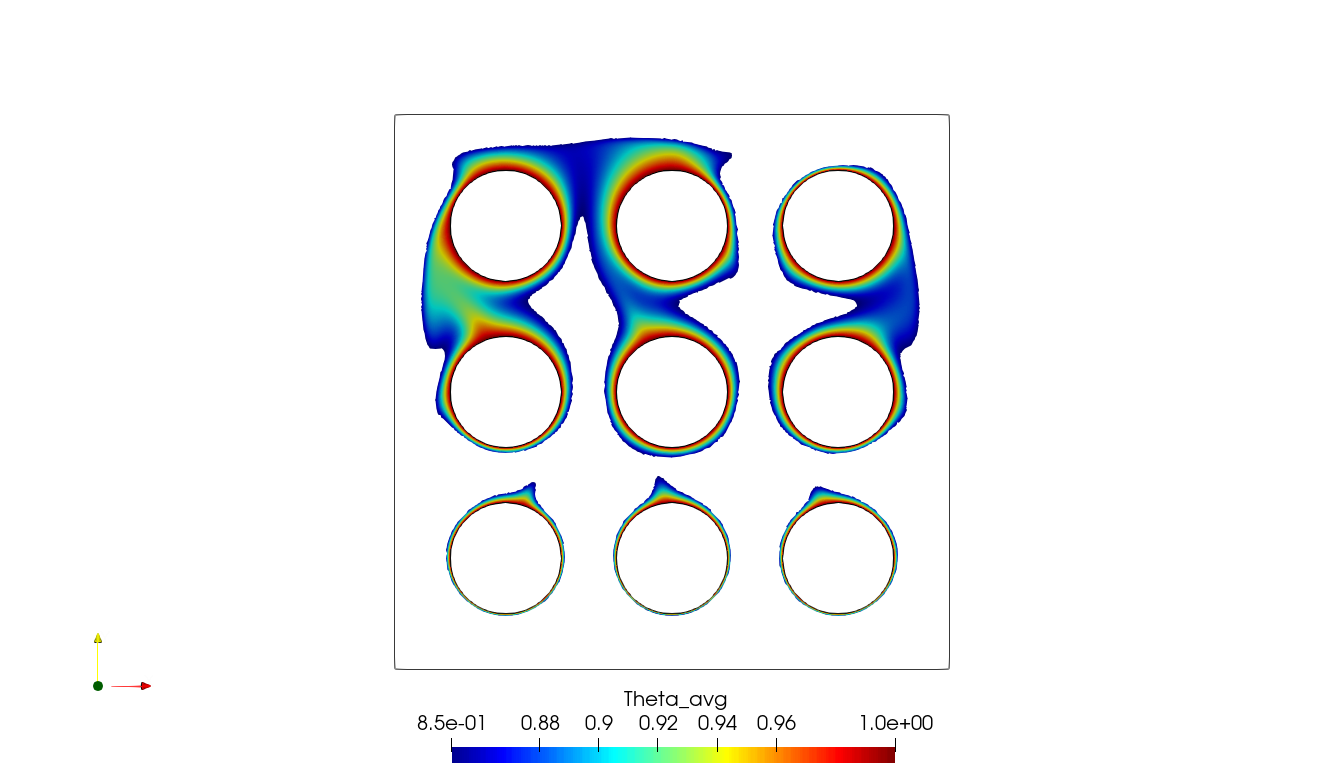}
    \caption{}
\end{subfigure}%
\begin{subfigure}{0.33\textwidth}
    \includegraphics[trim={12cm 0.5cm 12cm 1cm}, clip,width=\textwidth]{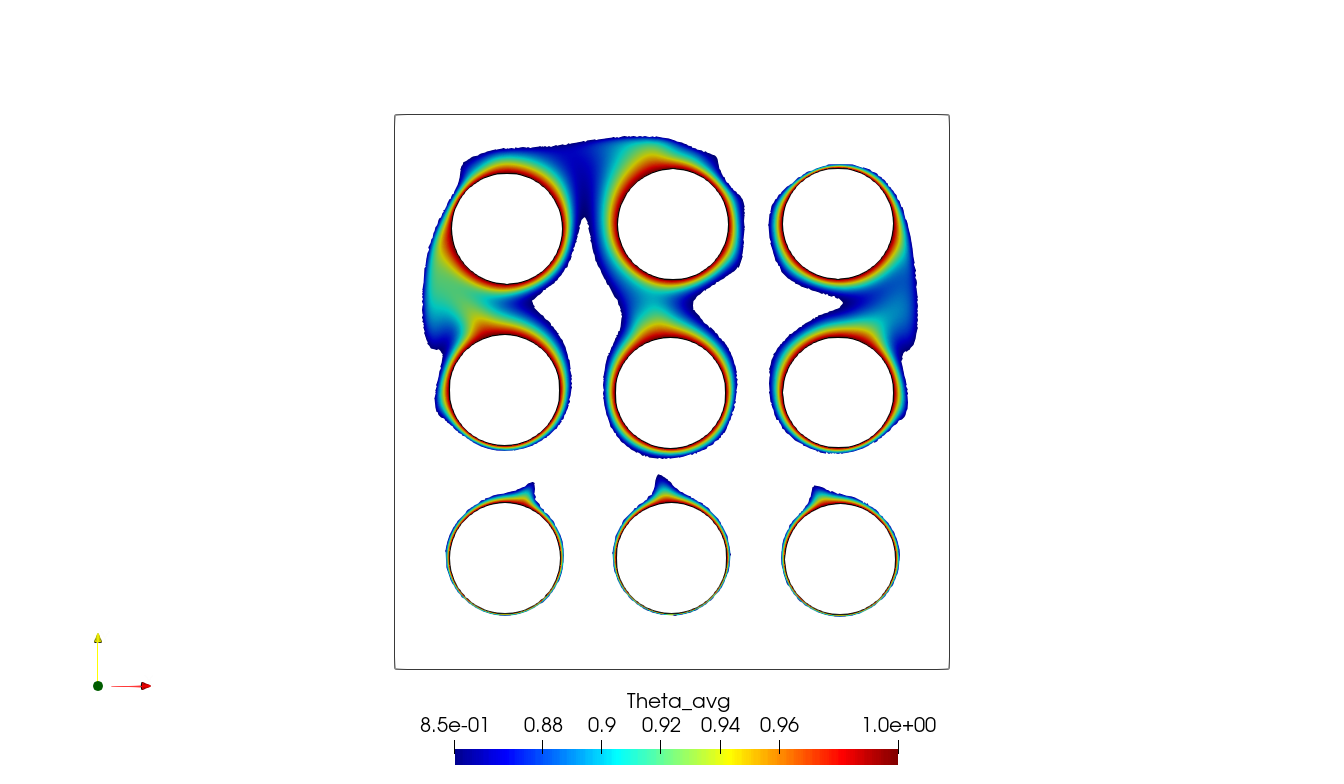}
    \caption{}
\end{subfigure}%
\begin{subfigure}{0.33\textwidth}
    \includegraphics[trim={12cm 0.5cm 12cm 1cm}, clip,width=\textwidth]{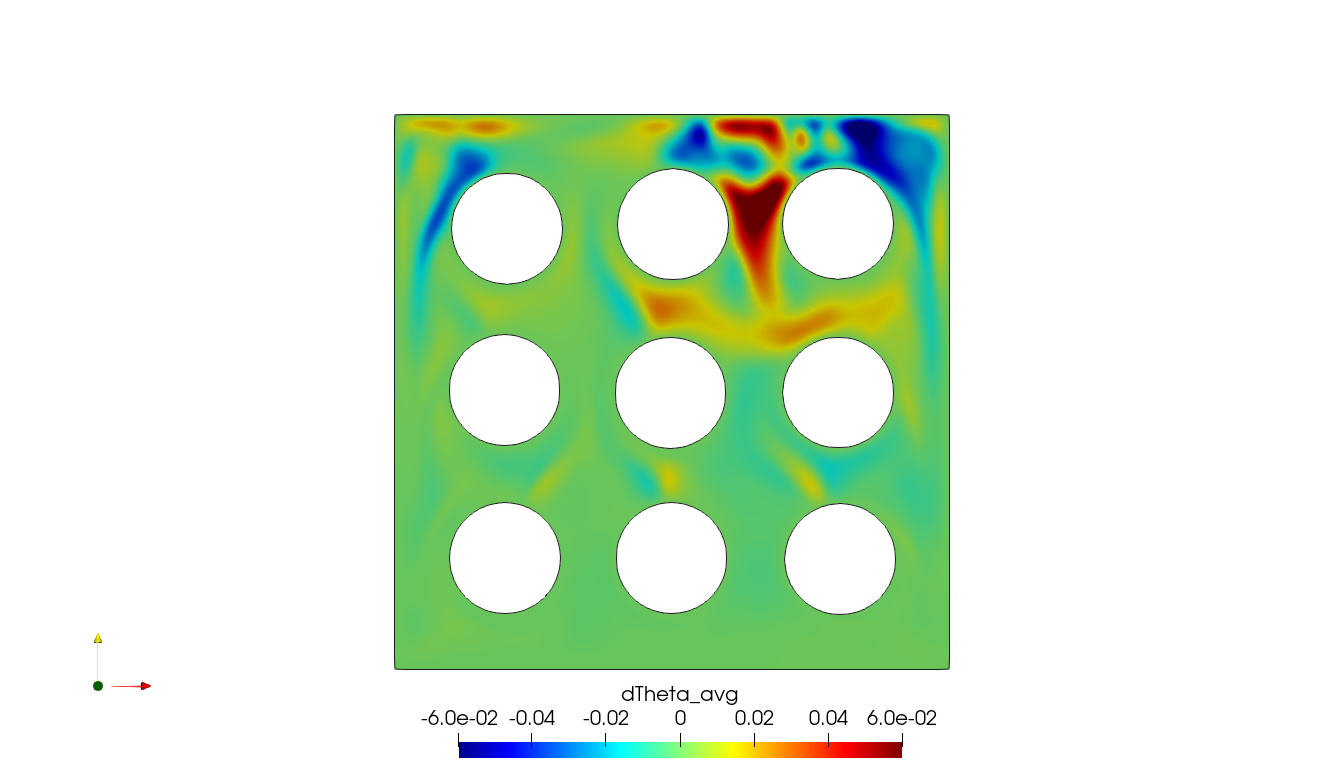}
    \caption{}
\end{subfigure}%
\caption{Time-averaged temperature distribution for the $3\times 3$ array: (a) $\theta_{0,\mathrm{mean}}$ for stationary array, (b) $\theta_\mathrm{mean}$ at $U_r \approx 12.2$, and (c) $\theta_\mathrm{mean} - \theta_{0,\mathrm{mean}}$ showing the change in the temperature distribution compared to stationary array. Contour in (a) and (b) have been truncated to show $\theta_\mathrm{mean} \in [0.85, 1]$.}
\label{fig:dTemp-stat-FSI}
\end{figure}

In this section, the effect of the motion of the cylinders due to fluid-induced forces on the heat transfer is analyzed for the two array configurations. The relative change in the time-averaged Nusselt number of each cylinder compared to the respective cylinder in the stationary array is quantified as $(Nu_\mathrm{mean}-Nu_{0,\mathrm{mean}})/Nu_{0,\mathrm{mean}}$, where $Nu_{0,\mathrm{mean}}$ is the time-averaged Nusselt number for a particular cylinder in the stationary array. The variation of this quantity with reduced velocity has been shown in Figs. \ref{fig:NuMean-FSI}(a) and \ref{fig:NuMean-FSI}(b), for $3\times 3$ and $4\times 4$ arrays, respectively. For $3\times 3$ array, the change in the heat transfer for the first row (Cyl-11, 12, 13) is negligible across $U_r$. There is a monotonic increase with $U_r$ within 5\% increase in Nusselt number for the second row (Cyl-21, 22, 23). The mean Nusselt number for Cyl-31 increases by 2-10\% while that for Cyl-32 and Cyl-33 decreases by 2-9\%, consistently in the $U_r$ range. The change in the Nusselt number appears across all $U_r$ and is unrelated to the synchronization of the cylinders with the wall-gap plume oscillation frequency at $f_{wgp}W/U_b = 0.183$. This behavior can be explained by observing the distribution of temperature for the stationary array and comparing it with a case where the cylinders move ($U_r \approx 12.2$), shown in Fig. \ref{fig:dTemp-stat-FSI}. The contours have been truncated to visualize the change in the plume inside the cavity. For the stationary cylinder array, the plume is more inclined towards the left direction compared to $U_r \approx 12.2$. The time-averaged temperature fields show that, compared with the stationary array (Fig. \ref{fig:dTemp-stat-FSI}(a)), the thermal plume rising from the centre column is displaced to the right in the moving cylinder case (Fig. \ref{fig:dTemp-stat-FSI}(b)), warming the fluid adjacent to the right face of Cyl-32 and the left and upper faces of Cyl-33, while cooler fluid reaches the upper and left faces of Cyl-31 (observed in Fig. \ref{fig:dTemp-stat-FSI}(c) where the change in temperature is quantified). The near-wall mean temperature consequently decreases around Cyl-31 and increases around Cyl-32 and Cyl-33, consistent with the observed increase in $Nu$ of Cyl-31 ($\approx +7\%$) and decrease in $Nu$ of Cyl-32 ($-5\%$) and Cyl-33 ($-9\%$) at $U_r \approx 12.2$. Vibration of the array thus partially weakens the left–right asymmetry of the stationary array configuration. This behavior is purely due to the motion of cylinders and not lock-in phenomenon and persists almost throughout the reduced velocity range.

For $4\times 4$ array, the time-averaged Nusselt number of the second and the third rows of cylinders is found to increase with reduced velocity. The first row (Cyl-11, 12, 13, 14) shows negligible change in the Nusselt number, while it decreases for the top row (Cyl-41, 42, 43, 44) within $8\%$. The increase in $Nu_\mathrm{mean}$ for the second and third rows follow the vibration amplitude and occurs without any synchronization. Therefore, the motion of the cylinders produces small but significant increase of time-averaged Nusselt number as the reduced velocity increases. In contrast to $3\times 3$ array where the total time-averaged Nusselt number for the whole array was negligible, the larger response of the $4\times 4$ array results in a measurable heat transfer enhancement. Furthermore, the fluctuations in the Nusselt number remain dominated by the slow plume sway. 

\section{Conclusions}
\label{conclusion}

In this work, the flow-induced oscillations of an array of circular cylinders inside a square enclosure at a high Prandtl number of 7.0 has been investigated via coupled thermal-fluid-structure interaction formulation based on finite element methodology.

First, thermo-fluid characteristics of stationary arrays ($2\times 2$, $3\times 3$ and $4\times 4$) of cylinders are analyzed with the following key findings:
\begin{enumerate}
    \item All the considered array configurations show a slow plume sway in the square enclosure which has a very small frequency, which is typical for buoyancy driven flows.
    \item The $2\times 2$ and $4\times 4$ arrays have mirror-symmetry about the central vertical axis which is observed in the mean force coefficients on the cylinders as well as time-averaged Nusselt numbers for each cylinder.
    \item The symmetry is broken for $3\times 3$ array. Furthermore, apart from the slow plume sway, a high frequency wall-gap plume oscillation at $f_{wgp}W/U_b = 0.183$ is observed for the left column of cylinders in the array. The source of this plume frequency is the periodic transfer of vortex core from the top corner to the bottom of the square enclosure. 
\end{enumerate}

Second, the elastically mounted cylinders are subjected to the plume oscillations, and their response is recorded. The key findings are as follows:
\begin{enumerate}
    \item The left column cylinder of the $3\times 3$ array show the signs of synchronization with the wall-gap plume oscillation frequency after $U_r \geq 15.7$. The wall-gap plume oscillation frequency is modified by the motion of the cylinder from $f_{wgp}W/U_b =0.183$ to $f_{wgp}W/U_b \approx 0.195$ showing mutual synchronization between a self-sustaining flow oscillation and the structure. This is also observed by effective added mass becoming negative and change in the phase between the force and displacements. The synchronized motion of the cylinders also drive the vortex cores in unison from the top-left corner of the square enclosure to the bottom.
    \item The remaining cylinders in the $3\times 3$ array do not undergo synchronization. Their response is forced buffeting driven by the slow circulation in the square enclosure where the forcing is broadband and below the natural frequencies of the cylinders in the tested range of reduced velocity. Thus, the amplitudes increase with reduced velocity monotonically.
    \item In the $4\times 4$ array configuration, no synchronization is observed for any of the cylinders in the considered reduced velocity range. The cylinders in the outer column respond with larger amplitude compared to the inner columns where the response becomes broadband with increase in reduced velocity and is dominated by the slow plume forcing. The amplitudes of the cylinders are observed to be symmetric about the centerline of the domain. 
    \item The motion of the cylinders due to the buoyancy-induced fluid forcing influences the heat transfer characteristics, especially for the upper row of cylinders which are subjected to the hot plume from the lower rows. For the $3\times 3$ array, the top-left cylinder has enhanced heat transfer of $\approx 7\%$ while the other two cylinders at the top row show reduced heat transfer compared to the stationary configuration. This effect is observed throughout the reduced velocity range and synchronization does not seem to play a role here. On the other hand, for $4\times 4$ array, the heat transfer of second and third row of cylinders is noticed to be enhanced with reduced velocity. The top row shows reduced heat transfer.
\end{enumerate}

The current study is based on Boussinesq approximation to model the thermally buoyant flow. Non-Boussinesq effects are neglected. Three-dimensional effects are not considered in the current study which could be important with spanwise correlations. The Rayleigh number is also kept constant throughout the study. Influence of Rayleigh number and various arrangements of the circular cylinders inside the enclosure on the flow-induced oscillations, will be an interesting prospective future work, especially with its applications to boilers and heat exchangers. 

\section*{Conflict of interest}
The authors have no conflicts to disclose.

\section*{Data Availability}
The data that support the findings of this study are available from the corresponding author upon reasonable request.

\appendix

\section{Numerical tests and benchmarking}
\label{benchmarking}
The formulation presented in the Section \ref{num_formulation} is subjected to several numerical tests and benchmarks and the results are compared with that from the literature. 

\subsection{Natural convection in square cavity}
\begin{figure}
    \centering
    \begin{subfigure}{0.5\textwidth}
    \includegraphics[trim={35cm 0.1cm 30cm 0.1cm}, clip,width=\textwidth]{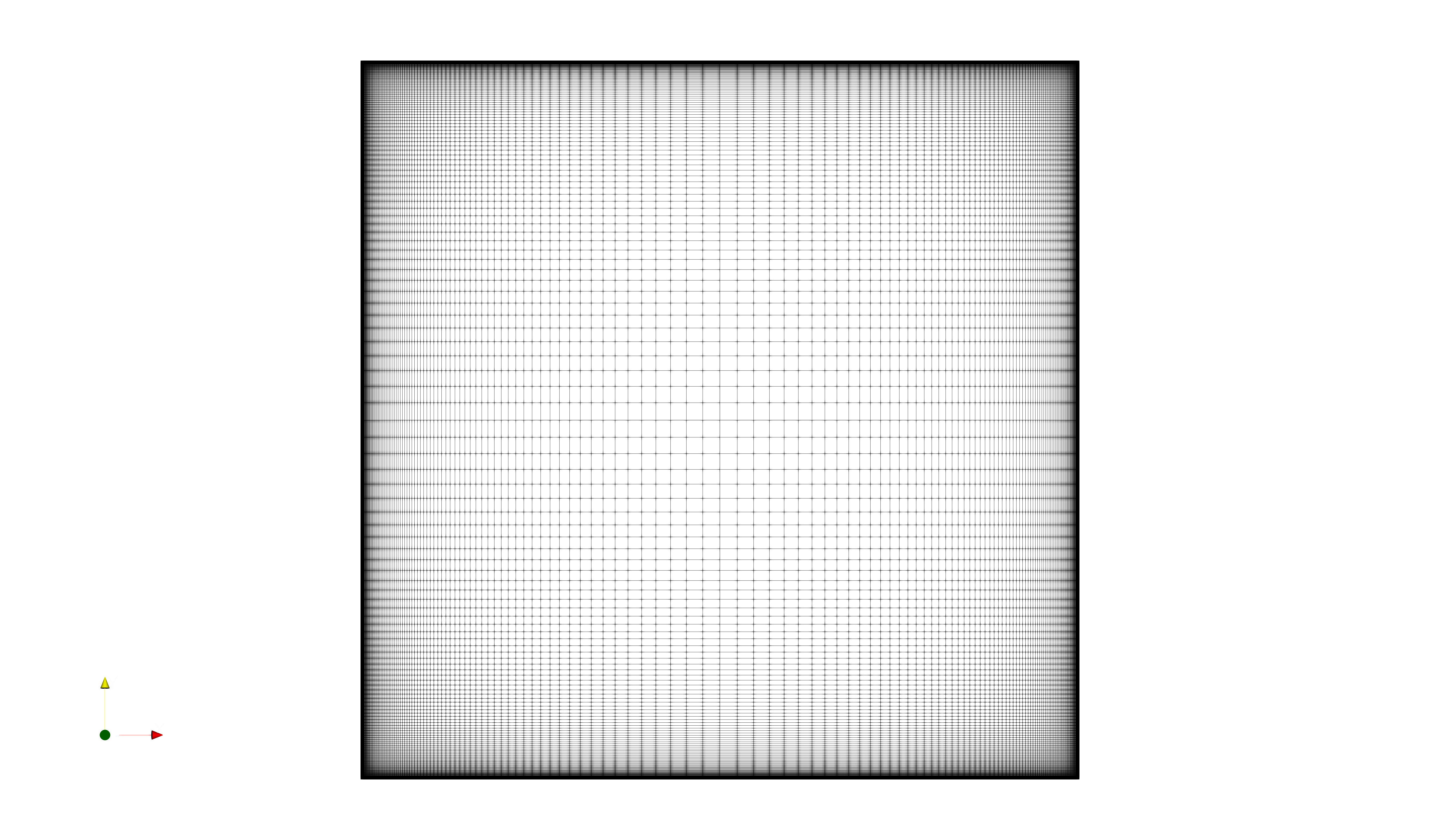}
    \caption{}
\end{subfigure}%
\begin{subfigure}{0.5\textwidth}
    \includegraphics[trim={10cm 0.1cm 10cm 0.1cm}, clip,width=\textwidth]{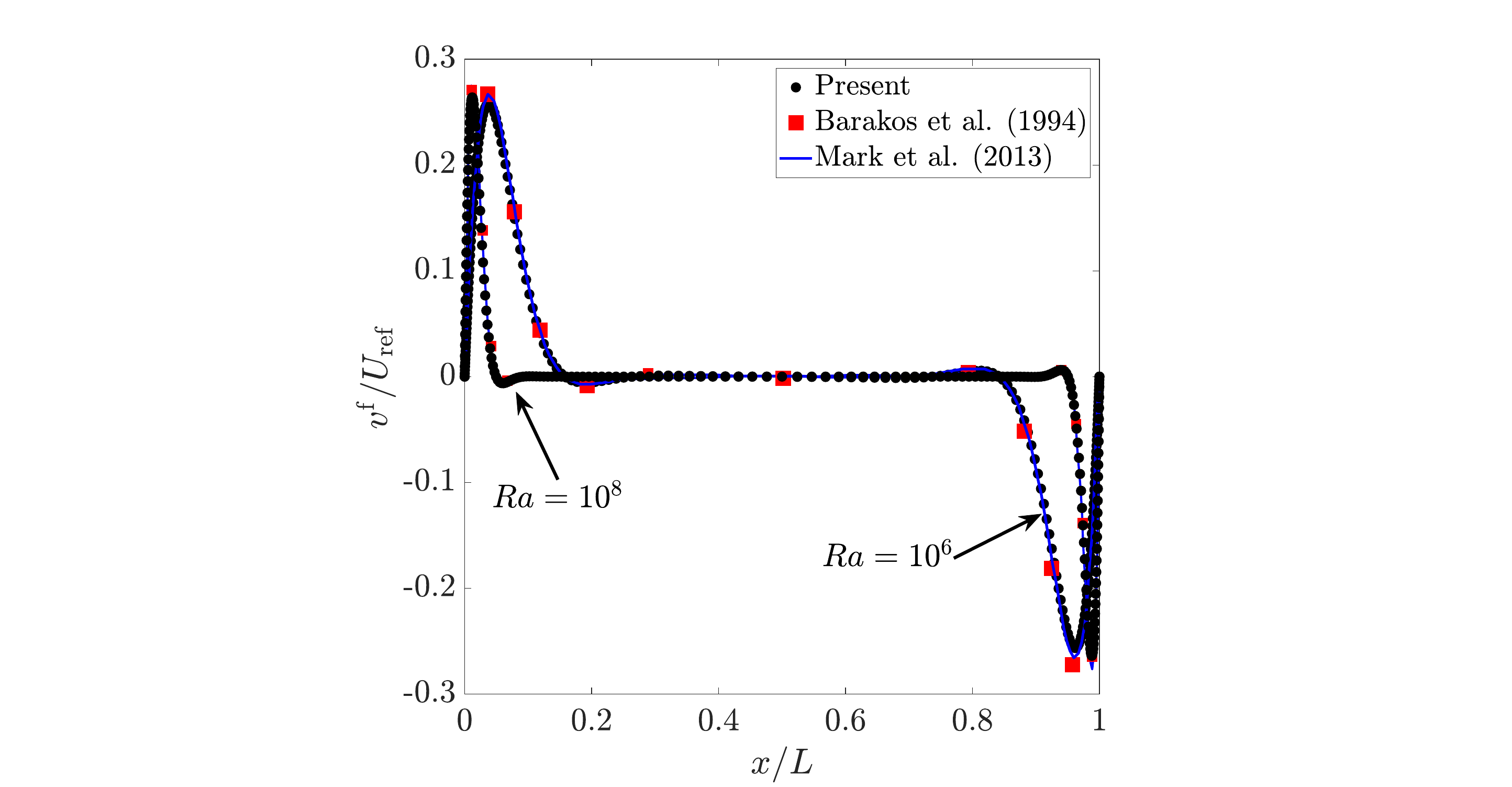}
    \caption{}
\end{subfigure}%
\caption{Differentially heated cavity test problem: (a) Computational mesh with refined boundary layer mesh, and (b) Comparison of the Y-component of fluid velocity along the horizontal centerline of the square cavity for $Ra = 10^6$ and $10^8$.}
    \label{fig:val_cavity}
\end{figure}

The coupling between the flow and energy equations is benchmarked by considering natural convection in a square cavity with differentially heated side walls. The computational setup and boundary conditions are identical to the studies by \cite{Barakos1994, MARK2013424}. The computations are carried out at Prandtl number $Pr = \mu^\mathrm{f}C_p/k^\mathrm{f} = 0.71$. Two different Rayleigh numbers (defined on the basis of the cavity length $L$ as $Ra = |\boldsymbol{g}|\beta \Delta T L^3\rho^\mathrm{f}/(\mu^\mathrm{f}\alpha)$) are chosen, viz., $Ra = 10^6$ and $10^8$. The computational mesh consists of 40401 nodes with 40000 four node quadrilateral elements. The mesh is constructed such that a boundary layer mesh is formed near to the walls of the cavity, as shown in Fig. \ref{fig:val_cavity}(a). This is accomplished by discretizing each side with 200 divisions with a progression ratio of 1.05 towards the walls. The computation is carried out by marching in time with a time step of $\Delta tU_b/L = 0.009$ and $0.09$ for $Ra = 10^6$ and $Ra=10^8$, respectively, for 10000 time steps, to ensure that the solution reaches steady state. The vertical velocity is extracted along the central horizontal line at the steady state and compared with the literature in Fig. \ref{fig:val_cavity}(b). An excellent agreement can be found for both the $Ra$ numbers.

\subsection{Natural convection around a circular cylinder within a square enclosure}

To further test the formulation to handle natural convection at various Rayleigh numbers, a heated circular body is kept inside the cavity of side length $L$ with cooled walls. This problem has been investigated thoroughly in the literature using immersed boundary techniques. However, presently, we employ the body-fitted grid approach which tracks the exact normal to the circular body resulting in more accurate results for the surface Nusselt number. The number of grid points for this problem is 73984 with 177472 three-node triangular elements. Apart from the cavity wall refinement, a boundary layer mesh is also constructed surrounding the cylinder of $0.02L$ thickness consisting of 60 layers with progression ratio of 1.2. The circular cylinder is circumferentially divided into 256 divisions. This boundary layer mesh ensures that the hydrodynamic and the thermal boundary layers are successfully captured.

\begin{figure}
\centering
\begin{subfigure}{0.45\textwidth}
    \includegraphics[trim={9cm 0.1cm 9cm 0.1cm}, clip,width=\textwidth]{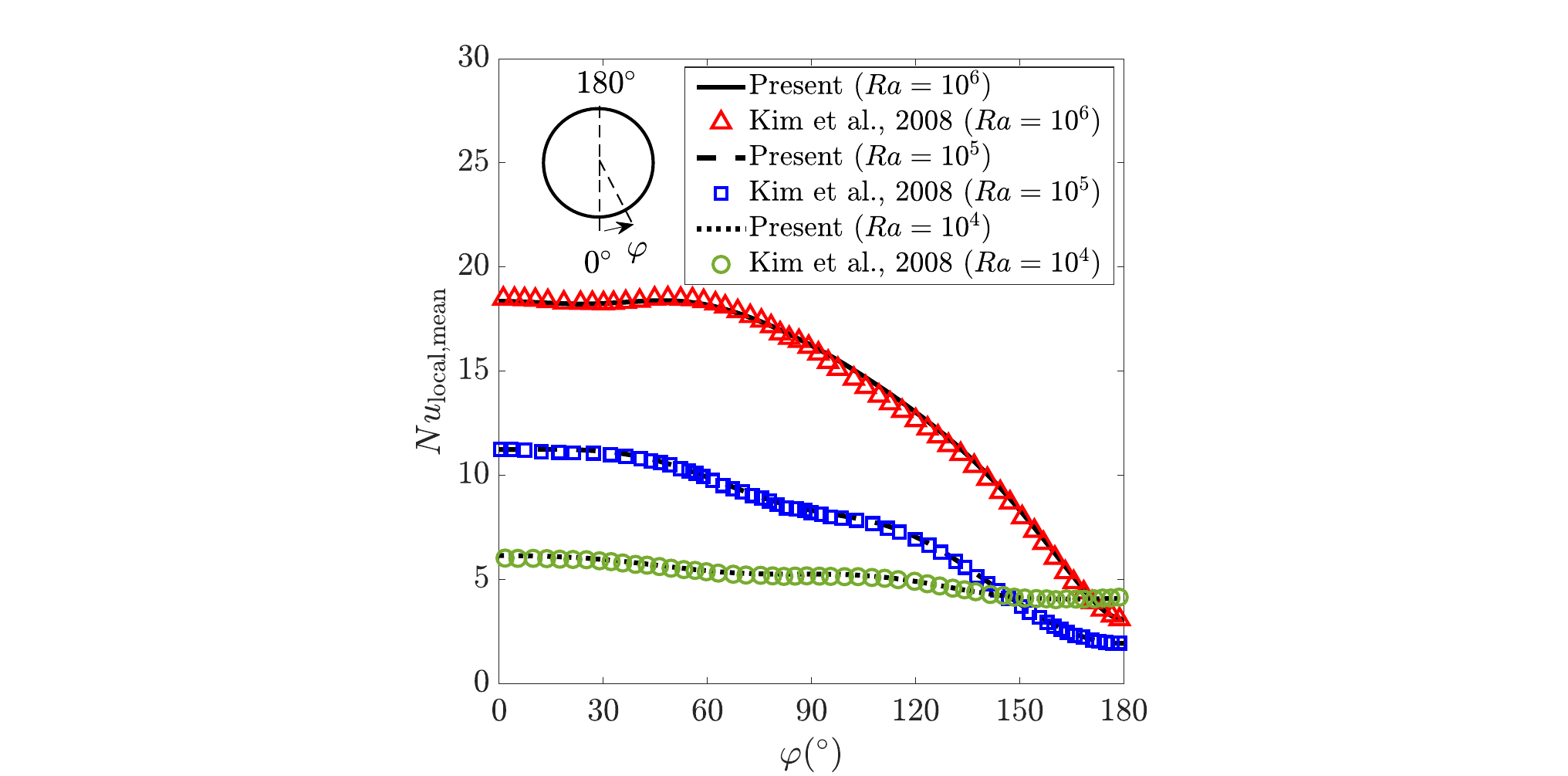}
    \caption{}
\end{subfigure}%
\begin{subfigure}{0.45\textwidth}
    \includegraphics[trim={9cm 0.1cm 9cm 0.1cm}, clip,width=\textwidth]{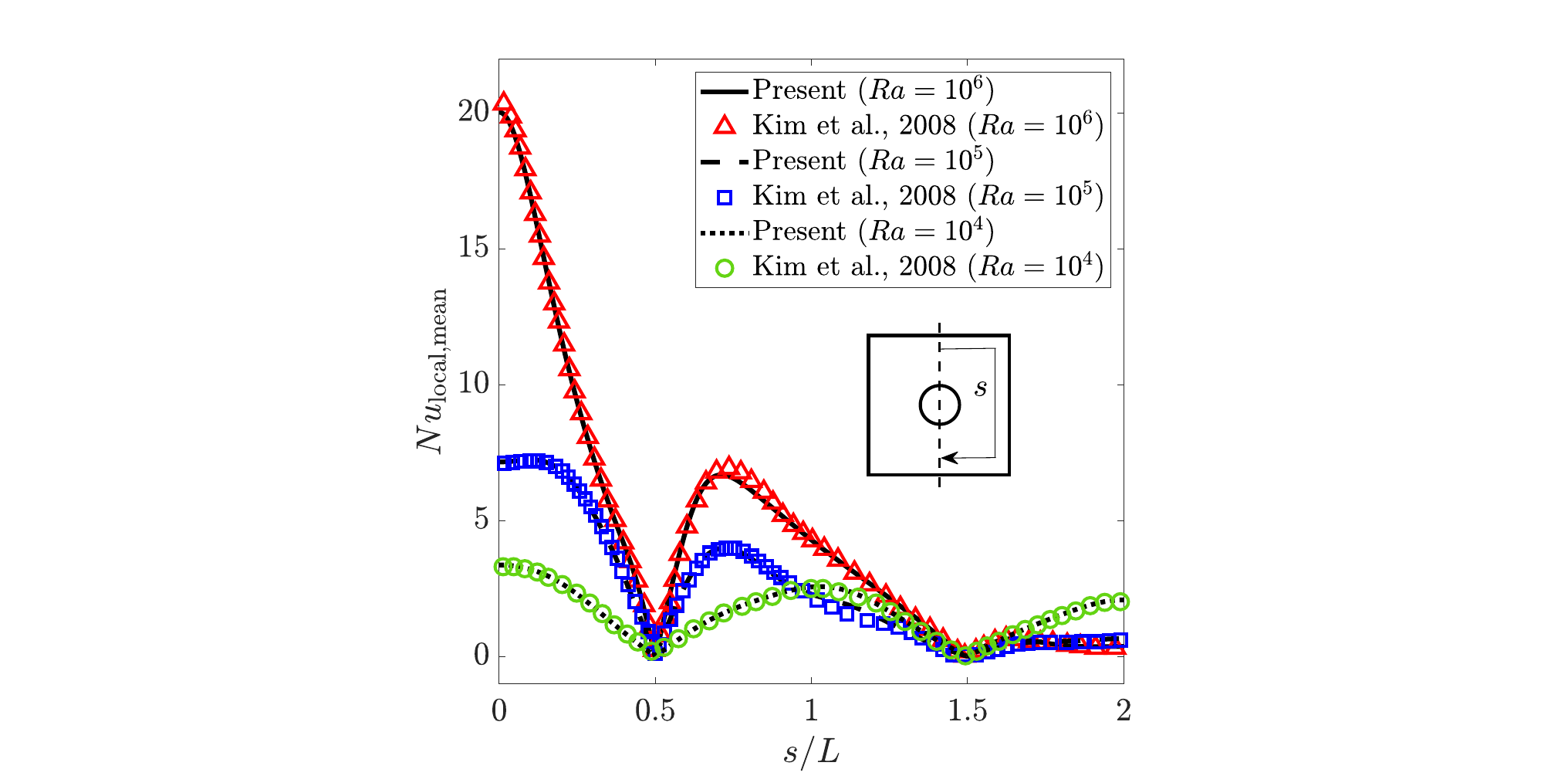}
    \caption{}
\end{subfigure}
\caption{Local variation in Nusselt number at various $Ra$ and comparison with the results by Kim et al. \cite{KIM20081888} over the surface of the (a) cylinder with $D/L=0.4$, and (b) walls.}
\label{fig:localNu_1Cyl_Ra}
\end{figure}

First, a circular cylinder with $D/L = 0.4$ positioned at the center of the cavity is considered to record the time-averaged Nusselt number integrated over the surface of the cylinder at various Rayleigh numbers (based on the cavity length scale $L$), viz., $Ra = 10^4-10^7$. The results have been tabulated in Table \ref{tab:1cyl_val} closely match with the literature for all the Rayleigh numbers. Furthermore, the variation of the local Nusselt number along the circumference of the cylinder and across the cavity walls is also depicted in Fig. \ref{fig:localNu_1Cyl_Ra} for $Ra=10^4-10^6$. A good agreement shows the mesh resolution is enough to capture the thermal physics at this range of Rayleigh numbers.

\begin{table}
\centering 
\caption{Time-averaged Nusselt number ($Nu_\mathrm{mean}$) integrated over the surface of the cylinder with $D/L=0.4$ at various Rayleigh numbers.} 
\label{tab:1cyl_val} 
\begin{tabularx}{\textwidth}{XXXXXXX}
\hline
\textbf{$Ra$} & \textbf{Present} & \textbf{Moukalled \& Acharya, 1996 \cite{Moukalled1996}} & \textbf{Kim et al., 2008 \cite{KIM20081888}} & \textbf{Park et al., 2015 \cite{PARK2015490}} & \textbf{Yang \& Ordonez, 2019 \cite{Yang2019}} & \textbf{Xu \& Choi, 2025 \cite{Xu2025}} \\ \hline
$10^4$ & 5.134 & 5.490 & 5.108 & 5.128 & 5.156 & 5.325 \\ 
$10^5$ & 7.806 & 8.377 & 7.767 & 7.836 & 8.06 & 8.095 \\ 
$10^6$ & 14.148 & 15.40 & 14.11 & 14.46 & 14.49 & 14.66 \\ 
$10^7$ & 23.795 & 26.03 & 23.50 & -- & 24.44 & 24.24 \\ \hline
\end{tabularx}
\end{table}

\begin{figure}
\centering
\begin{subfigure}{0.45\textwidth}
    \includegraphics[trim={9cm 0.1cm 9cm 0.1cm}, clip,width=\textwidth]{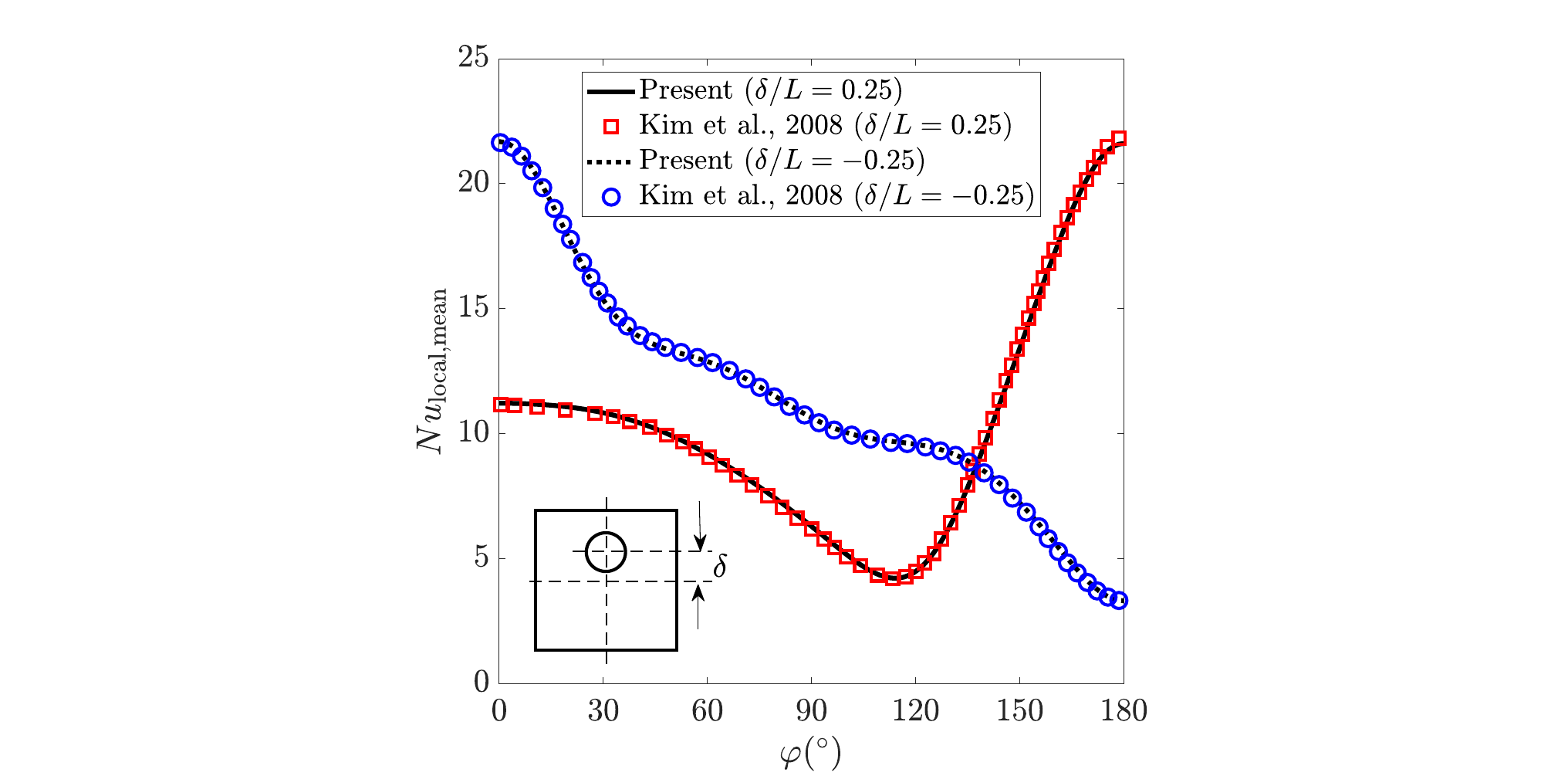}
    \caption{}
\end{subfigure}%
\begin{subfigure}{0.45\textwidth}
    \includegraphics[trim={9cm 0.1cm 9cm 0.1cm}, clip,width=\textwidth]{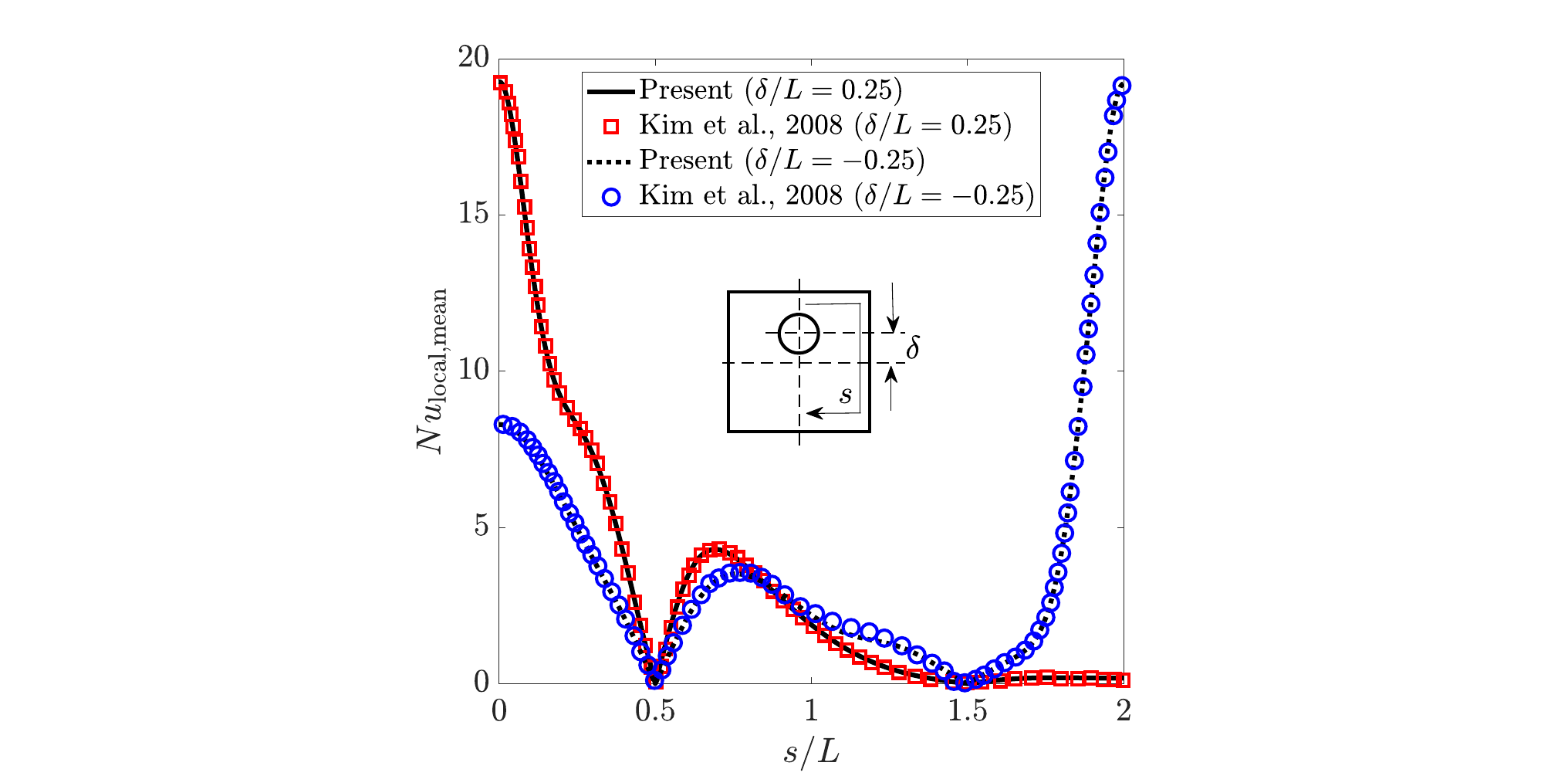}
    \caption{}
\end{subfigure}
\caption{Local variation in Nusselt number at $Ra = 10^5$ at different cylinder offset locations $(\delta/L)$ and comparison with the results by Kim et al. \cite{KIM20081888} over the surface of the (a) cylinder with $D/L=0.4$, and (b) walls.}
\label{fig:localNu_1Cyl_offset_Ra}
\end{figure}

Second, to compare the effect of the position of the cylinder inside the cavity on the local Nusselt number, the cylinder was placed at two different vertical offset locations, $\delta/L = 0.25$ and $-0.25$ at $Ra=10^5$ from the centre of the cavity. The local Nusselt number variations are shown in Fig. \ref{fig:localNu_1Cyl_offset_Ra} where excellent agreement is again found. Therefore, the fluid-thermal coupled formulation successfully captures the natural convection effects for a range of Rayleigh numbers.

\subsection{Vortex-induced vibrations of a circular cylinder subjected to mixed convection}

\begin{figure}
\centering
\includegraphics[trim =2cm 3cm 2cm  4cm, clip,width=1\textwidth]{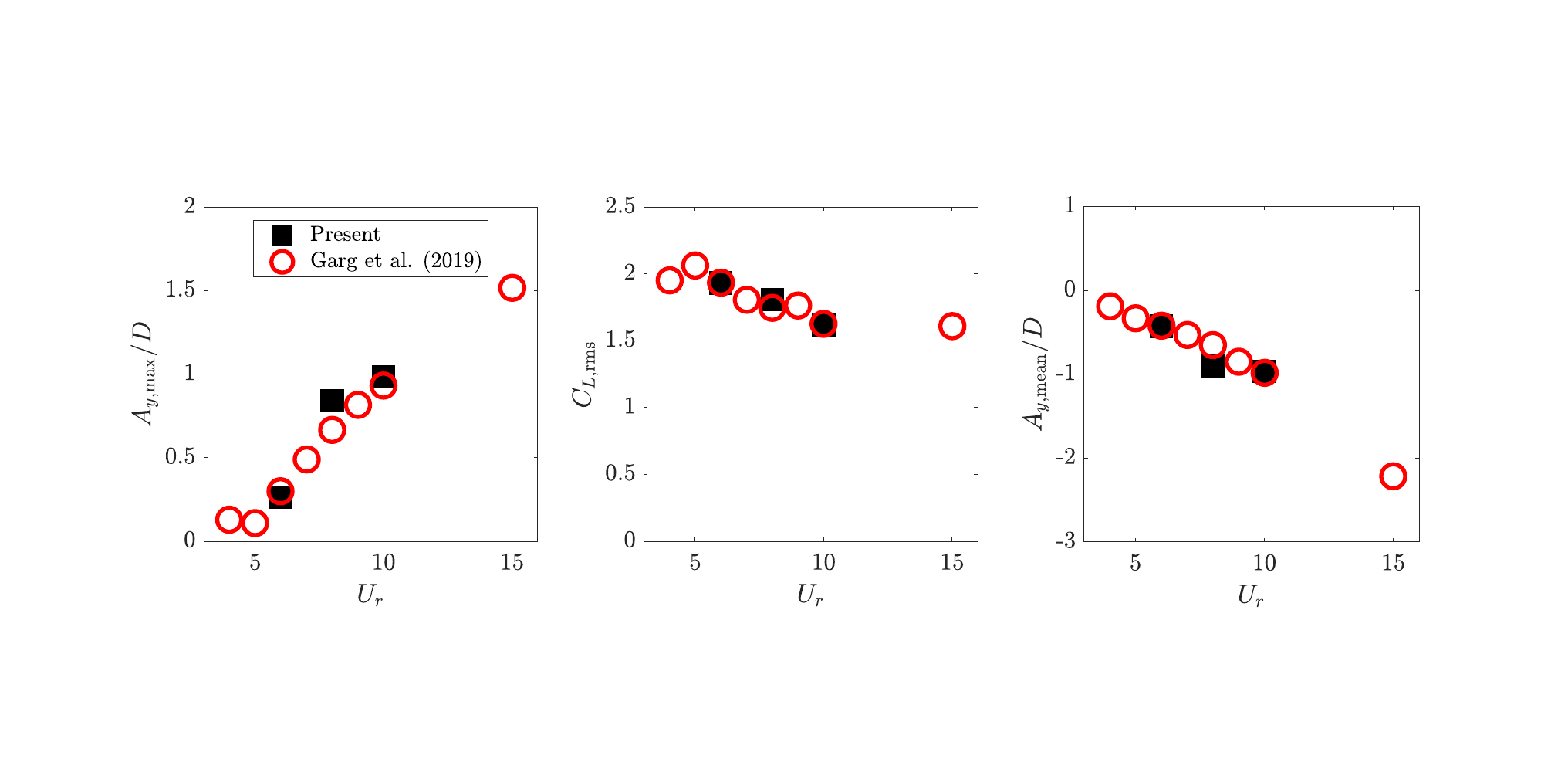}
\caption{Mixed convection and flow-induced vibrations of a circular cylinder at $Re=50$, $Pr = 7.1$, $Ri = 4.0$: Variation of maximum amplitude, root mean square of lift coefficient and mean amplitude with the reduced velocity $U_r = U_\infty/(f_n D)$.}
\label{val-Thermal-VIV}
\end{figure}

Finally, the thermal-fluid-structure coupling is benchmarked by the work carried out by Garg et al. \cite{Garg2019}. Here, an elastically mounted circular cylinder is subjected to a flow at $Re = \rho^\text{f}U_\infty D/\mu^\text{f} = 50$ and $Pr = 7.1$ with Richardson number $Ri = |\boldsymbol{g}|\beta\Delta T D/U_\infty^2 = 4.0$. The characteristic length scale here is the diameter of the circular cylinder $D$. The mass ratio is $m^* = m/(\rho^\text{f} D^2) = 2$. The reduced velocity is varied as $U_r = U_\infty/(f_n D) \in [6, 8, 10]$. Systematic mesh and time convergence study is performed by refining the circumference of the cylinder as well as the wake, as shown in Table \ref{mesh_details}. The boundary layer is refined such that the first-layer width is maintained to capture the thermal boundary layer successfully since $Pr>1$. The mesh convergence study is carried out at $U_r = 8$ with the results tabulated in Table \ref{mesh_conv}. The refinement for mesh M2 is found to have error $<2 \%$ compared to the finest mesh M4. Therefore, it is selected for the time convergence study where the time step size is varied as $\Delta tU_\infty/D = [0.02, 0.01, 0.005]$. The results are tabulated in Table \ref{time_conv} where the time step size of $\Delta t U_\infty/D=0.01$ is found to have error $<1\%$ compared to the finest time-step. Next, with mesh M2 and $\Delta tU_\infty/D=0.01$, the amplitude of the vibrations, the root mean square of the lift coefficient and the mean amplitude of the cross-flow vibrations at different reduced velocities have been compared with the literature in Fig. \ref{val-Thermal-VIV}, where the results are found to align closely with the data from Garg et al. \cite{Garg2019}. 

\begin{table}
\centering
\caption{Details about the mesh refinement carried out for the mesh convergence study.}
\begin{tabular}{lcccc}
\hline
Mesh & Wake mesh size & Nodes on the cylinder circumference & Total Elements & Total Nodes\\
\hline
M1 & 0.2  & 60  & 21,680   & 10,958\\
M2 & 0.1  & 100 & 54,380   & 27,328\\
M3 & 0.06 & 140 & 1,18,554 & 59,435\\
M4 & 0.05 & 160 & 1,58,274 & 79,305\\
\hline
\end{tabular}
\label{mesh_details}
\end{table}

\begin{table}
\centering
\caption{Mesh convergence study for flow across an elastically mounted cylinder with mixed convection. The time step size was fixed at $\Delta t U_\infty/D=0.01$.}
\begin{tabular}{lcccc}
\hline
Quantity & M1 & M2 & M3 & M4  \\
\hline
$C_{D,\mathrm{mean}}$ & 1.7756 (6.79\%) & 1.8697 (1.86\%) & 1.9021 (0.16\%) & 1.9051 \\
$C_{L,\mathrm{rms}}$ & 1.8682 (5.47\%) & 1.8083 (2.09\%)  & 1.7846 (0.76\%) & 1.7712 \\
$A_{y,\mathrm{max}}/D$ & 0.8295 (1.71\%) & 0.8402 (0.44\%) & 0.8459 (0.24\%) & 0.8439 \\
$A_{y,\mathrm{mean}}/D$ & -0.9156 (3.74\%) & -0.8979 (1.73\%) & -0.8889 (0.71\%) & -0.8826 \\
\hline
\end{tabular}
\label{mesh_conv}
\end{table}

\begin{table}[!h]
\centering
\caption{Time convergence study for flow across an elastically mounted cylinder with mixed convection for mesh M2.}
\begin{tabular}{lccc}
\hline
Quantity & $\Delta tU_\infty/D = 0.02$ & $\Delta tU_\infty/D = 0.01$ & $\Delta tU_\infty/D = 0.005$  \\
\hline
$C_{D,\mathrm{mean}}$ & 1.8709 (0.25\%) & 1.8697 (0.18\%) & 1.8663 \\
$C_{L,\mathrm{rms}}$ & 1.8031 (0.76\%) & 1.8083 (0.47\%)  & 1.8169  \\
$A_{y,\mathrm{max}}/D$ & 0.8379 (0.63\%) & 0.8402 (0.36\%) & 0.8432 \\
$A_{y,\mathrm{mean}}/D$ & -0.8958 (0.67\%) & -0.8979 (0.43\%) & -0.9018\\
\hline
\end{tabular}
\label{time_conv}
\end{table}


\bibliography{references}
\bibliographystyle{apalike}

\end{document}